\documentclass[range]{ar2e}
\usepackage{ulem}  
\usepackage{ARAstroBib,amssymb}
\usepackage{psfig}

\begin{document}

\input epsf.def   

\jname{Annual Revews of Astronomy and Astrophysics}
\jyear{2011}
\jvol{49}
\ARinfo{}

\title{Physical Properties of Galaxies from $z=2-4$}

\markboth{Shapley}{Galaxies at $z=2-4$}

\author{Alice E. Shapley
\affiliation{Department of Physics and Astronomy, University of California, Los Angeles, CA 90095-1547; aes@astro.ucla.edu}}


\begin{abstract}
The epoch of galaxy assembly from $2\leq z \leq 4$ marks a critical
stage during the evolution of today's galaxy population.
During this period the star-formation activity in the Universe
was at its peak level, and the structural patterns observed
among galaxies in the local Universe were
not yet in place. A variety of novel techniques
have been employed over the past decade to assemble
multiwavelength observations of galaxies during this important
epoch. In this primarily observational review, I present
a census of the methods used to find distant galaxies and the
empirical constraints on their
multiwavelength luminosities and colors. I then discuss what
is known about the stellar content and past histories
of star formation in high-redshift galaxies;
their interstellar contents including
dust, gas, and heavy elements; and their structural and
dynamical properties. I conclude by considering
some of the most pressing and open questions regarding the
physics of high-redshift galaxies, which
are to be addressed with future facilities.
\end{abstract}

\maketitle

\section{INTRODUCTION}
\label{sec:intro}


Understanding the detailed formation and evolution of the 
galaxies we observe today remains one of the great challenges of 
modern cosmology. An exceedingly rich variety of galaxy 
properties exists in terms of luminosity, mass, color, 
structure, gas content, heavy-element enrichment, and 
environment, many of which, in turn, are strongly correlated 
with each other. As reviewed by \citet{blanton2009}, based on 
the latest generation of wide-field surveys of the local 
Universe, astronomers have constructed an exquisitely detailed 
and statistically robust description of the galaxy population 
{\it today}. These results are crucial in terms of providing a 
boundary condition, or endpoint, for our description of the 
formation and evolution of galaxies. Ultimately, we 
strive to tell that story from beginning to end. To do so, we
must assemble additional observations and theories.


Multiple complementary approaches can be used to construct the 
history of galaxy formation. These include ab initio analytic 
models or numerical simulations; examinations of the fossil 
record contained in the ages, metallicities, and phase-space 
distributions of stars in nearby galaxies; and direct 
observations of distant galaxies, for which the cosmologically 
significant lookback time allows a probe of the Universe at an 
earlier time. In order to test theoretical models of galaxy 
formation at every time step, observations of galaxies over a 
wide range of lookback times are required. Furthermore, a robust 
translation must be performed between the observer's empirical 
quantities of luminosity, color, and velocity dispersion, and 
the theorist's physical quantities of stellar and dynamical 
mass, current star-formation rate and past star-formation 
history.


In the comparison between observations and theoretical models of 
galaxy evolution, an important recent development is the 
establishment of a precision cosmological framework. 
Observations of the cosmic microwave background radiation, 
large-scale structure, Type Ia supernova, the abundance of galaxy
clusters, and the expansion rate of the Universe, all appear to 
be well described by a cosmological model in which the Universe 
is spatially flat, with the dominant component of the 
mass-energy density in the form of dark energy, and the 
remainder consisting mostly of cold dark matter with a small 
fraction of baryons. The initial spectrum of density 
fluctuations in this model is adiabatic, Gaussian, and nearly 
scale invariant. The most recent determination of cosmological 
parameters from the {\it Wilkinson Microwave Anisotropy Probe} 
(WMAP) \citep{spergel2003} is presented in \citet{komatsu2011}, 
and highlights the fact that most parameters are determined with 
better than $5-10$\% precision.

Constraining the background cosmological parameters is a crucial 
part of understanding galaxy formation not only for converting 
apparent quantities such as flux and angular size, 
respectively, into intrinsic ones such as luminosity, 
star-formation rate, stellar mass, and physical size. Cosmological
parameters are also required for precise theoretical 
calculations whose predictions can be compared with observations, because, 
according to leading models, the underlying set of cosmological 
parameters determine how tiny primordial dark matter density 
fluctuations evolve under the influence of 
gravity into the large-scale spatial distribution of matter in 
the current Universe. In this framework, the collapsed 
perturbations of dark matter --  dark matter halos -- serve as the 
very sites of galaxy formation. Based on recent determinations 
of cosmological parameters, massive numerical simulations of the 
growth of dark matter structure have been performed 
\citep{springel2005,boylankolchin2009}. In order to compare more 
directly with observations of galaxy formation, models 
(either numerical hydrodynamic or semi-analytic)
describing the baryonic processes of gas cooling, star 
formation, and metal enrichment are also required, and these, 
too, are advancing with increased spatial resolution and 
complexity \citep[e.g.,][]{ceverino2010,somerville2008}.


In order to probe the origin of the global patterns 
observed in the current galaxy population, we must look back to 
a time before these trends were already in place. 
Furthermore, the old stellar populations of nearby early-type 
galaxies in dense environments suggest that the bulk of their stars 
formed at $z\geq 2$ \citep{thomas2005}. Therefore, catching the 
formation of spheroids ``in the act" requires observations at 
such early times. Given the strong correlation between the 
properties of the spheroidal components of galaxies and their 
central black holes, the epoch when the spheroids are forming 
holds special interest for explaining this connection. As 
described in more detail during the course of this review, the 
redshift range, $2\leq z\leq 4$, corresponding to a lookback 
time of $\sim 10-12$~Gyr, is ideal for directly observing the 
progenitors of today's fairly luminous spheroidal and disk 
galaxies while in the very process of attaining the properties 
that come to define them over the next 10 Gyr up to the present 
day. Towards the end of this redshift range, the overall level 
of ``activity" in the Universe -- both in terms of star 
formation and black hole accretion - was at its peak value. In 
contrast to what is observed in the present-day Universe, a 
significant fraction of the most massive galaxies still 
sustained active star formation. Furthermore, the Hubble 
Sequence of disk and elliptical galaxies was not yet in place, 
and the abundance of rich clusters was vanishingly small. The 
early Universe looked drastically different from its current state. 
Therefore, studying this epoch can yield important clues about the 
evolution of galaxies.


Within the past decade, there has been incredible progress in the 
study of galaxies in this important redshift range of galaxy assembly. A variety of 
novel techniques have been used to identify distant galaxies, 
and the sample of galaxies with spectroscopic redshifts at $2\leq z\leq 4$ 
now numbers  well into the thousands. Although we are far from approaching
the overwhelming statistical power of the giant local redshift 
surveys such as the Sloan Digital Sky Survey 
\citep[SDSS;][]{abazajian2009,blanton2009} and 2dF Galaxy
Redshift Survey \citep[2dFGRS;][]{colless2001}, in terms of 
number of galaxies surveyed spectroscopically, volume probed, 
and data quality, key features of the galaxy population at these 
early times are emerging. The physical properties inferred for 
distant galaxies have provided important inputs and daunting 
challenges to state of the art theoretical models of galaxy 
formation. As we look forward to the next generation of 
instrumentation on current and future large ground-based 
telescopes, as well as the James Webb Space Telescope ({\it JWST}) in 
space, and ever more sophisticated galaxy formation models, it 
is worth reviewing what is known about the physical properties 
of high-redshift galaxies at $2 \leq z \leq 4$.


This primarily observational review is constructed as follows. 
In Section~\ref{sec:technique}, we provide an overview of the 
many different techniques that have recently been employed for 
identifying high-redshift galaxies. We continue in 
Section~\ref{sec:empirical} by reviewing the global 
multiwavelength distributions in luminosity and color for 
galaxies in this redshift range. In 
Section~\ref{sec:stellarpop}, we delve into the techniques used 
to transform empirical quantities such as luminosity and color 
into physical ones relating to galaxy stellar populations. In 
particular, we focus here on stellar content and the history of 
star-formation activity, as well as the relationship between these 
quantities.  Sections~\ref{sec:ISM} and \ref{sec:structure} in turn 
consider what is known about the interstellar medium (ISM) of 
distant galaxies --  where the interstellar contents include gas, 
dust, and metals -- and their structural properties and dynamics. 
Although many gaps remain 
in our knowledge of these fundamental physical properties, as 
described in Section~\ref{sec:future}, future facilities and 
instrumentation will guide us in our quest to assemble a 
comprehensive picture of the galaxy population during this 
distant yet intriguing epoch in the history of galaxy formation.

\section{HIGH-REDSHIFT GALAXY SELECTION TECHNIQUES}
\label{sec:technique}

In this section, we provide a brief historical context for the 
recent dramatic developments in the study of high-redshift galaxies,
as well as reviewing several of the most common and complementary techniques
for identifying distant objects. 

\subsection{Historical Context}
\label{subsec:technique-history}

Over the past 10--15 years, the study of high-redshift galaxies 
has truly exploded, with an increasing number of surveys for 
systems at lookback times of order 10 Gyr. The number 
of galaxies with spectroscopic redshifts at $z>2$ is now well 
into the thousands, and the number whose multiwavelength 
photometric properties identify them as such is more than an 
order of magnitude larger. Due to the difficulties of obtaining 
optical and near-infrared (near-IR) spectra of faint objects, photometric redshifts 
have played an increasingly common role in describing the 
properties of distant galaxies -- bringing with them both the advantages
of much larger samples, and also the drawbacks of larger uncertainties in derived
galaxy properties. Also, with some exceptions like 
the VIRMOS VLT Deep Survey \citep[VVDS;][]{lefevre2005}, which 
targets galaxies for spectroscopy down to a given optical 
magnitude limit, these new results for the most part utilize 
several novel techniques for efficiently identifying distant 
galaxies with minimal contamination by systems at lower 
redshift. Although effective at isolating high-redshift galaxies, 
all of these selection methods suffer from incompleteness with 
respect to a sample defined in terms of physical quantities such 
as stellar or dynamical mass, or star-formation rate. In this 
section, we describe the landscape of galaxy surveys that have 
contributed thus far to our picture of the high-redshift galaxy 
population.

\subsection{Rest-frame Ultra-Violet Selection}
\label{subsec:technique-LBG}

\begin{figure}
\centerline{\psfig{figure=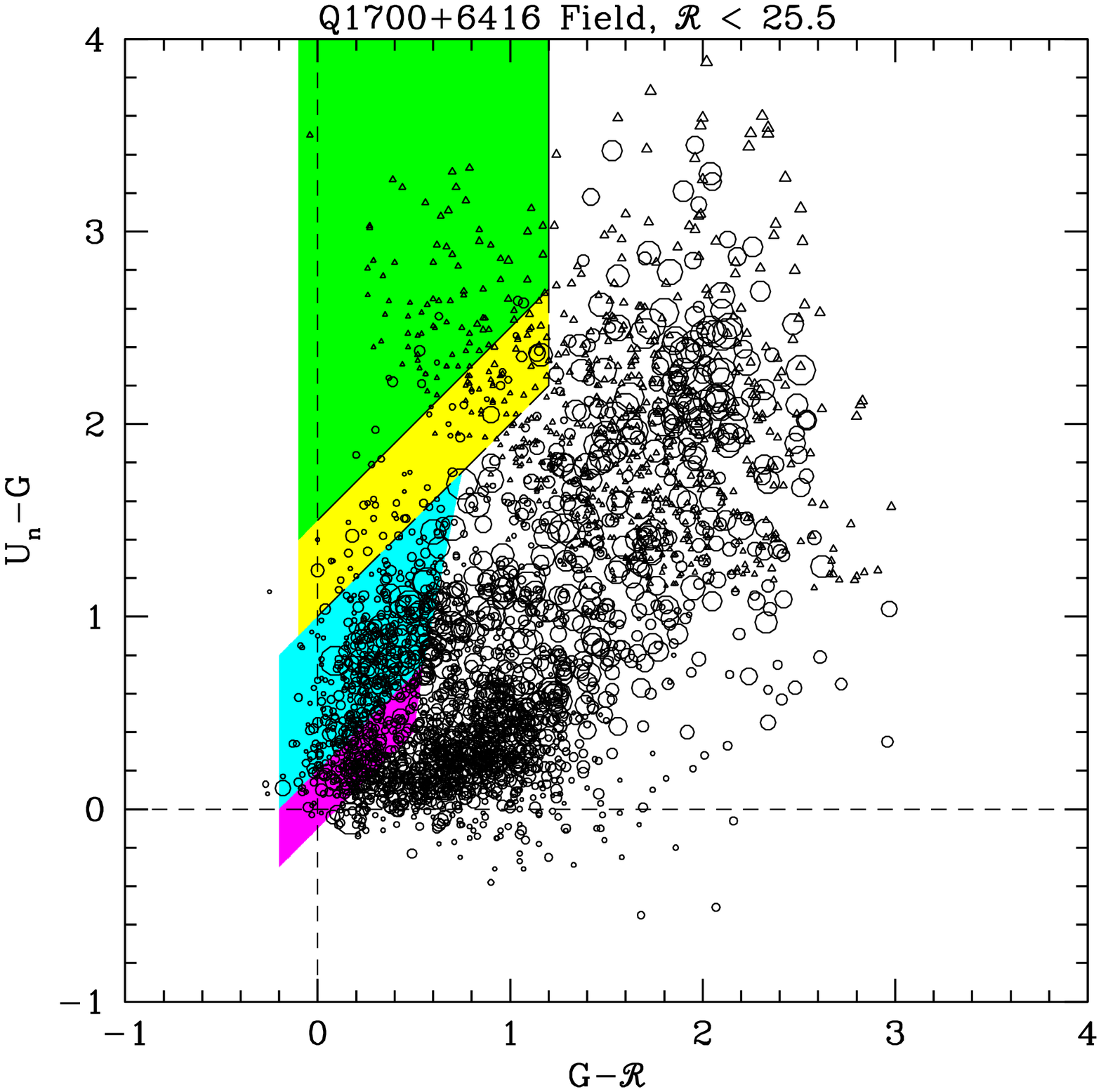,height=5in}}
\caption{\citep[From][]{steidel2004} 
Two-color ($U_n-G$ vs. $G-{\cal R}$) diagram from one of the UV-selected
survey fields, demonstrating the UV-selection technique described
in Section~\ref{subsec:technique-LBG}. 
The green and yellow shaded regions are the $z\sim 3$
LBG color selection windows, while the cyan and magenta regions
are used to select galaxies at $z\sim 2.0-2.5$ and $z\sim 1.5-2.0$,
respectively.}
\label{fig:technique-LBG-steidel2004}
\end{figure}

One method for selecting distant galaxies is based on their 
rest-frame UV colors. This method was first applied at $z\sim 
3$, and specifically exploits the combined effects of neutral 
hydrogen opacity within a star-forming galaxy and along the line 
of sight through the intergalactic medium (IGM). Accordingly, 
$z\sim 3$ star-forming galaxies with moderate amounts of dust 
extinction (less than a factor of 100 in the rest-frame UV) will 
have distinctive colors in a $UG{\cal R}$ filter system, with fairly 
flat $G-{\cal R}$ and extremely red $U-G$ colors. The ``Lyman Break 
Technique" has been used to identify thousands of galaxies  (so-called
Lyman Break Galaxies, or LBGs) at 
$z\sim 3$ \citep{steidel1996b,steidel2003}, and, using different 
filter sets, at $z\sim 4$ and $5$ \citep{ouchi2004a}. At $z\sim 
6$ and beyond, even redder sets of three filters have been used 
to identify star-forming galaxies 
\citep{bouwens2007,oesch2010,bouwens2010}. In these latter cases, 
however, the main spectral break between the bluest and middle 
filter arises due to hydrogen Ly$\alpha$ opacity in the IGM, as 
opposed to opacity at the Lyman limit. Rest-frame UV selection 
using the initial set of $UG{\cal R}$ filters has also been extended 
down to lower redshift \citep{steidel2004,adelberger2004}, where
galaxies at $1.4 \leq z \leq 2.5$ are isolated due to 
a {\it lack} of significant spectral break. Their fairly flat 
rest-frame UV colors, modulated at $2\leq z \leq 2.5 $ only by 
Ly$\alpha$ forest line blanketing in the $U$-band, also prove 
distinctive. Figure~\ref{fig:technique-LBG-steidel2004},
from \citet{steidel2004},
provides an illustration of the rest-frame UV selection criteria
in $UG{\cal R}$ color space, tuned to find galaxies at $1.5\leq z \leq 3.5$.

Given the criterion of detection in the rest-frame UV,
the techniques described above necessarily select galaxies with ongoing
star formation and are not sensitive to passive galaxies. 
Furthermore, the windows in color-color selection
space exclude galaxies whose rest-frame UV
continuum shape is indicative of significant dust reddening.
Ground-based rest-frame UV surveys are also typically characterized
by a rest-frame UV (observed optical) flux limit. Detection
in rest-frame UV (observed optical) bands results in objects
well suited to optical spectroscopic follow-up. Accordingly,
successful spectroscopic follow-up of high-redshift galaxies has been 
weighted towards rest-frame-UV-selected samples, probing
galaxies with active ongoing star formation.

\subsection{Rest-frame Optical/Infrared Selection}
\label{subsec:technique-DRGBzK}

\begin{figure}
\centerline{\psfig{file=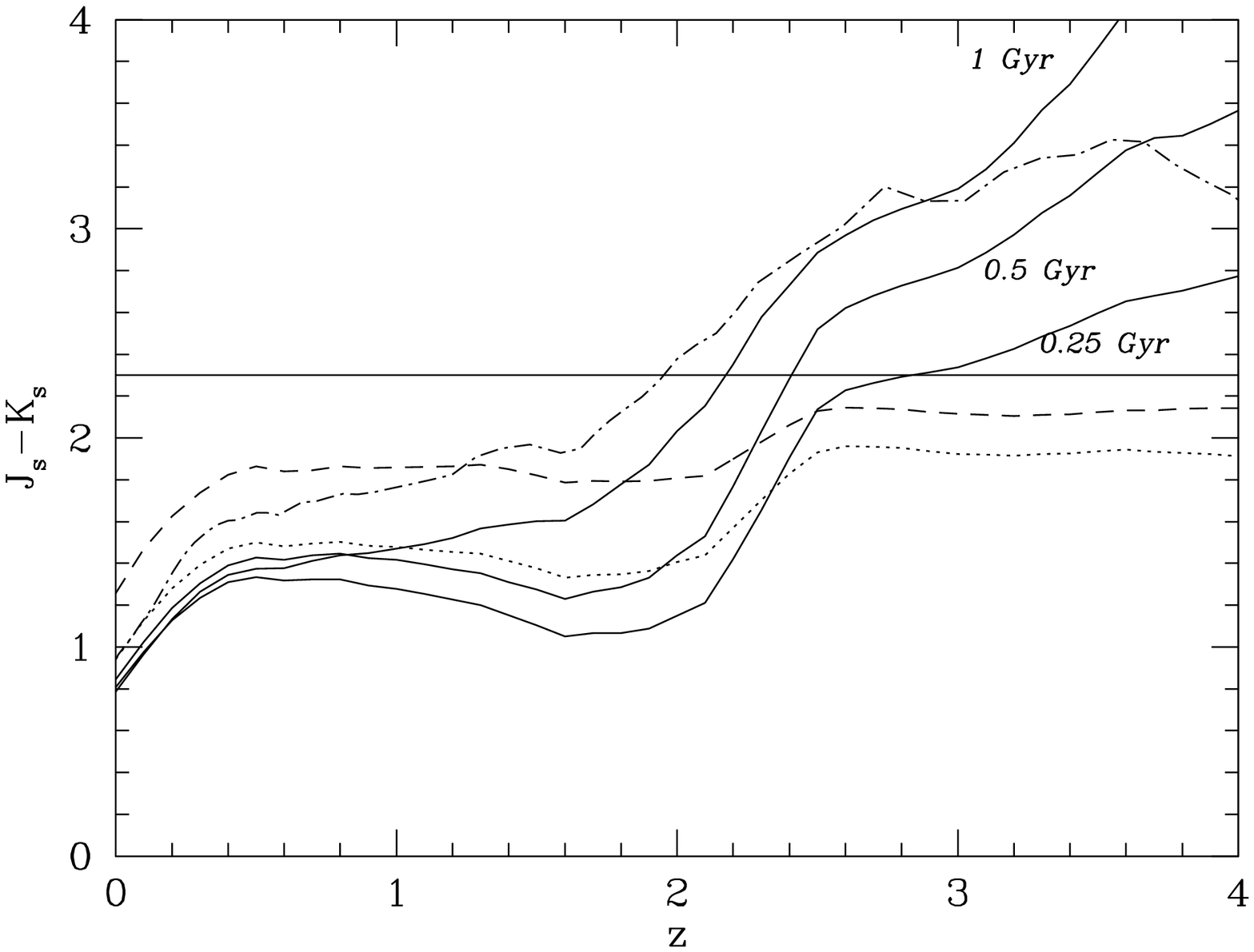,height=2.7in}\psfig{file=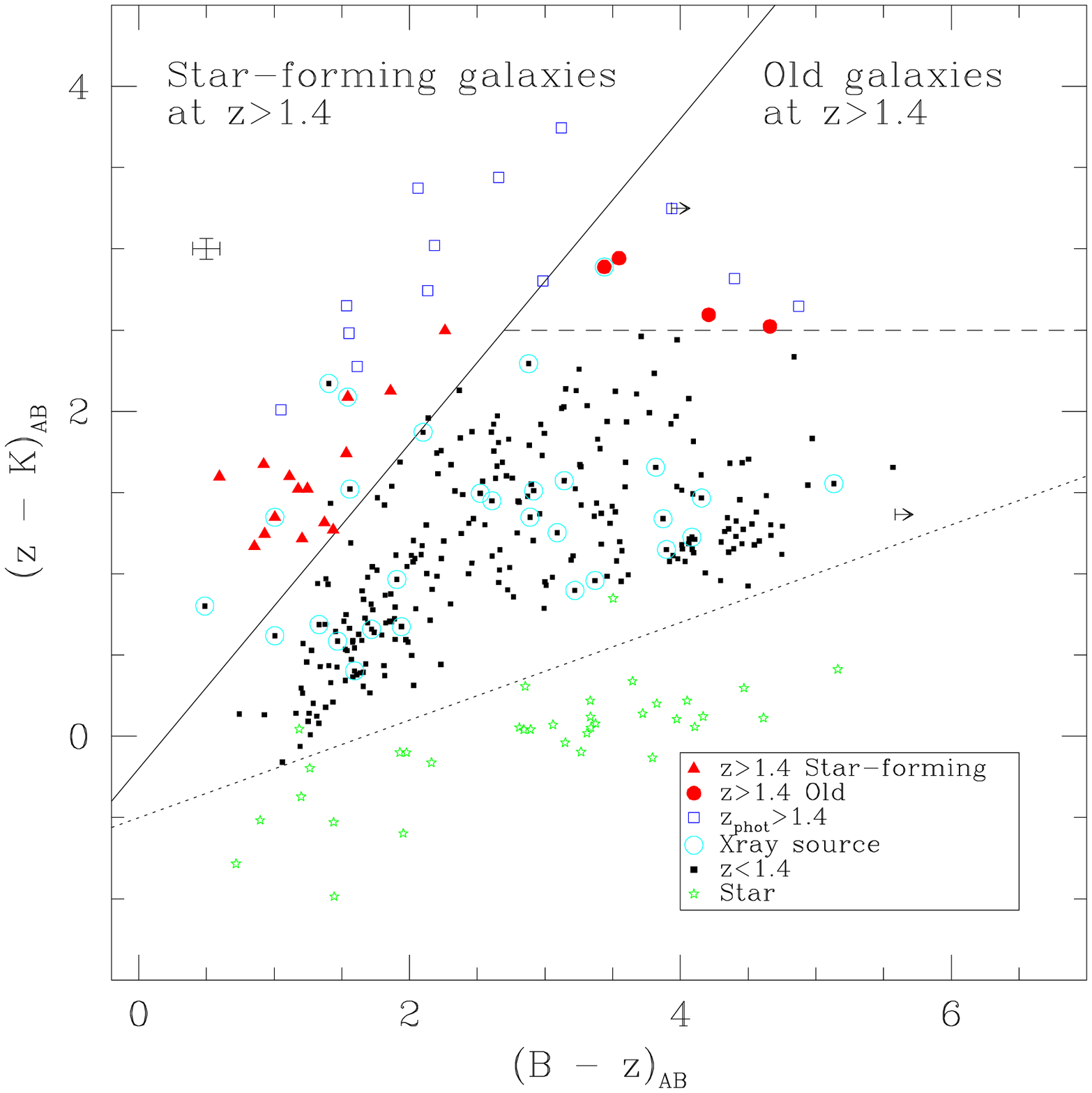,height=2.7in}}
\caption{{\bf Left:} \citep[From][]{franx2003}
$J_s-K_s$ (Vega) color as a function of redshift for
several different galaxy spectral types, illustrating the DRG selection technique.
Solid curves indicate
single-age stellar populations with ages of $0.25$, $0.5$, and
$1$~Gyr. $J_s-K_s$ colors exceed
a value of 2.3 at $z>2$ as a result of either the Balmer or
4000~\AA\ break moving into the $J_s$ band. Dotted and dashed
curves indicate models with continuous star formation with ages and reddenings
of 1 Gyr, $E(V-B)=0.15$, and 100 Myr, $E(B-V)=0.5$, respectively.
The dash-dotted curve
indicates the color evolution of a single-burst population that formed at 
$z=5$, and it also satisfies the color criterion above $z= 2$.
{\bf Right:} \citep[From][]{daddi2004}
Two-color ($z-K$ vs. $B-z$) diagram for galaxies
in the GOODS field, illustrating the $BzK$ selection technique.
Solid triangles represent star-forming
galaxies at $z>1.4$, solid circles represent passive
galaxies at $z>1.4$, empty squares are objects with no measured
spectroscopic redshift but with photometric redshift $z_{phot}>1.4$.
X-ray sources are circled, solid squares are objects at $z<1.4$,
and star symbols indicate Galactic objects.
The solid and dashed lines indicate the regions of $BzK$ color
space used to identify star-forming and passive galaxies at $1.4\leq z \leq 2.5$.
}
\label{fig:technique-DRGBzK-franx2003daddi2004}
\end{figure}

Other techniques are tuned to select galaxies on the basis of 
their rest-frame optical colors and are based on a detection at 
observed near- or mid-infrared (mid-IR) wavelengths. For these 
techniques, the relevant spectral break is either the Balmer 
break at $\sim 3650$\AA, which arises when the integrated 
stellar spectrum from a galaxy at these wavelengths indicates 
the overall spectral shape of A-stars, or else the 4000~\AA\
break, which reflects the absorption from ionized metals in the 
atmospheres of late-type stars. Although both breaks are 
indications of maturity in stellar populations, they are by no 
means equivalent in terms of the underlying stellar populations 
that cause them.  The Balmer break appears in stellar 
populations featuring ongoing star-formation over sustained 
timescales ($>100$~Myr), or post-starburst populations 
$0.3-1$~Gyr since the cessation of star formation. The 4000~\AA\ 
break is strongest in passive stellar populations in which the 
current level of star formation has been negligible for more 
than 1~Gyr. Isolating $z\geq 2$ galaxies on the basis of their 
rest-frame optical breaks requires deep, near-IR
photometry. As described in \citet{franx2003} and 
\cite{vandokkum2003}, a threshold of $J_{Vega}-K_{Vega} >2.3$ is 
effective at identifying objects that dominate the high-mass 
regime of the $z>2$ stellar mass function \citep{kriek2008a}. 
Figure~\ref{fig:technique-DRGBzK-franx2003daddi2004} \citep[left, from][]{franx2003}
demonstrates the sensitivity of the $J_{Vega}-K_{Vega}$ color
to mature stellar populations at $z>2$.
These Distant Red Galaxies (DRGs) in fact typically have significant 
dust obscuration ($A_V>1$) and active star-formation rates 
($\geq 100 M_{\odot}\mbox{ yr}^{-1}$) \citep{papovich2006}, although some show little 
evidence for ongoing star formation \citep{vandokkum2008b}. 
Regardless, the red rest-frame UV to optical colors of DRGs down 
to current near-IR limits indicate typical stellar masses in 
excess of $10^{11} M_{\odot}$. The main limitation of such 
studies is the limited amount of spectroscopic follow up, due to 
the optical faintness (typically $R\geq 25$) of DRGs and the 
difficulty of obtaining large samples of near-IR spectra 
\citep[but see, e.g.,][]{kriek2008a}. 

Another common technique for isolating galaxies at $1.4\leq z 
\leq 2.5$ consists of the so-called ``$BzK$" method \citep{daddi2004}. 
As shown in Figure~\ref{fig:technique-DRGBzK-franx2003daddi2004} (right), $BzK$ 
galaxies are identified in $K$-selected samples of galaxies on 
the basis of their colors in the $z-K$ versus $B-z$ plane. 
Galaxies with fairly blue $B-z$ colors and red $z-K$ colors are 
selected as star-forming $z\sim 2$ systems (``$sBzK$"), due to the presence 
of a Balmer break. At the same time, quiescent systems (``$pBzK$") at the 
same redshift are identified on the basis of red colors in both 
$B-z$ and $z-K$. Although fairly general in terms of selecting both 
star-forming and quiescent galaxies, this method misses the 
youngest star-forming galaxies at $z\sim 2$, which lack a 
significant Balmer or 4000~\AA\ break \citep{reddy2005}. While
the star-forming $sBzK$ galaxies have rest-frame UV colors that 
are redder on average than the corresponding $UV$-selected 
galaxies at $z\sim 2$, there is significant overlap between 
these two photometric selection technique down to a fixed 
$K$-band magnitude limit -- much more than between the 
UV-selected and DRG samples, which overlap at only the 10\% 
level \citep{reddy2005}. Other surveys tuned to find 
high-redshift galaxies based on their rest-frame optical or 
near-IR properties include the Gemini Deep Deep Survey 
\citep[GDDS; ][]{abraham2004} and the Galaxy Mass Assembly 
ultra-deep Spectroscopic Survey \citep[GMASS;][]{cimatti2008}.

\subsection{Submillimeter / Mid-Infrared Selection}
\label{subsec:technique-IR}

The overall increase in the level of 
star-formation activity in the Universe at earlier times results 
in an increased abundance of extreme, bolometrically 
ultra-luminous systems. These systems emit the bulk of their 
radiation at rest-frame far-IR wavelengths, because of copious 
amounts of dust obscuring their star-formation and AGN activity. 
Systems with bolometric luminosities greater than $L=10^{12} L_{\odot}$
are commonly referred to as Ultra-luminous Infrared Galaxies (ULIRGs).
Submillimeter and mid-IR instrumentation sensitive 
to cool and hot dust, respectively, have enabled the 
identification of these extreme high-redshift ULIRGs on the 
basis of their reprocessed emission, while multiwavelength 
imaging and spectroscopic follow-up has elucidated the range of 
their properties. The Submillimetre Common-user Bolometer Array 
(SCUBA) has been used to identify bolometrically-luminous 
submillimeter galaxies \citep[SMGs;][]{smail1997}, and the 
largest set of spectroscopically-confirmed such objects was 
obtained by following up the subset of sources with both $F_{850 
\mu m} \geq 5 $~mJy and {\it Very Large Array} (VLA)
1.4~GHz fluxes greater than $F_{1.4 
\mbox{ GHz}} \sim 30 \mu$Jy \citep{chapman2003,chapman2005}. The radio 
fluxes were used to obtain the precise ($1"-2$") positions 
required for spectroscopic observations, which were not 
achievable with the coarse (15") SCUBA beam. These luminous SMGs 
with radio counterparts have a median redshift of $z=2.2$, although, given that the 
requirement of a radio flux detection recovers $\sim 50$\% of 
the $F_{850 \mu m} \geq 5 $~mJy population, their redshift 
distribution may not be fully representative of the luminous SMG 
population as a whole. 

The Multiband Imaging Photometer for 
Spitzer (MIPS) onboard the {\it Spitzer Space Telescope} has 
been used to identify high-redshift dusty sources on the basis 
of their brightness at $24 \mu$m and faintness at optical 
wavelengths. \citet{yan2007} and \citet{dey2008} present such 
samples, identified using slightly different criteria, but both 
based on similar criteria of detection at $24 \mu$m, with a 
large ratio of mid-IR to optical flux. The majority of these 
objects are at $1.5 \leq z \leq 3$, with comparable space 
densities ($\sim 10^{-5} \mbox{ Mpc}^{-3}$) to those of SMGs. On 
the other hand, the $24 \mu$m-selected sources are characterized 
by warmer dust temperatures and a higher frequency of AGN 
signatures at mid-IR wavelengths than SMGs selected
at longer wavelengths (see Section~\ref{subsec:ISM-dustem}). Quantifying the relative
contributions of star-formation and AGN activity in powering
the extreme luminosities of both mid-IR and submillimeter-selected ULIRGs
will enable us to isolate the underlying nature of these sources.
Another key goal consists of constraining the
relative importance of major mergers and 
smooth mass accretion as triggers for the ULIRG phase.

\subsection{Narrowband Selection}
\label{subsec:technique-LAE}

\begin{figure}
\centerline{\psfig{figure=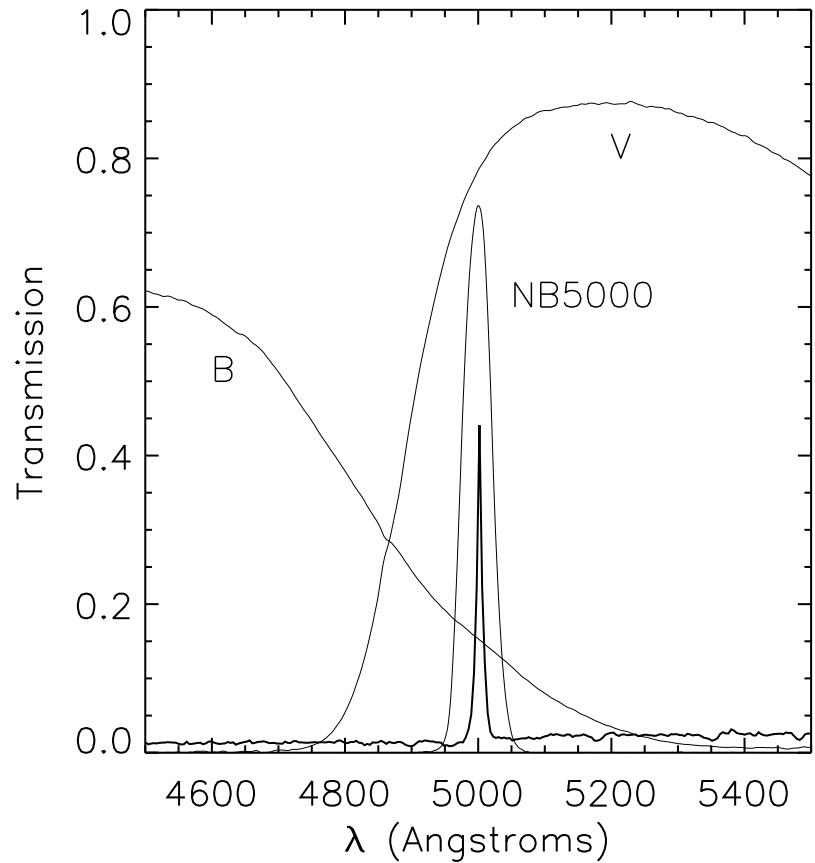,height=5in}}
\caption{\citep[From][]{gronwall2007}
Filter bandpasses used for
narrowband selection of LAEs at $z=3.1$, where
a spectrum of a typical 
$z=3.1$ Ly$\alpha$-emitting galaxy (overlaid for comparison)
would be detected as having a red broadband minus narrowband color.
This technique preferentially selects objects with bright line emission
(and, often, faint continua).
}
\label{fig:technique-LAE-gronwall2007}
\end{figure}

In contrast to the identification of high-redshift objects on 
the basis of broadband spectral shape, the use of a narrowband 
filter tuned to the redshifted wavelength of a specific emission 
line is effective at isolating objects with large emission-line 
equivalent widths. The most common emission line for which 
narrowband filters are designed is hydrogen Ly$\alpha$. As
shown in Figure~\ref{fig:technique-LAE-gronwall2007} \citep[from][]{gronwall2007},
based on images through both the narrowband filter and a broadband filter 
close in wavelength, objects with red broadband minus narrowband 
colors are flagged as Ly$\alpha$ line emitters (LAEs). More than 
2000 LAEs have been identified with 
ground-based facilities at $2\leq z \leq 8$ 
\citep[e.g.,][]{cowie1998,rhoads2000,ouchi2008,gronwall2007,nilsson2011}. 
The star-forming LAEs tend to be significantly fainter on 
average than the UV-continuum-selected objects described in 
Section~\ref{subsec:technique-LBG}, and therefore offer a probe of the 
faint end of the luminosity function. On the other hand, the 
faint nature of these objects leads to a challenge in assembling 
high signal-to-noise (S/N) multiwavelength imaging and spectra 
for individual LAEs (with information 
other than a measurement of Ly$\alpha$ emission),
hindering the determination of their relationship to other galaxy 
populations at similar redshifts.

\section{EMPIRICAL PROBES OF THE HIGH-REDSHIFT GALAXY POPULATION}
\label{sec:empirical}

Before reviewing what is known about the stellar and 
interstellar content of distant galaxies, we must consider the 
empirical measurements from which these physical properties are 
inferred. The observables here are distributions in 
luminosity and color, which offer some of the most basic and 
fundamental probes of a galaxy population. In addition to 
traditional optical photometry tracing the rest-frame 
ultraviolet, our view of the global photometric properties of 
galaxies at $z\geq 2$ is now based on deep, near-IR surveys from 
the ground using wide-field imagers on $4$-meter-class 
telescopes (e.g., KPNO/NEWFIRM, UKIRT/WFCAM, CTIO/ISPI, 
Palomar/WIRC) and narrow- and wide-field imagers on $8-10$-meter 
class telescopes (e.g., Keck/NIRC, VLT/ISAAC, Subaru/MOIRCS, 
VLT/HAWK-I), and mid- and far-IR surveys using {\it Spitzer} and 
{\it Herschel} in space. Multiwavelength observations have, 
therefore, granted us a window into the luminosity and color 
distributions of high-redshift galaxies spanning from the 
rest-frame UV through the rest-frame far-IR.

\subsection{Luminosity Functions}
\label{subsec:empirical-LF}

The galaxy luminosity function offers constraints
on the overall abundance of objects, as well as the integrated
luminosity density at a given wavelength. As such, the luminosity
function provides a key observational baseline for each
redshift at which it is measured.

\subsubsection{REST-FRAME UV LUMINOSITY FUNCTIONS}
\label{subsubsec:empirical-LF-UV}
Some of the first luminosity functions to be measured for high-redshift 
galaxies were based on optical observations of 
rest-frame UV-selected $z\sim 3$ and $z\sim 4$ LBGs 
\citep{steidel1999}, probing rest-frame wavelengths of 
$\lambda\sim 1700$\AA. This work highlighted
the importance of dust extinction for converting 
the rest-frame UV luminosities of star-forming galaxies into 
unobscured values, in order to obtain dust-corrected 
star-formation rates. We will return to this point in 
Sections~\ref{sec:stellarpop} and \ref{subsec:ISM-dustext}, when we 
consider the star-formation rates and dust content of 
high-redshift galaxies. Rest-frame UV luminosity functions for 
large samples of distant galaxies have now been estimated 
by several different groups, from $z\sim 2$ all the way 
out to $z\sim 8$ \citep{bouwens2010}. As in the local Universe,
the galaxy luminosity function at high redshift
is typically parameterized using the 
Schechter form of a power law multiplying an exponential function \citep{schechter1976}.
The associated free parameters to constrain are the characteristic
luminosity, $L^*$ (or $M^*$ in the space of absolute magnitude),
the faint-end slope, $\alpha$, and the overall normalization,
$\Phi^*$.

\begin{figure}
\centerline{\psfig{figure=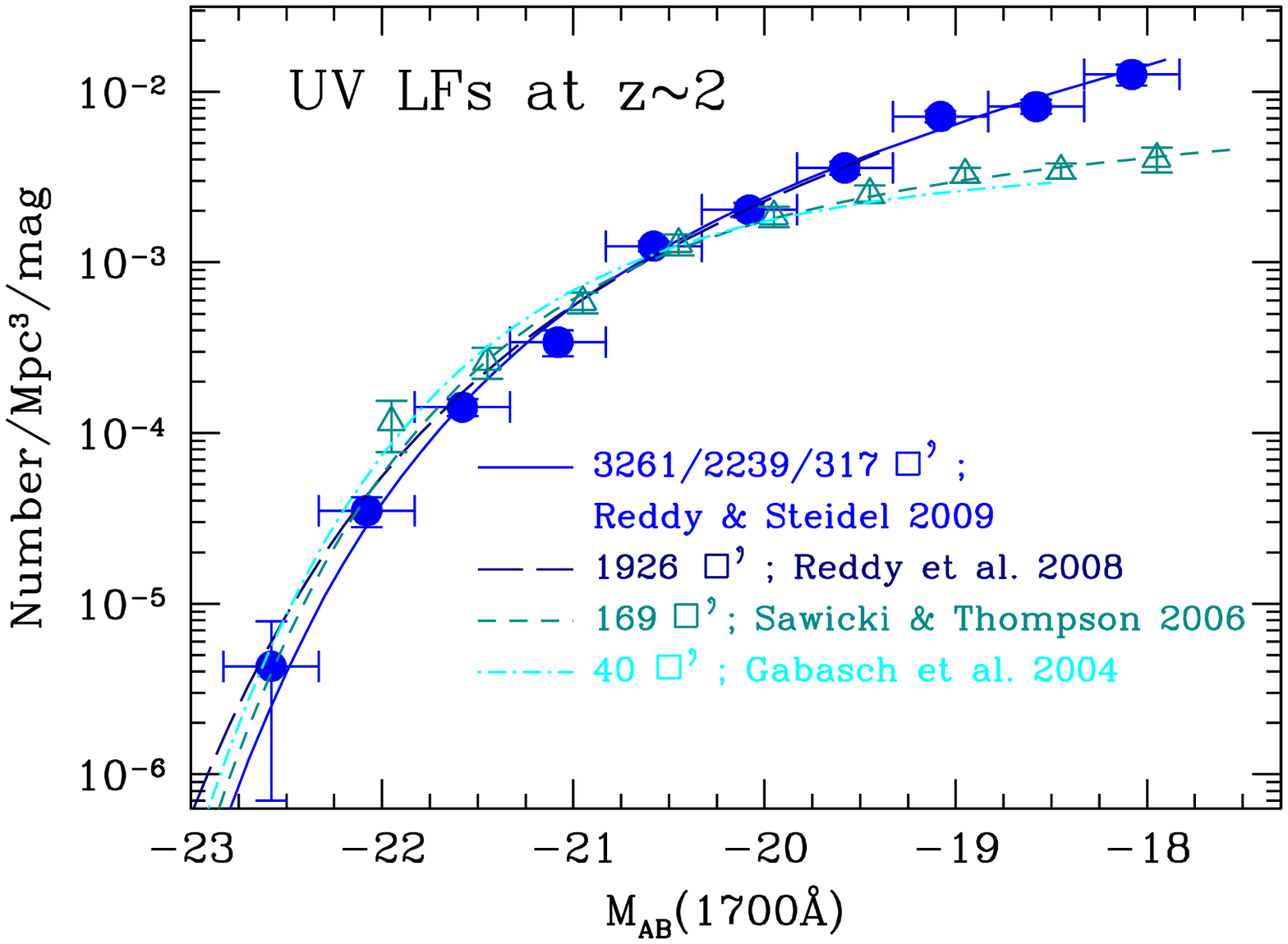,height=1.8in}
\psfig{figure=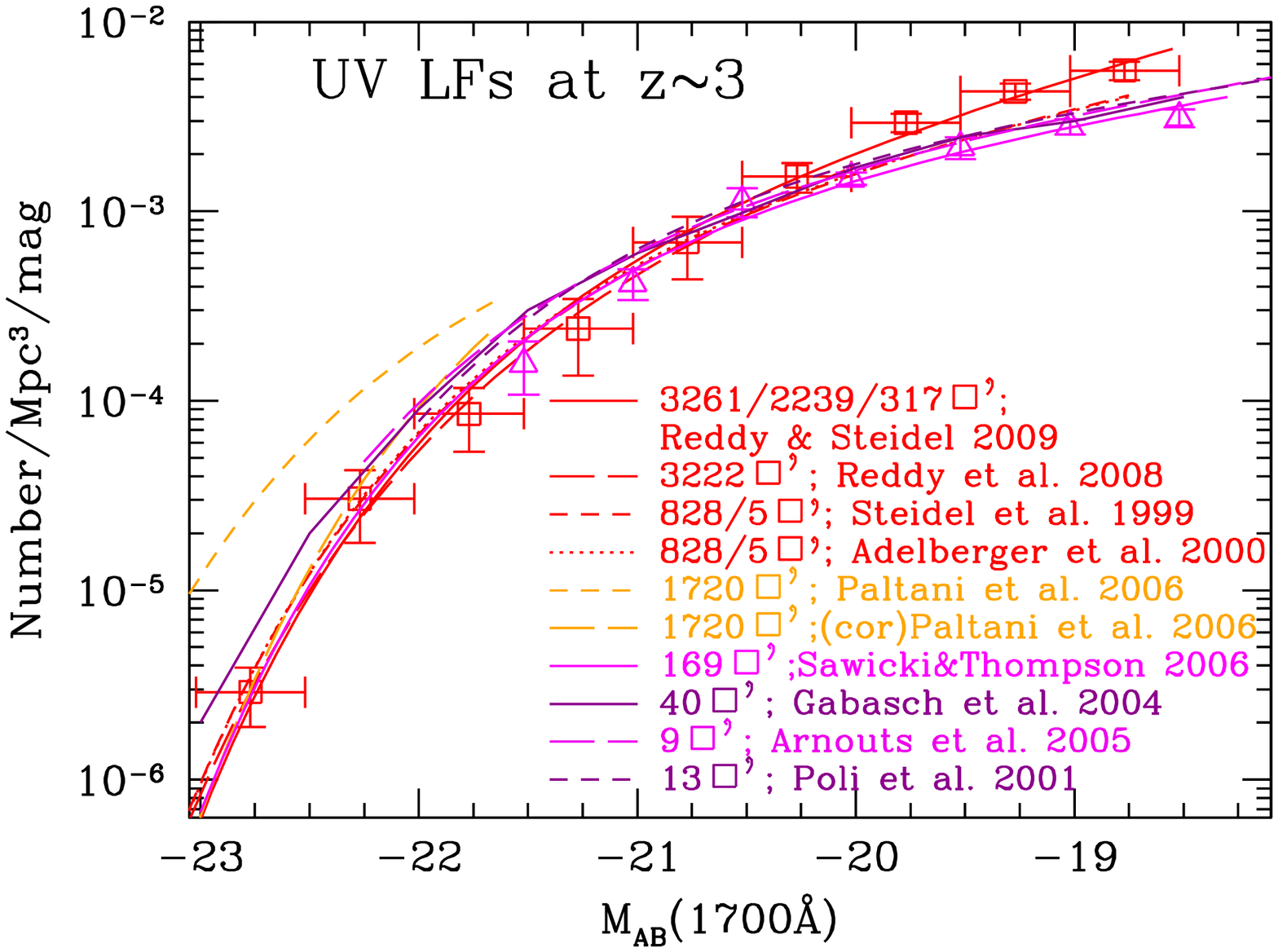,height=1.8in}
\psfig{figure=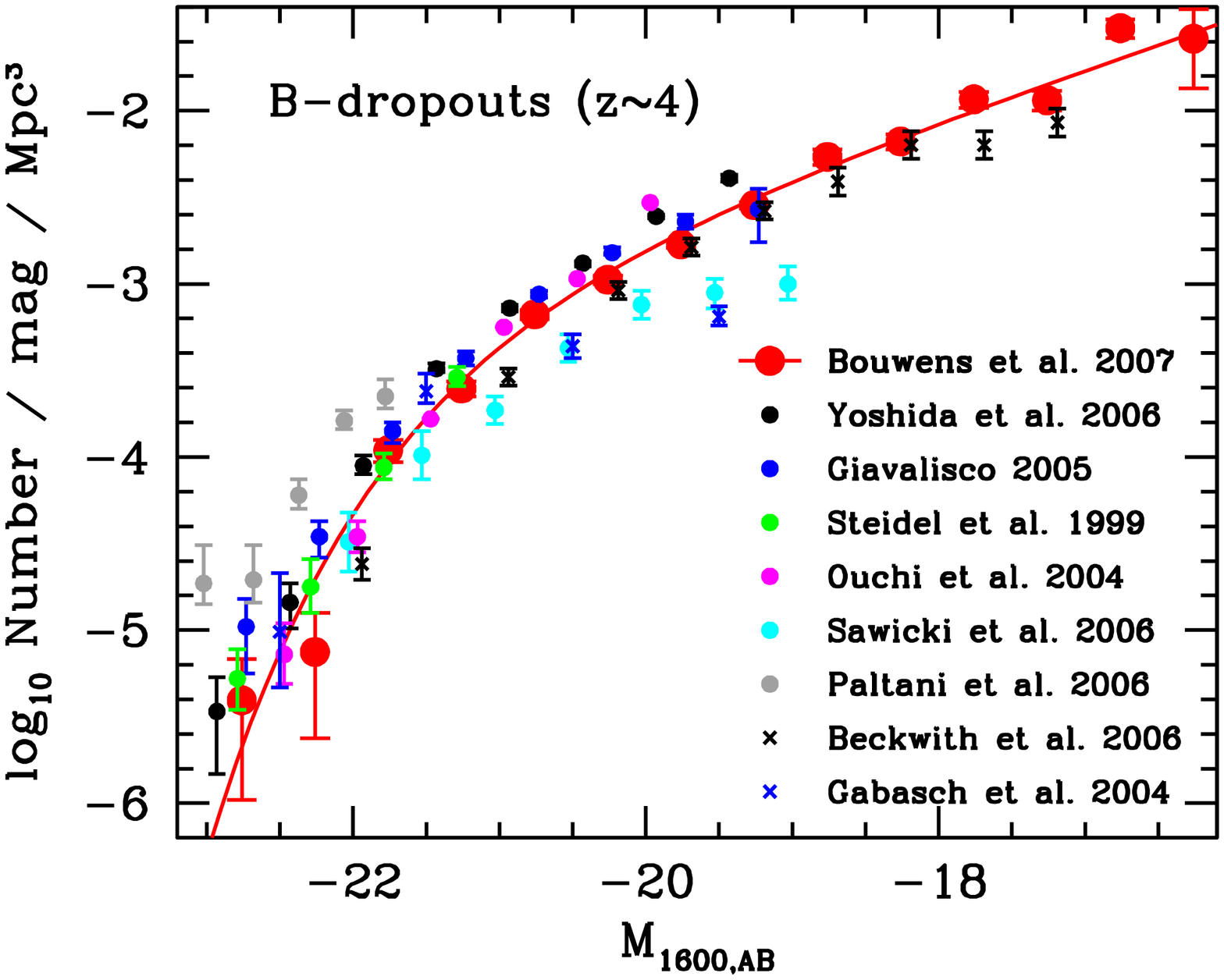,height=1.8in}}
\caption{Multiple determinations of the $z\sim 2$,
$z\sim 3$, and $z\sim 4$ rest-frame UV luminosity functions.
{\bf Left:} $z\sim 2$: \citep[From][]{reddysteidel2009} 
Points from \citet{reddysteidel2009} are shown
as solid blue circles, while the results from other groups
are indicated as in the legend, including survey areas.
{\bf Middle:} $z\sim 3$: \citep[From][]{reddysteidel2009} 
Points from \citet{reddysteidel2009} are shown as empty squares
while the results from other groups are as indicated in the legend,
including survey area.
{\bf Right:} $z\sim 4$: \citep[From][]{bouwens2007} 
Points from \citet{bouwens2007} are indicated as 
red circles. Results from other groups are as indicated in the legend.}
\label{fig:empirical-LF-UV-reddysteidel2009bouwens2007}
\end{figure}

At $z\sim 2-3$, the work of 
\citet{reddy2008} and \citet{reddysteidel2009} is based on the 
largest set of spectroscopic redshifts in the literature, and, including
a consideration of the systematic variation of dust reddening with
UV luminosity, is accordingly the most robust. Specifically, 
\citet{reddysteidel2009} utilizes $>2000$ spectroscopic 
redshifts, and $\sim 31,000$ $z\sim 2-3$ photometric candidates in 
31 independent fields over $0.9\mbox{ deg}^2$. In this paper, 
the standard LBG selection limit of ${\cal R}=25.5$ was extended 
to fainter magnitudes $(\sim 0.1 L^*$) to obtain tighter 
constraints on the faint-end slope, $\alpha$. Recent luminosity 
function determinations at $z\sim 4$ include those by 
\citet{bouwens2007}, using 4671 $B$-dropout galaxies selected 
over $580\mbox{ arcmin}^2$ with deep {\it Hubble Space Telescope}
({\it HST}) Advanced Camera for Surveys (ACS) imaging, down 
to $M_{UV,\mbox{AB}}=-16$, and by \cite{vanderburg2010}, based 
on $\sim 36,000$ $g-$dropout galaxies selected in the CFHT 
Legacy Survey in four independent $1\mbox{ deg}^2$ fields down to 
$M_{UV,\mbox{AB}}=-18.7$. Both of these $z\sim 4$ surveys, however, are 
solely based on photometric selection, without spectroscopic 
confirmation. Figure~\ref{fig:empirical-LF-UV-reddysteidel2009bouwens2007}
reviews recent determinations of the rest-frame UV luminosity
function at $2 \leq z \leq 4$.

The rest-frame UV luminosity functions of star-forming galaxies 
at $z\sim 2-4$ are characterized by steep faint-end slopes 
($\alpha\sim -1.6 - -1.7 $) and characteristic luminosities of 
$M^*_{\mbox{AB}}=-21$. These steep faint-end slopes are in 
contrast to the flatter one ($\alpha= -1.22$) determined from 
the local far-UV (FUV) luminosity function using {\it Galaxy Evolution Explorer}
(GALEX) data \citep{wyder2005}. 
Also, the characteristic luminosities at $z\sim 2-4$ are roughly 
three magnitudes brighter than those determined from the local 
GALEX FUV luminosity function. \citet{reddysteidel2009} review 
other recent luminosity function determinations at $z\sim 2-3$, 
highlighting some of the discrepancies in the literature, 
including those determinations with significantly flatter 
faint-end slopes \citep{sawicki2006a,gabasch2004}, and/or larger 
counts at the brightest luminosities \citep{paltani2007}. To 
address some of these differences, \citet{reddysteidel2009} 
point out the pitfalls associated with attempting to measure the luminosity 
function over small areas $(<50 \mbox{ arcmin}^2)$, making 
incorrect assumptions about the nature of the intrinsic mean and 
dispersion in colors of the faintest galaxies, and insufficiently
accounting for contamination by low-redshift interlopers. Again, 
we emphasize the importance of spectroscopy for understanding 
the redshift selection function of the objects for which the 
luminosity function is being constructed, and the proper 
characterization of systematic effects modulating the volume 
probed by a given galaxy survey.

\subsubsection{REST-FRAME OPTICAL AND NEAR-INFRARED LUMINOSITY FUNCTIONS}
\label{subsubsec:empirical-LF-optnearir}

While the rest-frame UV luminosity of distant galaxies reflects 
the emission from massive stars,
longer wavelengths probe different aspects of galaxy stellar 
populations and dust content. Specifically, the rest-frame 
optical luminosity function is more reflective of older stars, 
although the extent to which emission at these wavelengths reflects the 
integrated stellar mass depends in detail on the star-formation 
history of the galaxy \citep{shapley2001,shapley2005}. The 
rest-frame optical ($V$-band) luminosity function was first 
determined at $z\sim 3$ by \citet{shapley2001}, based on the LBG 
${\cal R}$-band luminosity function and the distribution of 
${\cal R}-K_s$ colors for a sample of 118 LBGs, 81 of which had 
spectroscopic redshifts. These measurements yielded a steep 
faint-end slope of $\alpha=-1.85$, which is significantly different from 
the local determinations with $\alpha \sim -1$, and a 
characteristic luminosity of $M^*_V=-22.98$, which is 1.5 
magnitudes brighter than the local value \citep{blanton2003}.

\begin{figure}
\centerline{\psfig{figure=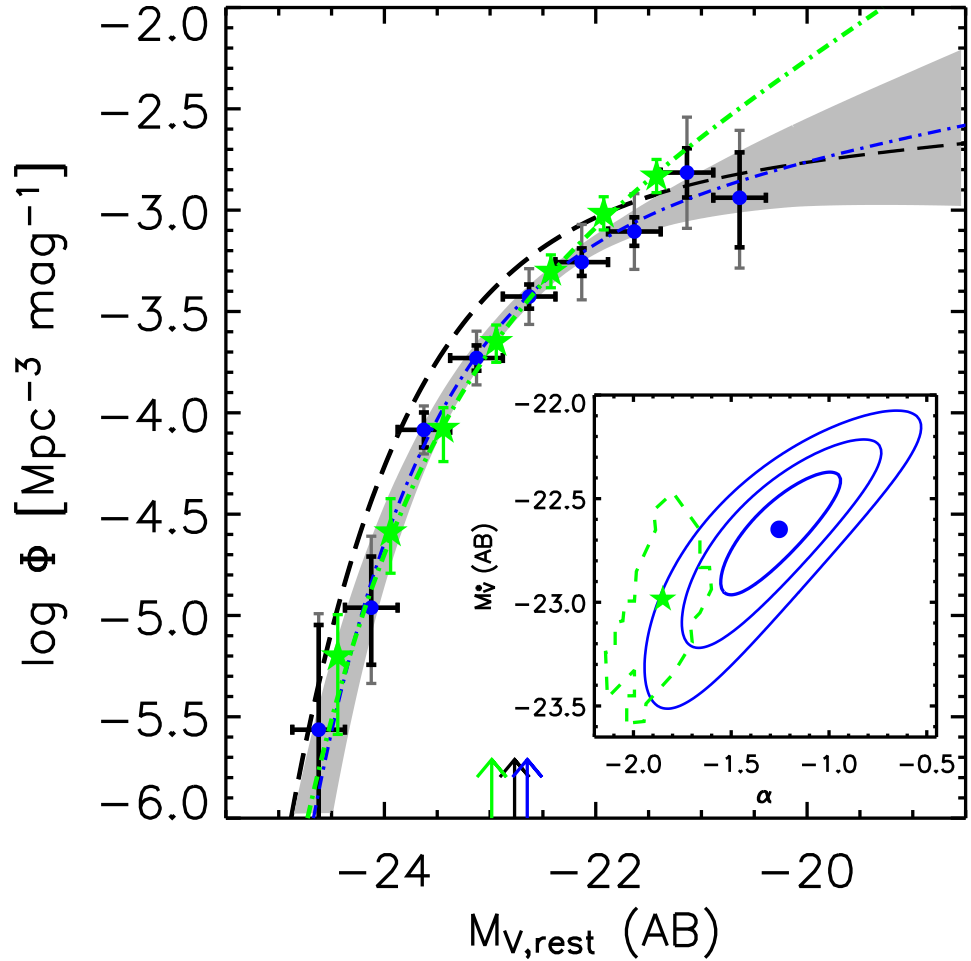,height=5in}}
\caption{\citep[From][]{marchesini2007} Rest-frame Optical Luminosity Functions.
Comparison of rest-frame $V$-band luminosity functions at $z\sim 3$.
Total rest-frame $V$-band luminosity function of 
$K$-selected galaxies at $2.7\leq z \leq 3.3$ (black dashed line). Also
shown are the $V$-band luminosity function of ``blue" ($J-K\leq 2.3$) $K$-selected
galaxies (blue circles), and the $z\sim 3$ LBG $V$-band luminosity function
from \citet{shapley2001} (green stars).  Black error bars on the blue points represent
Poisson uncertainties, whereas gray error bars also include field-to-field
variations. The gray shaded area represents 
the $1\sigma$ uncertainties of the luminosity function of the blue, K-selected 
galaxies. The inset shows the best-fit value and 1, 2, and $3\sigma$ confidence 
intervals on $\alpha$ and $M^*$ for the blue, $K$-selected galaxies (blue, solid curve), 
and LBGs (green, dashed curve).}
\label{fig:empirical-LF-optnearir-marchesini2007}
\end{figure}

Most recently, based on a much larger set of $\sim 1000$ 
$K$-band measurements at $2\leq z\leq 3.5$, selected over a 
total area of $378\mbox{ arcmin}^2$ from multiple near-IR 
surveys of varying depths, \citet{marchesini2007} constructed 
$B$, $V$, and $R$-band luminosity functions. Using mainly photometric redshifts,
\citet{marchesini2007} find that the faint-end slopes of these 
luminosity functions are, within the errors, consistent with the 
fairly shallow slope determined for the local optical luminosity 
function, whereas the characteristic magnitudes are significantly 
brighter ($\gtrsim 1$~magnitude). In contrast to 
\citet{shapley2001}, \citet{marchesini2007} find values of 
$\alpha$ ranging from $-1.0 - -1.4$. However, it is worth 
pointing out that, in the region of overlap, the 
\citeauthor{marchesini2007} rest-frame $V$-band luminosity function
for ``blue" ($J-K\leq2.3$) galaxies and the \citeauthor{shapley2001} 
rest-frame $V$-band LBG luminosity function are entirely 
consistent. Both total and ``blue" $V$-band luminosity functions from
\citet{marchesini2007} are shown along with the luminosity
function from \citet{shapley2001}
in Figure~\ref{fig:empirical-LF-optnearir-marchesini2007}, reproduced
from \citet{marchesini2007}. Potential limitations of the \citet{marchesini2007} 
analysis are in the small fraction of spectroscopic redshifts 
($\sim 4$\%) and corresponding reliance on photometric redshifts, 
and the small area over which the photometry is deep enough to robustly
probe the faint end ($\sim 25 \mbox { arcmin}^2$). With ultra-deep $K$-band 
[$K_{limit}\sim 23$(Vega)] surveys over significantly larger 
areas ($\sim 1000$s of arcmin$^2$), such as the UKIDSS 
Ultra-Deep Survey (UDS), the much-needed robust constraints on 
the faint-end slope of the rest-frame optical luminosity 
function will be within reach. Extensive spectroscopic follow-up 
is also necessary to avoid some of the biases related to 
photometric redshifts that are described in \citet{reddy2008} and 
\citet{reddysteidel2009}.

In principle, the rest-frame near-IR luminosity is even more 
closely tied to stellar mass than the rest-frame optical
(despite some of the uncertainties 
we will discuss in Section~\ref{subsec:stellarpop-SPS}). The 
largest study to date of the evolving rest-frame $K$-band 
luminosity function is based on the UKIDSS UDS First Data 
release, presented by \citet{cirasuolo2010}, updating 
their earlier work \citep{cirasuolo2007}. This study features a 
$K$+$z$-band-selected catalog of $\sim 50,000$ galaxies over 
$0.7 \mbox{ deg}^2$ ($\sim 10,000$ of which are at $z>1.5$), 
which is complete down to $K=23$ (AB). Given the location of the 
UKIDSS UDS in the Subaru/XMM-Newton Deep Survey field, this 
survey also benefits from extensive multiwavelength coverage 
spanning from FUV (GALEX) to mid-IR ({\it Spitzer}) 
wavelengths. \citet{cirasuolo2010} estimate photometric 
redshifts and rest-frame $K$-band luminosities based on 
spectral-energy distribution (SED) fits 
to the multiwavelength photometry of their sources. The depth 
of the \citet{cirasuolo2010} sample is not sufficient to trace 
the faint-end slope and its evolution past $z\sim 1$, so the 
evolution to high redshift is simply quantified in terms of the 
luminosity function normalization and characteristic luminosity. 
From $z\sim 0$ to $z\sim 2$, \citet{cirasuolo2010} report a 
brightening in $M^*$ by $\sim 1$~magnitude and a decrease by a factor of 
$\sim 3.5$ in normalization, $\Phi^*$. Again, deeper near-IR photometry 
and extensive spectroscopic follow-up for near-IR-selected 
catalogs will be required to probe the full evolution of the 
rest-frame near-IR luminosity function at high redshift.

\subsubsection{REST-FRAME MID-INFRARED AND BOLOMETRIC LUMINOSITY FUNCTIONS}
\label{subsubsec:empirical-LF-ir}

At even longer rest-frame wavelengths, we begin to probe the 
direct emission from dust. Building on the work of earlier 
missions such as the {\it Infrared Astronomical Satellite} (IRAS) 
and the {\it Infrared Space Observatory} (ISO), which traced the 
evolution of infrared-luminous (IR-luminous) sources out to $z\sim 1$, {\it 
Spitzer} has played a crucial role in tracing the global dust 
emission properties for large samples of $z>1$ star-forming 
galaxies. The most widely-used tool in this endeavor is the $24 
\mu\mbox{m}$ channel of the MIPS instrument, which probes a 
rest-frame wavelength of $\lambda \sim 8\mu\mbox{m}$ at $z\sim 
2$, sensitive to the emission from polycyclic aromatic 
hydrocarbon (PAH) emission. Ground-based submillimeter 
observatories such as SCUBA have also played a key role in 
understanding the evolution of most IR-luminous sources, but for 
much smaller galaxy samples. Overall, out to $z\sim 1$, the 
evolution of the mid-IR and total IR luminosity function is 
characterized by both significant luminosity and density 
evolution \citep{lefloch2005}. IR-luminous galaxies were more 
numerous in the past, and the IR luminosity density at $z\sim 1$ 
is dominated by luminous infrared galaxies (LIRGs), with 
$L_{IR}>10^{11}L_{\odot}$.

\begin{figure}
\centerline{\psfig{figure=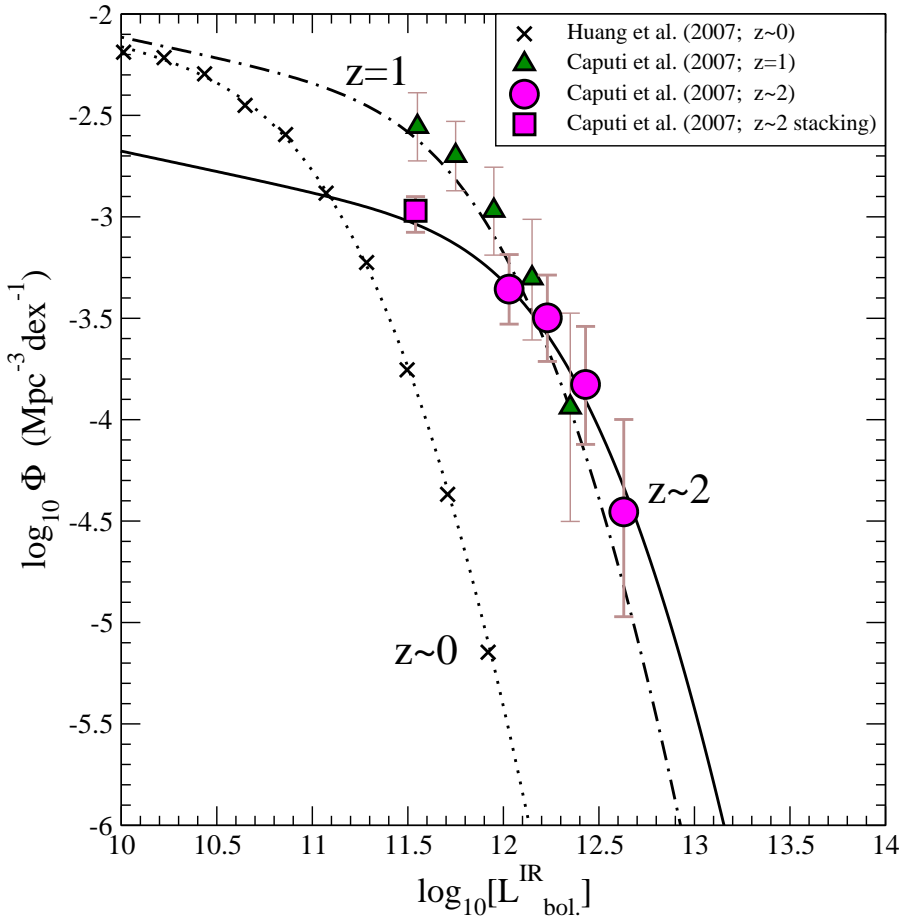,height=5in}}
\caption{\citep[From][]{caputi2007} IR Luminosity Functions.
Bolometric IR luminosity functions for star-forming
galaxies at $z\sim 2$ (individual detections: circles;
stacked result for fainter galaxies: square) 
and $z\sim 1$ (triangles) in the GOODS fields.
The local IR luminosity function is shown for comparison (crosses).
}
\label{fig:empirical-LF-ir-caputi2007}
\end{figure}

Recent work by \citet{perezgonzalez2005}, \citet{caputi2007}, 
and \citet{rodighiero2010} have characterized the luminosity 
functions of MIPS 24$\mu$m-selected galaxies at $z>1$. We focus 
here on the results of \citet{caputi2007} and 
\citet{rodighiero2010}, who used more conservative criteria for 
excluding AGNs from their samples, and adopted conversions 
between rest-frame $8\mu$m and total IR luminosities that are 
most consistent with observed constraints (and agree well with 
each other). These studies select galaxies down to a flux limit 
of $S(24 \mu\mbox{m})=80 \mu\mbox{Jy}$ in the $0.08 \mbox{ 
deg}^2$ of the Great Observatories Origins Deep
Survey (GOODS) North and South fields, while 
\citet{rodighiero2010} additionally include a shallower catalog 
down to a limit of $S(24 \mu\mbox{m})=400 \mu\mbox{Jy}$ in the 
$0.85 \mbox{ deg}^2$ of the VVDS-SWIRE area. 
As shown in Figure~\ref{fig:empirical-LF-ir-caputi2007}, \citet{caputi2007} 
determine mid- and total IR luminosity functions at $z\sim 1$ 
and $z\sim 2$, while \citet{rodighiero2010} construct the 
corresponding luminosity functions in nine redshift bins from 
$z\sim 0$ to $z\sim 2.5$.

Both of these works echo the previously determined increase of 
an order of magnitude in the bolometric IR luminosity density 
from $z\sim 0$ to $z\sim 1$, and the heightened abundance of 
both LIRGs and ULIRGs.
\citet{rodighiero2010} find that the IR luminosity density is 
roughly constant from $z\sim 1$ to $z\sim 2.5$, while 
\citet{caputi2007} find evidence for a slight decline. For both 
of these studies, it is worth pointing out that their 
completeness limits at $z\sim 2$ in $8\mu$m luminosity translate 
into bolometric IR luminosities of $\sim 10^{12}L_{\odot}$. 
Therefore, these luminosity functions only directly probe the 
ULIRG regime. Furthermore, neither study can place constraints 
on the faint-end slope of the luminosity function, and instead adopt a 
fixed parameter of $\alpha=-1.2$ based on local observations. 
Using indirect estimates of the rest-frame $8\mu$m and 
bolometric IR luminosity functions at $z\sim 2$, inferred from the 
extinction-corrected luminosities of UV-selected star-forming 
galaxies, \citet{reddy2008} extend the determination of the IR 
luminosity function into the LIRG regime, and suggest that a 
significantly steeper faint-end slope may be required. Accordingly,
ULIRGs make a sub-dominant ($\sim 25$\%) contribution 
to the far-IR luminosity density at $z\sim 2$. Future direct
observations of this fainter IR regime with {\it Herschel},
and even more sensitive planned facilities such as the {\it Single Aperture
Far-Infrared} (SAFIR) observatory,
will be crucial in untangling the evolution of the IR luminosity 
density and the history of obscured star formation in the 
Universe.

\subsection{Color-Magnitude Diagrams}
\label{subsec:empirical-CMD}

A standard tool for describing the galaxy population in the 
local Universe is the  optical  ``color-magnitude" 
diagram, in which rest-frame optical colors are plotted as a 
function of rest-frame optical absolute magnitude. The added 
dimension of color proves very useful for separating different 
types of galaxies. Indeed, while the division of the galaxy 
population into different types (e.g., late-type spirals and 
irregulars, and early-type ellipticals and lenticulars), and the 
correlation between galaxy structure and color have been 
well-known for a long time, the advent of large spectroscopic 
surveys such as SDSS have allowed for a quantitative description 
of the bimodality in galaxy photometric properties, based on 
incredibly robust statistics. Both \citet{strateva2001} and 
\citet{baldry2004} have presented striking evidence that local 
galaxies occupy a bimodal distribution in the space of $u-r$ 
colors, and that the fraction of bluer galaxies increases at 
fainter optical absolute magnitudes. Galaxies occupying the 
so-called ``red-sequence" part of the bimodal distribution 
primarily consist of morphologically early-type galaxies and
follow a very tight relationship between color and magnitude in
which more luminous galaxies have redder colors. At the same time,
objects residing in the ``blue cloud" are predominantly late-type 
galaxies, and show a looser, though still systematic, variation of color
and luminosity in the same sense. The bimodal distribution of galaxy colors as a 
function of luminosity, along with the correlation of color and other 
galaxy properties, suggests two distinct types of formation 
histories. It is therefore of critical interest to trace the 
galaxy color-magnitude diagram to earlier times, and to 
determine how far back the bimodality in the galaxy distribution 
persists.

\begin{figure}
\centerline{\psfig{figure=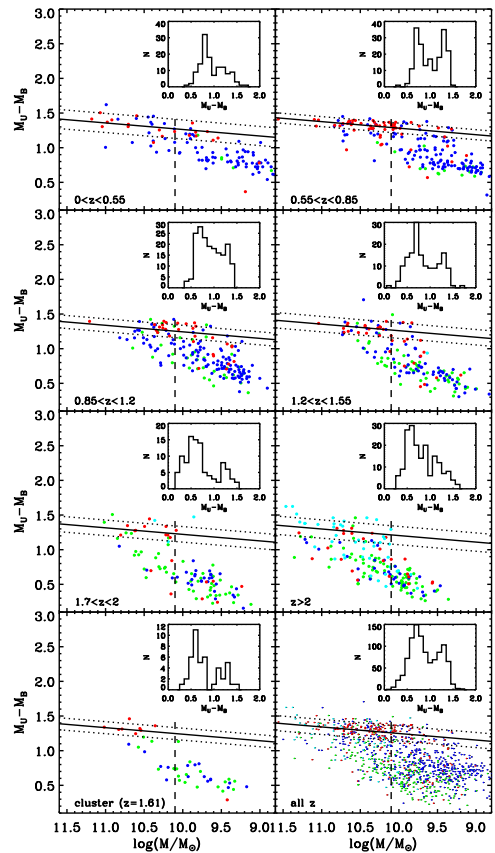,height=5.5in}}
\caption{\citep[From][]{cassata2008} 
Rest-frame $U-B$ color versus stellar mass in six
bins of redshift, from $z\sim 0$ to $z\sim 2$. 
The bottom left panel indicates the color-magnitude
diagram in a protocluster identified at $z=1.61$, whereas the bottom right
panel shows the color-magnitude diagram for all galaxies, regardless of redshift.
In each panel, the diagonal continuous line indicates a fit to the red
sequence. The dashed line indicates $\log(M/M_{\odot})=10.1$, the 
stellar mass completeness
limit for galaxies on the red sequence. Color-coded symbols indicate galaxies in 
different morphological classes: red, blue,
green, and cyan symbols respectively represent early-types, spirals, irregulars,
and undetected objects. In each panel, the inset shows the one-dimensional
distribution in $U-B$ color. A clear bimodality in the $U-B$ color distribution
is observed to $z\sim 2$.}
\label{fig:empirical-CMD-cassata2008}
\end{figure}

Based on the COMBO-17 survey, \citet{bell2004} demonstrate that bimodality
is detected in the color-magnitude diagram out to $z\sim 1$. Now, using
the results from recent near- and mid-IR-selected surveys, 
other groups have considered the question of galaxy bimodality
at even higher redshifts. As shown in Figure~\ref{fig:empirical-CMD-cassata2008},
with a sample of  1021
{\it Spitzer}/IRAC $4.5\mu$m-selected objects from the GMASS survey
with optical through mid-IR SEDs, 190 of which have spectroscopic redshifts 
above $z=1.4$, \citet{cassata2008}
demonstrate that a bimodality in galaxy rest-frame $U-B$ colors
persists up to $z=2$. \citet{brammer2009} uses $\sim 25,000$ objects with
$K<22.8$ (AB) selected from the NEWFIRM Medium-Band Survey (NMBS),
and detect a bimodality in rest-frame $U-V$ colors
out to $z\sim 2.5$. 
The increasing importance of obscured star formation
at higher redshifts tends to cause contamination of the red
sequence by dusty, star-forming galaxies,
and \citet{brammer2009} show a much cleaner division 
in dust-corrected colors, or when considering only galaxies whose
mid-IR flux limits of $S(24\mu\mbox{m})<20 \mu$Jy
indicate the presence of little or no dust.
Based on a smaller sample of 28 $K$-selected
galaxies with spectroscopic redshifts $2 < z < 3$,
\citet{kriek2008b} detect a red sequence in rest-frame $U-B$
colors, quantified in terms of a significant overdensity
of galaxies within a narrow bin of red rest-frame $U-B$ color.
These high-redshift red-sequence galaxies are characterized by
little or no ongoing star formation, with strong Balmer breaks
indicating that they are likely in a post-starburst phase.
While suggestive of the existence of a red sequence, the sample 
in \citet{kriek2008b} is too small to test for the 
presence of bimodality in galaxy colors. 

Both \citet{kriek2008b} and
\citet{cassata2008} stress the need for precise (and
ideally spectroscopic) redshift information when attempting to 
detect features in the galaxy rest-frame color distribution such
as bimodality or the presence of a red sequence.  While they lack 
actual spectroscopic confirmation, galaxies in the NMBS presented
by \citet{brammer2009} have extremely
accurate photometric redshifts, with $\Delta z / (1+z) < 0.02$
at $z>1.7$. Photometric redshifts typical of other studies,
with errors of $\Delta z/(1+z)\sim 0.1$ or worse, will lead to random
and systematic errors in inferred rest-frame UV and optical colors
that will wash out these trends when a single color is considered. 
However, using only photometric redshifts but considering {\it two} rest-frame
colors together, \citet{williams2009} discern a bimodal
behavior in the space of rest-frame $U-V$ versus $V-J$ color-color
space out to $z\sim 2.5$. The rest-frame colors for this 
study are inferred from optical, near-IR, and 
{\it Spitzer}/IRAC photometry drawn from the UKIDSS UDS, the
Subaru-{\it XMM} Deep Survey (SXDS), and the
{\it Spitzer} Wide-Area Infrared Extragalactic Survey (SWIRE),
and effectively separate quiescent, non-star-forming galaxies
from their actively star-forming counterparts. The evolving
locations in color space and very existence
of the ``red sequence" and ``blue cloud" provide important
constraints on models of galaxy formation.

\section{THE STAR-FORMATION RATES AND STELLAR CONTENT OF HIGH-REDSHIFT GALAXIES}
\label{sec:stellarpop}

While the luminosities and colors of galaxies represent
basic and fundamental observables, we seek to translate these
measurements into physical quantities. Specifically, the SEDs of galaxies
are commonly interpreted in terms of the current rate of star formation
and its past history, as well as the integrated stellar content of galaxies.
In this section, we summarize both the simple methods used to infer
star-formation rates from specific luminosities, as well as the techniques used
to model the stellar populations of high-redshift galaxies based on
multiwavelength SEDs. Along the way, we highlight
the results of applying these methods to determine the global history
of star formation, the build-up of stellar mass density, and the relationship
between star-formation rate and stellar mass in distant galaxies.

\subsection{Star-formation Rate Indicators}
\label{subsec:stellarpop-SFindicator}

As reviewed by \citet{kennicutt1998}, there are several diagnostics
of star-formation activity in external galaxies. In the study
of distant galaxies at $z>2$, integrated light measurements
tracing young stellar populations are used to infer the rate
at which stars are being produced. These measurements include rest-frame UV
and IR luminosities, hydrogen recombination emission-line luminosities,
and stacked X-ray and radio luminosities (due to the sensitivity
limits of current X-ray and radio facilities, $z>2$ star-forming
galaxies are typically only detected in a statistical sense).
In general, these diagnostics are only sensitive to the presence
of massive stars. Therefore, an estimate of the total star-formation
rate requires the assumption of a particular form for the stellar
initial mass function (IMF), which is then used to extrapolate 
down to the low stellar masses that dominate the integrated stellar mass.

The most commonly used star-formation diagnostic for high-redshift galaxies
is the ultraviolet continuum luminosity over the wavelength
range $1500-2800$~\AA, which is dominated by O and B stars
and directly related to the star-formation
rate in galaxies where star formation has been proceeding at
a roughly constant rate on $\sim 10^8$-year timescales. The rest-frame
UV luminosity at $2\leq z \leq 4$ is based on observed-frame
optical photometry. As described in Section~\ref{subsec:ISM-dustext},
the colors of star-forming galaxies at high redshift suggest
that dust extinction leads to a significant attenuation of
the rest-frame UV luminosity. Therefore, the rest-frame UV
luminosities based on optical apparent fluxes must be corrected
for dust when calculating intrinsic star-formation rates.

Dust absorbing the radiation from massive stars
is heated and re-radiates this absorbed energy
in the far-IR region of the spectrum. Therefore, the 
far-IR luminosity from dust also provides a tracer
of the rate of massive star formation. The range
``far-IR" is commonly defined as $8-1000 \mu$m \citep{kennicutt1998},
and the luminosity spanning this wavelength range is adopted
as a tracer of star formation for systems in which star formation
has proceeded continuously for at least $10^7-10^8$-year timescales.
For galaxies with bolometric IR luminosities large enough
($L_{IR}\gtrsim 10^{12}L_{\odot}$)
to be detected with ground-based submillimeter telescopes
such as SCUBA, the observed 850$\mu$m flux can be converted
into a far-IR luminosity with an assumption of dust temperature
(see Section~\ref{subsec:ISM-dustem}).
For $z\sim 2$ galaxies with smaller bolometric luminosities
(down to $L_{IR}\sim 10^{11} L_{\odot}$), deep {\it Spitzer}/MIPS
$24 \mu$m observations have provided a proxy for far-IR luminosity. At 
$z\sim 2$, the  $24 \mu$m channel probes a rest wavelength of $8 \mu$m,
the location of strong PAH emission. Local galaxy templates from
\citet{dale2002}, \citet{chary2001}, \citet{elbaz2002},
and \citet{rieke2009} are used to convert from
the mid-IR to total IR luminosity, which corresponds to
an increase by an order of magnitude.
Recent {\it Herschel} Photodetector Array Camera and Spectrometer (PACS)
$160 \mu$m measurements of the far-IR luminosities of star-forming
galaxies at $z\sim 2$ suggest that MIPS $24 \mu$m proxies
for far-IR luminosities tend to yield overestimates by factors
of $\sim 4-7.5$ \citep{nordon2010}. This apparent discrepancy
will require additional study.  {\it Spitzer}/MIPS
$70 \mu$m data probing rest-frame $20-25 \mu$m at $z\sim 2$
have not been used as widely because of the limit of the
depth of the existing {\it Spitzer} data, in which very few
individual $z\geq 2$ galaxies are detected.

As photons from massive stars ionize nearby interstellar gas, 
the emission from recombining
hydrogen gas in these star-forming regions serves as a
proxy for the rate of production of ionizing photons, and,
by extension, the formation rate of massive stars. The ionizing
flux is dominated by emission from the most massive ($>10M_{\odot}$)
stars and provides an estimate of the instantaneous rate
of star formation. With ground-based observations,
H$\alpha$ emission lines have been measured
for star-forming galaxies up to $z \sim 2.6$ (at which redshift
the thermal background begins to dominate the noise) and used
to estimate star formation rates \citep{erb2006c,forsterschreiber2009}. 
The weaker H$\beta$
feature has also been used to estimate star-formation
rates at redshifts beyond $z=2.6$ \citep{pettini2001,mannucci2009}.  
In addition to Balmer lines,
the Ly$\alpha$ feature has been used as a proxy for star-formation
rate. However, due to the resonant nature of the Ly$\alpha$ transition,
this line is especially sensitive to the effects of dust extinction
and scattering, which can both lead to the preferential destruction
of Ly$\alpha$ photons and their diffusion over a large area with
reduced surface brightness.
Accordingly, there is evidence that even dust-corrected observations of Ly$\alpha$
from individual galaxies tend to underpredict the star formation
rate, when compared with other indicators such as rest-frame UV
luminosity or H$\alpha$ emission \citep{hayes2011}.

Additional star-formation rate diagnostics are based on X-ray and radio
luminosities. High-mass X-ray binaries, young supernova remnants,
and hot interstellar gas contribute to the X-ray luminosity
in star-forming galaxies. \citet{ranalli2003} calibrate the relation between
rest-frame $2-10$~keV and far-IR luminosities for such systems, leading
to a relation between hard X-ray luminosity and star-formation rate.
Based on the tight linear correlation between far-IR and $1.4$~GHz
radio luminosity among nearby star-forming galaxies, \citet{yun2001}
derive the relation between star-formation rate and radio luminosity.
The deepest current {\it Chandra} $0.25-2.0$~keV (observed) and 
VLA 1.4~GHz radio data (e.g.,
in the GOODS-N region) are still not sufficient for detecting
all but the most luminous star-forming galaxies and AGNs
at $z\geq 2$. Therefore, most studies that make use of X-ray
or radio estimates of star-formation rates use stacking techniques
\citep[e.g.,][]{reddy2004,daddi2007,pannella2009}, yielding only
sample-averaged properties.

\subsection{Evolution of the Star-formation Rate Density}
\label{subsec:stellarpop-SFD}
The evolution of the star-formation-rate density is considered
one of the most fundamental observational descriptions of the galaxy population
as a whole. Matching and explaining this observed evolution is often 
used as a benchmark of success for theoretical models of galaxy formation.
The evolution of the star-formation-rate density is constructed by integrating the
luminosity function at a specific wavelength sensitive
to star formation (e.g., rest-frame UV, H$\alpha$, far-IR,
or radio), in order to obtain
the associated luminosity density. Then, a conversion between
luminosity and star-formation rate (as described in 
Section~\ref{subsec:stellarpop-SFindicator})
is used to obtain the associated star-formation-rate density.
In such studies, it is crucial to specify the limits down to which the luminosity
function is integrated at each redshift \citep[e.g.,][]{bouwens2007}.
Also, at shorter wavelengths, such as rest-frame UV and H$\alpha$, corrections
for dust extinction must be applied in order to obtain the unobscured star-formation
rate density \citep[see Section~\ref{subsec:ISM-dustext};][]{calzetti2000}. 
These corrections are sometimes applied in
an average sense to the integrated luminosity density, or
else in a luminosity-dependent fashion, taking into account
the relationship between luminosity and dust obscuration \citep{hopkins2004}.

\begin{figure}
\centerline{\psfig{figure=shapley_fig8a.eps,height=2.5in,angle=270}\psfig{figure=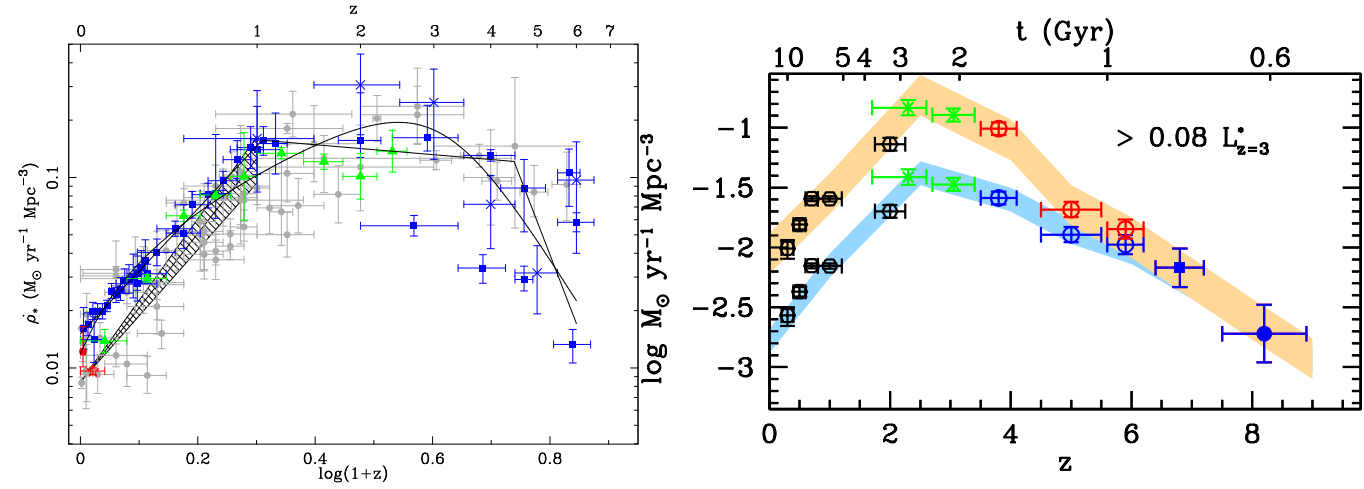,height=2.5in}}
\caption{Star-formation History of the Universe.
{\bf Left:} \citep[From][]{hopkinsbeacom2006} Evolution of the
star-formation rate density with redshift. Shown here is the multiwavelength compilation
from \citet{hopkins2004} (gray points) and \citet{hopkinsbeacom2006} (all
other symbols), and references therein. All data points have been
scaled to a common IMF and dust correction. The hatched region is the FIR
star-formation history from \citet{lefloch2005}. The solid lines represent
best-fitting parametric forms to the data compilation from \citet{hopkinsbeacom2006}.
{\bf Right:} \citep[From][]{bouwens2010} Evolution of the star-formation
rate density, based on rest-frame UV luminosity functions. At each
redshift, the luminosity function is integrated down to $M_{UV}=-18.3$ AB mag,
which is $0.08 L^*$ at $z=3$. The lower set of points (blue region) shows
the star-formation rate density determination inferred directly from
UV light with no dust correction applied. The upper set of points (orange
region) shows the inferred star-formation rate density using dust
corrections inferred from the measurement of UV continuum slopes. 
}
\label{fig:stellarpop-SFD-hopkinsbeacom2006bouwens2010}
\end{figure}

The first attempts to chart the star-formation history of the
Universe were presented by \citet{madau1996} and \citet{lilly1996}.
At the present, an extensive set of measurements has been
compiled by \citet{hopkins2004} and \citet{hopkinsbeacom2006} from $z\sim 0$
to $z\sim 6$, using a variety
of star-formation rate indicators including rest-frame
UV, far-IR, radio, and X-ray continuum luminosities, and
both Balmer and forbidden [OII] emission lines. 
These measurements were all
converted to a common cosmology ($\Omega_{m}=0.3$, $\Omega_{\Lambda}=0.7$,
$H_0=70 \mbox{ km s}^{-1}\mbox{ Mpc}^{-1}$), and rest-frame
UV and optical measurements were corrected for dust extinction.
In \citet{bouwens2007,bouwens2010}, rest-frame UV luminosity functions, 
both corrected and uncorrected for dust extinction, are used to map the evolution
of the star-formation-rate density to $z\sim 8$.
As shown in Figure~\ref{fig:stellarpop-SFD-hopkinsbeacom2006bouwens2010},
in both of these compilations,
the evolution of the star-formation-rate density is characterized
by an order-of-magnitude increase from $z\sim 0$ to $z\sim 2$. In more detail,
the compilations by \citet{hopkins2004} and \citet{hopkinsbeacom2006} show
the dust-corrected star-formation-rate density increasing by an order of
magnitude from its local value already by $z\sim 1$ and remaining roughly
flat to $z\sim 2$. The evolution in \citet{bouwens2007,bouwens2010} is
characterized by a smooth rise all the way from $z\sim 0$ to $z\sim 2$,
due to a different
extinction correction adopted for the rest-frame UV data at $z<2$.
In both versions of the cosmic star-formation history, the global
star-formation-rate density remains roughly constant between $z\sim 2$
and $z\sim 4$, and then declines towards higher redshift.
\citet{hopkinsbeacom2006} propose that the star-formation
history is constrained to within $\sim 30-50$\% up to
$z\sim 1$ and within a factor of $\sim 3$ at higher redshifts.
Additionally, recent results from \citet{reddy2008} suggest that
$\sim 70-80$\% of the star-formation-rate density at $z\sim 2-3$
is produced by galaxies with bolometric luminosities $L_{bol}\leq 10^{12} L_{\odot}$,
and that the highly-obscured ULIRGs selected by submillimeter surveys 
\citep[e.g.,][]{chapman2005}, although individually luminous,
do not dominate the star-formation-rate density.

\subsection{Stellar Population Synthesis Models}
\label{subsec:stellarpop-SPS}
While luminosities tied to individual specific wavelength ranges 
are commonly used to infer the current rate of star formation,
the multiwavelength SED of a galaxy can be interpreted in terms
of its integrated stellar, dust and metal content,
and its past history of star formation.
This technique is referred to as stellar population synthesis
modeling, and was first employed by \citet{tinsley1968}.
In the intervening decades, stellar population synthesis models
have become increasingly sophisticated, and have played
a crucial role in inferring the physical properties of galaxies both
near and far based on their photometric and spectroscopic properties. 
Population synthesis models are also used to
estimate photometric redshifts, in so far as the observed
multiwavelength photometry of a galaxy is fit in terms of not only 
best-fitting stellar population parameters, but also redshift.

The basic ingredients of stellar population synthesis models
consist of a theoretical prescription for all stages
of stellar evolution as a function of mass and metallicity,
as well as a stellar spectral library of the observed properties
of stars at different positions in the Hertzsprung-Russell
diagram. The time evolution of the integrated spectrum from a coeval population
of stars described by a specific stellar IMF and metallicity can then
be predicted. For a given star-formation history,
the evolution of the integrated spectrum of stars is computed
by summing the contributions of simple stellar populations formed
at successive time steps, with the normalization at each time step
determined by the corresponding star-formation rate. Thus, stellar population
synthesis models can predict the evolution of the integrated
spectrum of a stellar population for an arbitrary star-formation history,
IMF, and stellar metallicity. Stellar population synthesis models
are often coupled with theoretical models \citep{charlotfall2000}
or empirical parameterizations \citep{calzetti2000} of dust
extinction to explain the observed properties of galaxies
in terms of both stellar and dust content. 
The resulting model galaxy spectrum can be passed through photometric
filters in order to predict the evolution of luminosities, colors,
and mass-to-light ratios in specific photometric bands. Additional
properties of the stellar population can also be calculated, 
including the rate of Type Ia and Type II supernovae explosions,
the mass in stellar remnants and recycled gas, and
the ionizing photon luminosity.

Currently, there are several different stellar population synthesis
models employed for fitting the SEDs of distant galaxies. Perhaps most widely
used are the models of \citet{bruzualcharlot2003} and their
current version (Charlot \& Bruzual 2011, in preparation), though other
significant theoretical efforts include the PEGASE and
Starburst99 models \citep{fioc1997,leitherer1999}, and, more recently,
the models of \citet{maraston2005} and \citet{conroy2009}.
There is much ongoing discussion in the literature
regarding the systematic uncertainties
and differences among these different stellar population synthesis
codes, in terms of their descriptions of various stages of stellar evolution
off of the main sequence.

\begin{figure}
\centerline{\psfig{figure=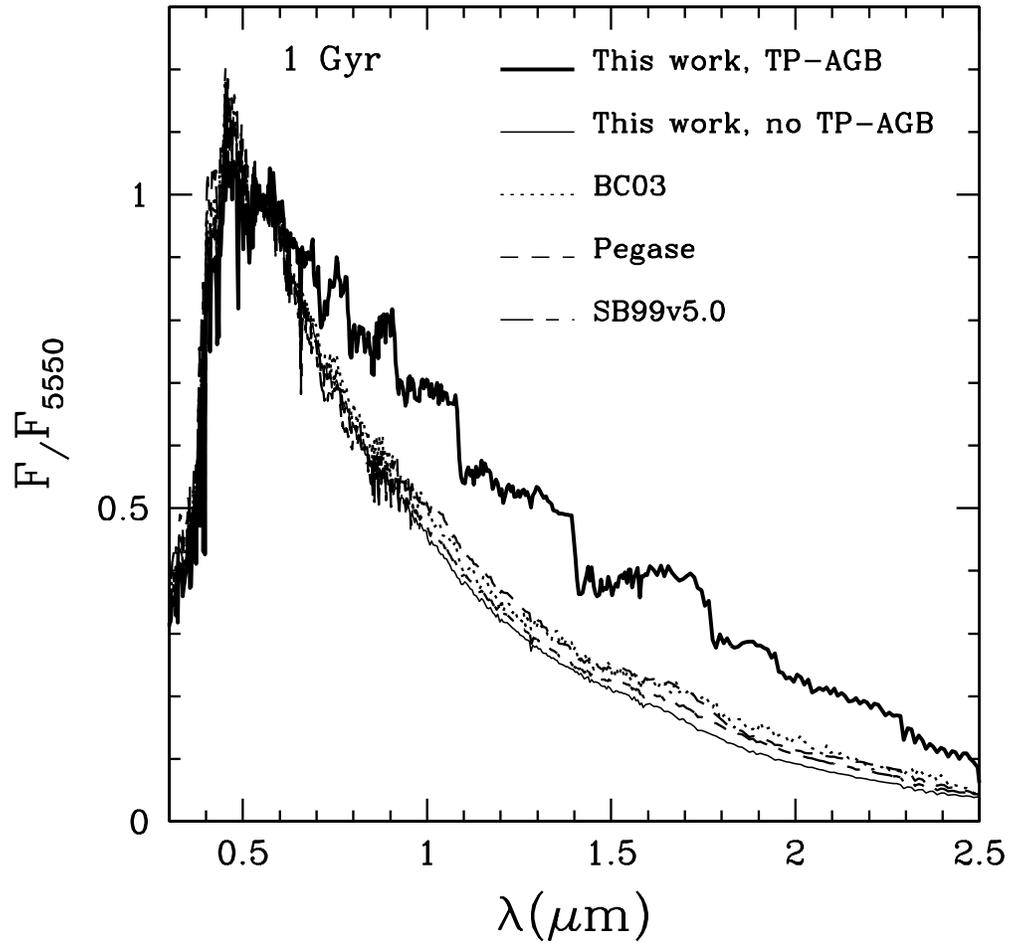,height=5in}}
\caption{\citep[From][]{maraston2005} The effects of TP-AGB Stars on Stellar
Population Synthesis Models. Shown here is a comparison of a 1~Gyr,
solar metallicity stellar population model from \citet{maraston2005}
both with (solid, thick line) and without (solid, thin line)
the contributions of TP-AGB stars. Also shown are other models
from the literature, including those from \citet{bruzualcharlot2003}, PEGASE,
and Starburst99, as indicated in the legend. Differences in the treatment
of the TP-AGB phase among stellar population synthesis codes
lead to significant discrepancies in the predicted
rest-frame near-IR spectra and luminosities of simple stellar populations
with ages between $0.5$ and $2.0$ Gyr.
}
\label{fig:stellarpop-SPS-maraston2005}
\end{figure}

There has been a particular focus on the treatment of the thermally-pulsating
asymptotic giant branch (TP-AGB). TP-AGB stars are red giants with
low- to intermediate-mass main sequence progenitors, which make
a significant contribution to the near-IR luminosity of a simple
stellar population at ages  between $0.5$ and $2.0$ Gyr. As emphasized
by \citet{maraston2005}, derived
properties such as the age and near-IR mass-to-light ratio, and, correspondingly,
the inferred stellar mass, will be very sensitive
to the treatment of TP-AGB stars in systems in an evolutionary
state that prominently features these stars. As a result, 
the stellar populations in which the correct treatment of the TP-AGB
matters the most are post-starburst systems, as opposed to those
in which star formation is proceeding at a roughly constant rate
\citep{daddi2007}. As shown in Figure~\ref{fig:stellarpop-SPS-maraston2005},
in the models of \citet{maraston2005},
TP-AGB stars make a much more significant contribution to
the integrated galaxy luminosity at $\sim 1$~Gyr than
in the models of \citet{bruzualcharlot2003}, resulting in systematically
lower derived stellar masses and ages, in particular for post-starburst
systems \citep{maraston2006}. Rather than adopting a specific prescription for
the contribution of the TP-AGB phase, \citet{conroy2009}
parameterize the luminosities and effective temperatures
of TP-AGB stars as variables to be constrained by the actual data.
Systematic uncertainties in TP-AGB parameters therefore translate
into systematic errors in other derived properties such
as stellar mass and past star-formation history. Recently,
\citet{kriek2010} constructed a composite SED over the rest-frame
range $1200-40000$~\AA\ for a
sample of 62 post-starburst galaxies at $0.7\leq z \leq 2.0$
drawn from the NEWFIRM Medium-Band Survey. 
In contrast to the results of \citet{maraston2006},
this entire SED is fit well by the \citet{bruzualcharlot2003}
models, whereas the rest-frame optical and near-IR regions of the spectrum
are not simultaneously fit by the models of \citeauthor{maraston2005}.
The SED-fitting results of \citet{kriek2010}
suggest that the \citeauthor{maraston2005} models 
give too much weight to the TP-AGB phase. Clearly, consensus has
yet to be reached about the influence of TP-AGB stars in the integrated
spectra of galaxies. 

Uncertainties in the detailed nature of the stellar IMF also 
translate into systematic uncertainties in the conversion between
luminosity and stellar mass. For the range of stellar populations
typically observed at high redshift (i.e., when the Universe was less
than a few Gyr old), and over the rest-frame UV to near-IR wavelength range
where current observations probe, models with the same star-formation
history and the assumption of either a \cite{chabrier2003}
or \citet{salpeter1955} IMF over the mass range $0.1-100 M_{\odot}$
produce virtually identical colors. However, the model assuming
a \citet{chabrier2003} IMF corresponds to a stellar mass a factor
of $\sim 1.8$ lower. Although both IMFs are described by power-law functions
at $M\geq 1 M_{\odot}$, the \citet{chabrier2003} IMF follows a log-normal
distribution below $1 M_{\odot}$, turning over at the so-called
``characteristic mass," whereas the \citet{salpeter1955} IMF
continues to increase as a power-law all the way down  to $0.1 M_{\odot}$.

Evidence in the local and low-redshift Universe suggests 
that the \citet{chabrier2003} IMF yields stellar $M/L$ ratios
for lower-mass elliptical galaxies (with velocity dispersions of $\sigma \sim 200 \mbox{ km s}^{-1}$)
in agreement with their luminosities and
dynamical mass estimates. On the other hand, elliptical galaxies with
larger dynamical masses appear to be characterized by either steeper 
(i.e., more Salpeter-like) IMFs or higher dark-matter fractions \citep{treu2010,graves2010}. 
There are also
extreme environments such as the Galactic Center and surrounding starburst
clusters such as the Arches, in which direct evidence for IMF variations
has been reported \citep{stolte2002}. At higher redshift, based on
more indirect methods such as measuring the simultaneous $U-V$ color and
stellar $B$-band $M/L$ ratio evolution for elliptical galaxies
at $z\sim 0$ to $z\sim 0.8$ (with the latter estimated from the observed evolution
in the fundamental plane), \citet{vandokkum2008a} infers that the
stellar IMF for elliptical galaxy progenitors had a flatter slope at $z\sim 4$
in the regime near $1 M_{\odot}$. Furthermore, the characteristic mass
at which the log-normal portion of the IMF turns over is inferred
to shift to $\sim 2 M_{\odot}$. Adopting a completely independent approach
based on a comparison of the star-formation rates and stellar masses of
vigorously star-forming galaxies at $z\sim 2$,
\citet{dave2008} also infers a higher characteristic mass at this redshift.
An upward shift in characteristic mass results in an IMF that is more ``bottom light"
than the present-day \citet{chabrier2003} function, and a lower conversion factor
from light to stellar mass. 
In contrast, recently \citet{vandokkum2010_nature} analyzed the spectra
of four elliptical galaxies in the Virgo cluster and, based
on the strength of stellar absorption features tracing
$M< 0.3 M_{\odot}$ stars, concluded that the IMF in the massive
star-forming progenitors of these systems had a ``bottom heavy" IMF {\it steeper}
than Salpeter between $0.1$ and $1.0 M_{\odot}$.

\citet{bastian2010} review evidence
for IMF variations both in the local Universe and at
significant cosmological distances for the most part with skepticism.
Clearly additional study is required to quantify or rule out the
evidence for IMF variations as a function of environment.
Stellar population synthesis models featuring explicit parameterizations of the most
unconstrained phases of stellar evolution, as well as the form
of the stellar IMF, will yield confidence intervals for derived stellar population
parameters that more accurately reflect the systematic uncertainties associated
with stellar population modeling \citep{conroy2009}.

\begin{figure}
\centerline{\psfig{figure=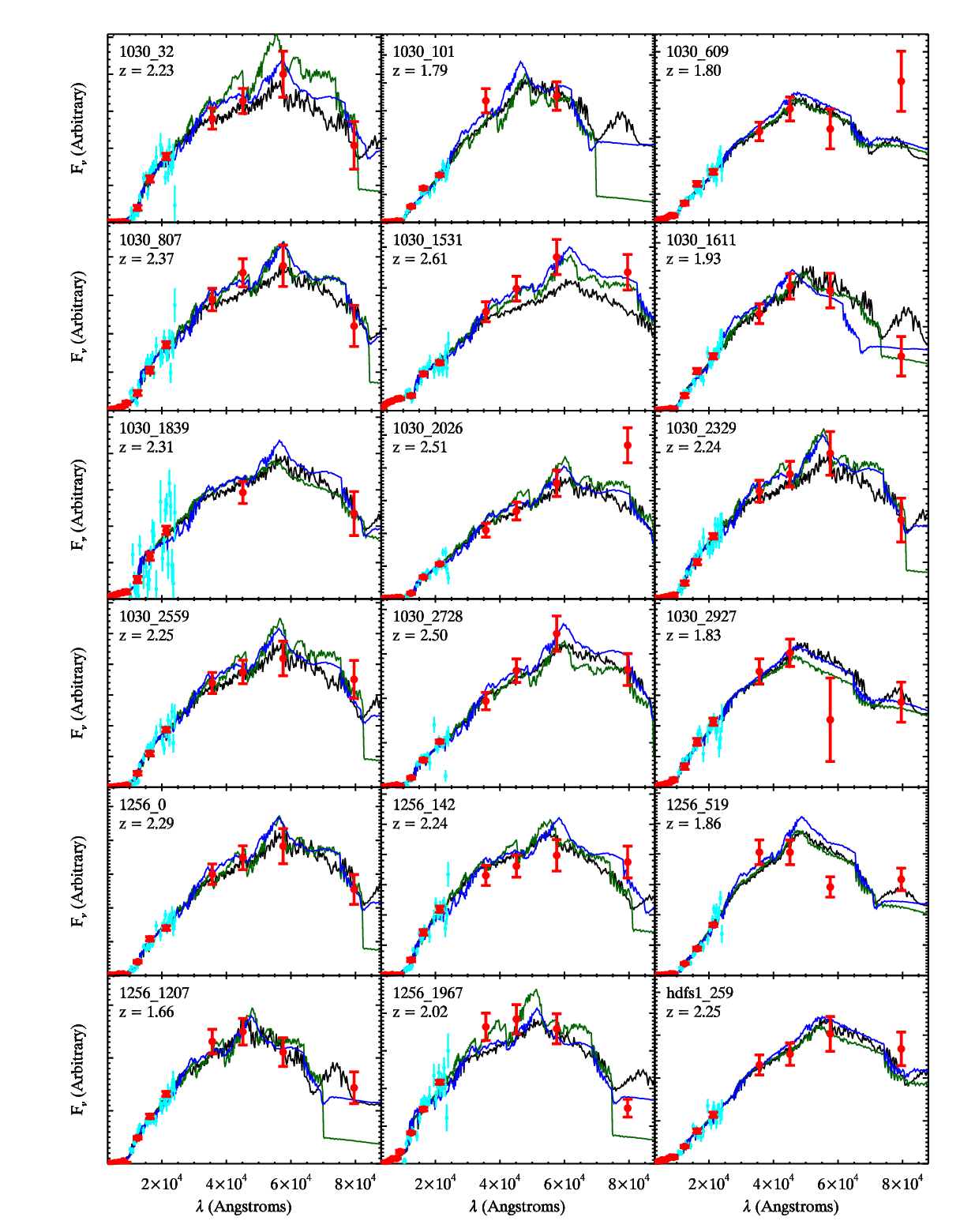,height=6in}}
\caption{\citep[From][]{muzzin2009} Sample population synthesis fits to
optical through mid-IR SEDs, and using different stellar population
synthesis codes. Shown here are well-sampled multiwavelength
SEDs including broadband optical and near-IR 
($UBVRIz'JHK$) and {\it Spitzer}/IRAC photometry (red circles), 
as well as GNIRS near-IR spectroscopy (cyan points) for a sample
of spectroscopically-confirmed galaxies at $z\sim 2$.
Also shown are a comparison of \citet{bruzualcharlot2003} (black curve),
updated Charlot \& Bruzual (blue curve), and \citet{maraston2005} (green curve) models.
}
\label{fig:stellarpop-SPS-muzzin2009}
\end{figure}

In spite of the significant uncertainties associated with stellar population
synthesis modeling, it is now standard practice to use such models to infer
basic physical properties of galaxies over a wide range of redshifts.
In the local Universe for a sample of $>10^5$ SDSS galaxies,
a combination of optical broadband
photometric properties and spectral indices (the 4000~\AA\ 
spectral break and the strength
of Balmer absorption lines) have been modeled using
the population synthesis code of \citet{bruzualcharlot2003} to infer 
galaxy properties such as star-formation history, dust attenuation, 
and stellar mass \citep{kauffmann2003a}. At higher redshifts,
stellar population modeling is typically tuned to broadband
photometry alone, as rest-frame optical spectra are not
of sufficient quality to measure stellar absorption features
or detailed continuum shape at high S/N. At $z\geq2$, stellar population
synthesis modeling is only possible if the rest-frame SED is probed
at wavelengths both above and below age-sensitive spectral
discontinuities such as the Balmer or 4000~\AA\ break. Therefore,
optical photometry probing the rest-frame UV must be, at the very least, combined
with data probing the rest-frame optical regime (i.e. near-IR observed
wavelengths), and, preferably, with additional {\it Spitzer}/IRAC photometry
probing the rest-frame near-IR. Early models of $z>2$ stellar populations
were featured in \citet{sawicki1998}, \citet{shapley2001}, and \citet{papovich2001}
(LBGs) and \citet{forsterschreiber2004} (DRGs). A fairly
recent example of $z \sim 2$ stellar population modeling from
\citet{muzzin2009} is shown in Figure~\ref{fig:stellarpop-SPS-muzzin2009}. 
This analysis is based on extremely well-sampled SEDs for spectroscopically-confirmed
galaxies including optical, near-IR, and {\it Spitzer}/IRAC photometry
as well as binned Gemini/GNIRS near-IR spectra, and features a
systematic comparison of the parameters derived 
from \citet{maraston2005}, \citet{bruzualcharlot2003},
and updated Charlot \& Bruzual models.
Stellar population synthesis models are now routinely
used to derive high-redshift galaxy physical properties.
Such modeling is a standard component of the derivation
of global distributions such as the galaxy stellar mass function
(Section~\ref{subsec:stellarpop-MstarD}).

For the modeling of high-redshift galaxy stellar populations, in addition
to the adoption of a particular stellar population synthesis code,
the stellar IMF and metallicity are assumed parameters. A specific,
wavelength-dependent dust extinction law (e.g., starburst, SMC, Milky Way) 
is also assumed.  The star-formation history, $SFR(t)$, is commonly 
parameterized in the form of an exponential decline,
with $SFR(t)=SFR_0 \times \exp(-t/\tau)$. In this case, $\tau$,  
the e-folding time, and $t$, the time since the onset of star formation,
are both parameters to constrain.  With such a parameterization,
a continuous star-formation history corresponds to 
$\tau=\infty$ and $SFR(t)=SFR_0$.  In addition to single episodes 
describing the star-formation rate as a function of  time,
more complex functions for the history of star formation
have been considered. In particular, two-component
models consisting of the linear combination of an old, 
high mass-to-light ratio component and a younger population
with ongoing star formation, have been used to constrain
how much stellar mass from the old stellar population
could be ``hiding" under the glare of a younger
burst of star formation. In addition to $\tau$ and $t$,
the parameters commonly derived
for high-redshift galaxies are indicators of the degree
of dust extinction [$E(B-V)$ or $A_V$], the current star-formation
rate, and the integrated stellar mass. For a given stellar population
synthesis code, the stellar mass has been demonstrated to be
the best-constrained parameter \citep{papovich2001,shapley2001,shapley2005},
whereas other parameters are more subject to uncertainties in the
nature of the star-formation history (i.e., $\tau$), which
is difficult to constrain in the absence of external multiwavelength
information. Recently,
\citet{maraston2010} have in fact argued that so-called ``inverted-$\tau$"
models [i.e. models in which the star-formation rate
increases with time as $SFR(t)=SFR_0 \times \exp(+t/\tau)$]
provide a better description of the extinction and star-formation
rate based on rest-frame UV data alone, as well as the star-formation
rates and stellar masses of mock high-redshift galaxies constructed
from semi-analytic models. In addition, \citet{papovich2011}
demonstrate that the globally averaged relations between the star-formation rates
and stellar masses of galaxies at high redshift (see Section~\ref{subsec:stellarpop-SFRM*}) appear
to favor rising star-formation histories, as opposed to ones that are constant
or declining. The best parameterization of the star-formation
histories of high-redshift galaxies is clearly still a matter of debate.

\subsection{The Diversity of High-Redshift Stellar Populations}
\label{subsec:stellarpop-diversity}

The methods described above have been used to investigate
the stellar populations of high-redshift galaxies selected
using the various techniques discussed in Section~\ref{sec:technique}.
Based on the results of stellar population syntehsis modeling, we can make some
general comments about the range of stellar populations observed, while
keeping in mind the uncertainties inherent to the modeling process,
and the biases that different selection techniques
impose. If the current samples of UV-selected galaxies
are roughly characterized by star-formation rates of $10-100 M_{\odot} 
\mbox{ yr}^{-1}$, typical stellar masses of $1-5 \times 10^{10} M_{\odot}$
\citep{erb2006b,shapley2001,reddy2004}, and moderate
amounts of extinction in the rest-frame UV (factor of $\sim 5$), the very complementary
sample of rest-frame optically selected DRGs are typically characterized
by higher star-formation rates ($\geq 100 M_{\odot}\mbox{ yr}^{-1}$),
stellar masses (down to the typical survey limits of $K\sim 21$ Vega) 
of $\sim 10^{11} M_{\odot}$, and larger amounts of dust extinction in 
the rest-frame UV and optical ranges of the spectrum
\citep{forsterschreiber2004,papovich2006}. At the same time, a minority of DRGs
show little evidence for ongoing star formation \citep{vandokkum2008b}. In addition
to their prodigious star-formation rates, SMGs appear to be characterized
by stellar masses that are comparable to those of the typical
DRGs, and several times larger on average
than those of UV-selected galaxies \citep{michalowski2010,borys2005}.
However, it is worth keeping in mind that
the effects of dust extinction and AGN contamination
on the broadband SED lead to larger systematic
uncertainties in the derived stellar populations of SMGs.
At the other end of the spectrum, so to speak, the stellar populations
of LAEs are characterized by less dust extinction on average
even than those of the UV-selected galaxies \citep{gawiser2007}. 
Due to the typically faint rest-frame UV luminosities of 
emission-line selected galaxies, the LAEs also tend to be faint
in the rest-frame UV continuum, with lower star-formation rates
than those of UV-selected galaxies, which were targeted down
to a brighter continuum limit \citep{kornei2010}.

In addition to considering the typical properties of galaxies
as a function of selection method -- which may not have anything other
than historical value -- it is also worth mentioning the range
of star-formation histories observed as a function of mass. In particular,
at high stellar masses $(M>10^{11} M_{\odot}$, assuming a Salpeter
IMF from $0.1-100 M_{\odot}$), there exist at $z\sim 2$ not 
only active star-forming galaxies with $SFR>100 M_{\odot} \mbox{ yr}^{-1}$,
but also passive, evolved galaxies with little evidence for ongoing star formation
\citep{kriek2008b,vandokkum2008b}. Quiescence and mature stellar 
populations constitute the physical interpretation of the empirically-derived
red sequence reported by \citet{kriek2008b} and described in 
Section~\ref{subsec:empirical-CMD}. These quiescent galaxies
appear to consitute $\sim 40-50$\% of the most massive galaxies
($M_{star}\geq 10^{11} M_{\odot}$ at $z\sim 2-3$). Based on these
results, it is worth
noting that, in contrast to the patterns observed in the local
Universe, at least half of most massive galaxies at $z\sim 2$
are still actively in the process of forming \citep{daddi2007,papovich2006}, 
and that there
is an incredible diversity observed among the star-formation histories
of these massive galaxies. At the same time, it is a challenge to
explain massive galaxies at early times
with little evidence for ongoing star formation,
given  that theoretical models predict copious rates
of mass accretion at $z\sim 2$ for such massive systems
\citep[e.g.,][]{dekel2009}.

\subsection{Evolution of the Stellar Mass Density}
\label{subsec:stellarpop-MstarD}
Stellar population synthesis modeling provides a powerful
tool for estimating the global evolution of the stellar content
in galaxies, which reflects the combined processes
of star formation and mergers. This evolution is described
by constructing the galaxy stellar mass function
at a range of redshifts. Analogous to
the galaxy luminosity function, the stellar mass
function is parameterized in terms of a characteristic
mass, $M_{star}^*$, low-mass slope, $\alpha$, and overall
normalization, $\Phi^*$. At
each redshift, the stellar mass function can be integrated
to determine the corresponding stellar mass density. Alternatively,
the growth in stellar mass as a function of galaxy mass
can provide important insights into galaxy formation models.
The cosmic stellar mass density should also reflect the integral of past
star formation in the Universe, and therefore
the integral of the star-formation-rate density described
in Section~\ref{subsec:stellarpop-SFD}
can be compared for consistency with the independently-derived
stellar-mass density.

In the local Universe, the galaxy stellar mass function has
been determined from large samples of galaxies drawn
from the 2dFGRS matched to the 2 Micron All-Sky Survey (2MASS), 
and SDSS \citep{cole2001,baldry2008}.
Presently, stellar mass functions have been measured
out to $z\sim 5$, suggesting that roughly half of the local
stellar mass density appears to be in place at $z\sim 1$.
Measurements of the stellar mass function at $z>1$
are based on samples with multiwavelength (optical
and IR) photometry and primarily photometric redshifts. The first determination
of the stellar mass function at $z\sim 2-3$ was presented
in \citet{dickinson2003}, based on a sample of rest-frame $B$-band-selected
objects in the Hubble Deep Field North (HDF-N) with both
optical and near-IR photometry. Subsequently, many other groups have 
measured stellar mass functions at $z\geq 2$, selecting
galaxies at optical, near-IR and mid-IR wavelengths
\citep[e.g.,][]{fontana2004,fontana2006,drory2005,pozzetti2007,elsner2008}.

\begin{figure}
\centerline{\psfig{figure=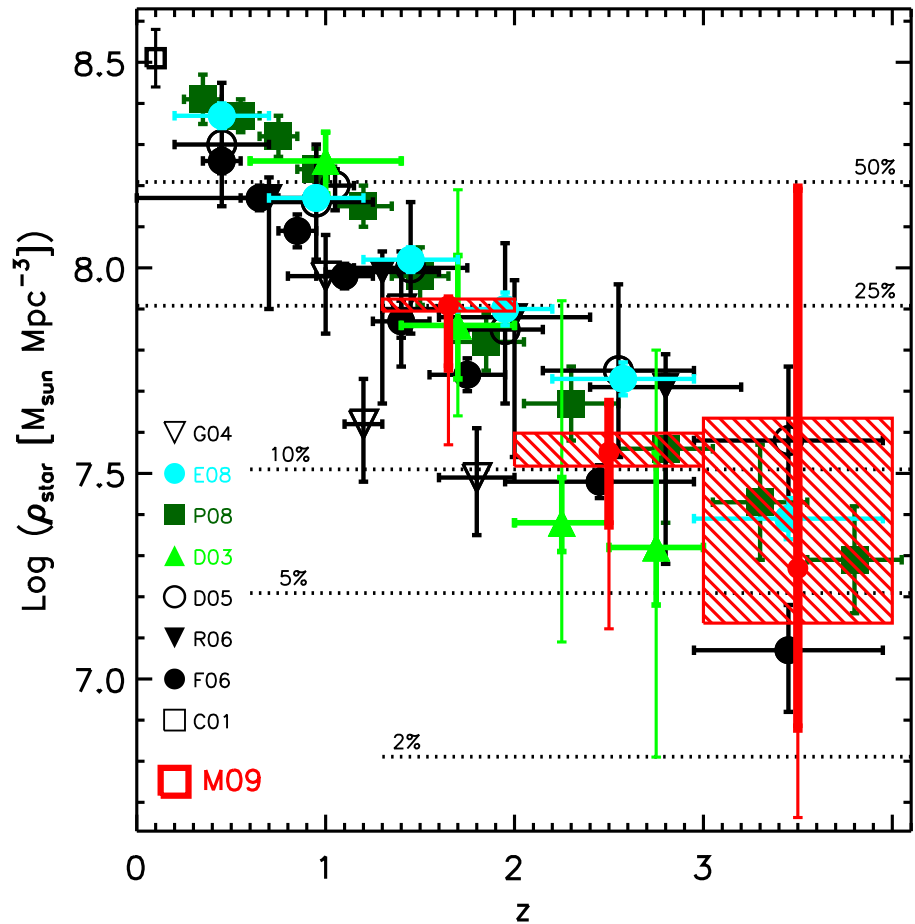,height=5in}}
\caption{\citep[From][]{marchesini2009}
The results from various surveys on the evolution
of the global stellar mass density. Mass-density estimates were
obtained by integrating stellar mass functions over the mass
range, $10^8 \leq M_{star}/M_{\odot} < 10^{13}$. Red symbols
represent the total stellar mass densities estimated from
\citet{marchesini2009} (i.e., ``M09"), where shaded boxes
do not include the systematic uncertainties, and error bars do.
Other estimates of stellar mass densities come from the literature,
with references as in \citet{marchesini2009}. Horizontal
dotted lines represent 50\%, 25\%, 10\%, 5\%, and 2\% of
the total stellar mass density at $z=0.1$.
}
\label{fig:stellarpop-MstarD-marchesini2009}
\end{figure}

Both \citet{perezgonzalez2008} and \citet{marchesini2009}
construct stellar mass functions at $z\geq 2$ based on fairly deep
(near-IR and mid-IR magnitude limits of $23-25$ AB)
and wide-area ($500-700$~arcmin$^2$) surveys, selected, respectively,
with IRAC and $K$-band data. The larger areas of these surveys,
compared to previous determinations, provide
results that are less susceptible to cosmic variance.
As shown in \citet{marchesini2009}, 
the $2\leq z\leq 3$ stellar mass functions in the literature,
when integrated over a fixed range in stellar mass 
($10^8 \leq M_{star}/M_{\odot} < 10^{13}$),
vary by a factor of $\sim 3$ in stellar mass density.
Figure~\ref{fig:stellarpop-MstarD-marchesini2009} 
\citep[from][]{marchesini2009} features a compilation of
global stellar mass density estimates as a function of redshift.
At $z\sim 2$, the reported fraction of the local stellar mass density
that is in place ranges from $\sim 8-25$\%, while that
number drops to $\sim 4-12$\% at $z\sim 3.5$.
\citet{marchesini2009} also offer an in-depth analysis
of the random and systematic uncertainties involved in constructing
a stellar mass function from the modeling of multiwavelength
photometry. These include the errors associated
with photometric redshifts, cosmic variance, differences among
stellar population synthesis codes \citep[as parameterized by, e.g.,][]{conroy2009},
choice of stellar IMF, stellar metallicity, and
extinction law. A proper accounting for these sources
of random and systematic error significantly increases
the uncertainties on derived quantities such as the 
evolution of the global stellar mass density,
as well as the evolution of galaxies as a function of
stellar mass. For example, when only taking into account
random uncertainties, \citet{marchesini2009} find that the abundance of
galaxies below the characteristic mass evolves more strongly
with redshift than that of the most massive galaxies $(M_{star}>10^{11.5})$,
which show a lack of strong evolution in number density. However, 
when the full random and systematic error budget is accounted for,
more significant evolution in the abundance of massive galaxies
cannot be ruled out.

Additional uncertainties in the stellar mass density at high redshift result from
uncertainties in the low-mass slope of the stellar mass function.
While most works adopt $\alpha$ in the 
range to $-1.0$ to $-1.4$, the stellar mass regime crucial
for constraining this parameter is not well probed with current
observations. For example, the dataset featured in \citet{marchesini2009}
suffers from incompleteness below $\sim 10^{10}M_{\odot}$. According
to \citet{reddysteidel2009}, if the stellar mass function
has a steeper low-mass slope, as suggested by the steep faint-end
slope of the rest-frame UV luminosity function and the relationship
between UV luminosity and stellar mass, up to $\sim 50$\% of the stellar
mass density may be contained in galaxies with stellar masses 
$\leq 10^{10} M_{\odot}$, as opposed to the $\sim 10-20$\%
inferred from extrapolating the Schechter fits of \citet{marchesini2009}.
In fact, using a $K$-selected sample based on significantly deeper
near-IR imaging with Subaru/MOIRCS ($K=24.1$ Vega),
\citet{kajisawa2009} estimate a steeper low-mass slope
of $\alpha=-1.5$ at $z\sim 2$ and $\alpha=-1.6$ at $z\sim 3$,
as well as perhaps a trend of $\alpha$ steepening with redshift, also
suggested by \citet{fontana2006}.
This result is intriguing, but the area over which the low-mass
slope is adequately probed is only $28 \mbox{ arcmin}^2$.
Clearly, data of this depth must be collected over a significantly
wider area to minimize the effects of cosmic variance
and obtain more robust constraints on the stellar mass density
at $z>2$. In particular, constraining the abundance and dust-extinction
properties of faint, low-mass galaxies at high redshift will prove very important
for comparisons of the past integral of the star-formation-rate density
with the stellar mass density at each redshift \citep{reddysteidel2009}. 
Careful comparisons of this sort are crucial for determining 
whether or not the integral of global past star formation indicates a discrepancy
with the global stellar mass density \citep{wilkins2008}.

\subsection{$SFR-M_{star}$ Scaling Relations}
\label{subsec:stellarpop-SFRM*}

\begin{figure}
\centerline{\psfig{figure=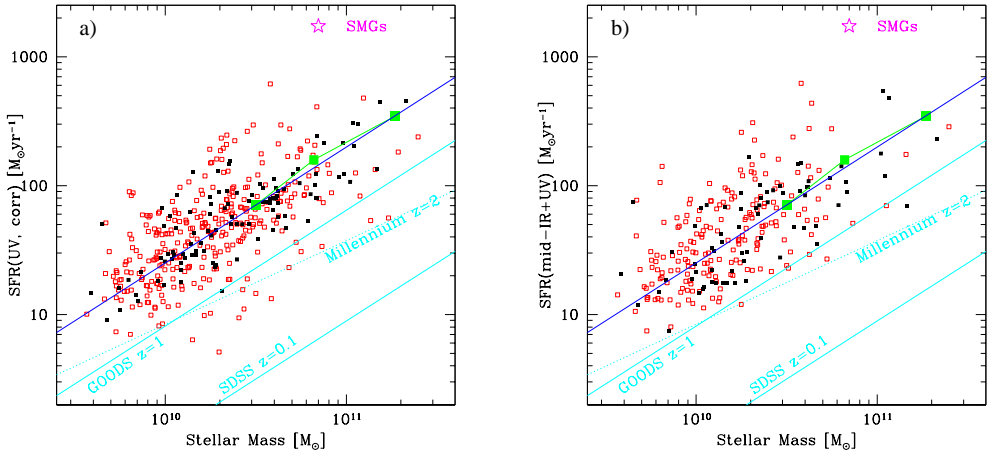,height=2.5in}}
\caption{\citep[From][]{daddi2007}
The relationship between SFR and stellar mass
for $z\sim 2$ star-forming galaxies in the GOODS fields.
Only galaxies with $24 \mu$m detections are included.
{\bf Left:} Star-formation rates are derived from UV luminosities,
corrected for dust extinction. {\bf Right:} Star-formation rates
are derived from the sum of mid-IR and UV-uncorrected luminosities.
In both panels, filled and empty squares are objects with and without
spectroscopic redshifts, respectively. The large green squares are the result
of the average $SFR-M_{star}$ relation in GOODS-N based
on radio stacking of $K<20.5$ galaxies in three bins of stellar mass.
The blue solid line is the functional form: 
$SFR = 200 (M_{star}/10^{11} M_{\odot})^{0.9} (M_{\odot}\mbox{ yr}^{-1})$.
The cyan solid lines are the $z=1$ and $z=0.1$ correlations from \citet{elbaz2007}.
The cyan dashed line is a prediction for $z=2$ from the Millennium simulation
and semi-analytic models. The magenta star indicates the location of typical
SMGs in this diagram. According to \citet{daddi2007}, 
star-forming galaxies at $z\sim 2$ appear
to follow a strong correlation between star-formation rate and stellar
mass. This trend is offset towards higher star-formation rates at
fixed stellar mass, relative to the correlations observed
among star-forming galaxies at lower redshift. The theoretical
model shown here for $z\sim 2$ galaxies tends to underpredict the active
rates of star formation for a given stellar mass.}
\label{fig:stellarpop-SFRM*-daddi2007}
\end{figure}

In addition to considering the evolution of star-formation rates
and stellar masses in galaxies separately, the evolution of the relationship
between these quantities provides important clues as to how stellar
mass builds up in galaxies as a function of redshift {\it and} mass.
At $z\sim 2$, this relationship is highlighted by \citet{daddi2007}
in a sample of star-forming $sBzK$ galaxies (passive $pBzK$ galaxies
and galaxies with no MIPS $24\mu$m detection were excluded from this analysis).
Using star-formation rates estimated from either dust-corrected
UV luminosity, or the sum of mid-IR and uncorrected UV luminosities, 
\citet{daddi2007} find a strong
correlation between star-formation rate (SFR) and stellar mass, described
by the relation, $SFR\propto M_{star}^{0.9}$, and shown in 
Figure~\ref{fig:stellarpop-SFRM*-daddi2007}. The ultra-luminous SMGs
are outliers to the $SFR-M_{star}$ trend, with star-formation rates a factor of
$\sim 10$ higher than expected, given their stellar masses. 
\citet{pannella2009} used 1.4 GHz radio stacking observations to estimate
average star-formation rates for 
non-AGN $sBzK$ objects in the Cosmic Evolution Survey (COSMOS) 
field, binned by stellar mass. The radio stacking analysis reveals
the  trend $SFR \propto M_{star}^{0.95}$, 
consistent with \citet{daddi2007}.
Accordingly, the specific star-formation rate, i.e. the
star-formation rate divided by stellar mass, appears to be roughly
constant over an order of magnitude in stellar mass ($10^{10} - 10^{11} M_{\odot}$,
assuming a Salpeter IMF from $0.1-100 M_{\odot}$). Similar trends
between star-formation rate and stellar mass have been observed among
star-forming galaxies at $z\sim 1$ by \citet{elbaz2007} and \citet{noeske2007},
but with a lower overall normalization, such that, at a given
fixed stellar mass, the expected star-formation rate is a factor of $\sim 4$
lower.  A comparison with the trend observed in
the local Universe \citep{brinchmann2004,elbaz2007} indicates an evolution
by a factor of $\sim 30-40$ \citep{daddi2007,pannella2009}.

On the other hand, perhaps the tightness of the $SFR-M_{star}$ correlation
has been overemphasized, as a result of selection effects. Using the same ultra-deep,
$K$-selected sample described in Section~\ref{subsec:stellarpop-MstarD},
\citet{kajisawa2010} investigate the $SFR-M_{star}$ relation 
from $0.5 \leq z \leq 3.5$. At $z\sim 2$, these authors find significantly
more scatter at the high-stellar-mass end of the relation. In this study,
UV dust-corrected and mid-IR+UV star-formation rates 
are estimated using the same techniques as in \citet{daddi2007},
but there is no requirement for galaxies to be detected at $24 \mu$m.
At $M_{star}=10^{11} M_{\odot}$,
for example, UV dust-corrected star-formation rates range from 
$1-1000 M_{\odot} \mbox{ yr}^{-1}$. Furthermore, the power-law
slope between star-formation rate and mass, and therefore, the specific
star-formation rate, tend to decrease at $M_{star}>10^{10.5} M_{\odot}$.
Given that the slope and small scatter of the $SFR-M_{star}$ relation
has been interpreted in terms of lending support to galaxy
formation models in which smooth gas accretion dominates the growth
of galaxies at high redshift \citep{dave2008}, it is crucial to
characterize these quantities accurately and in an unbiased manner.
The above studies are largely based on photometric redshifts, which tend
to increase the uncertainties in all derived physical properties. Therefore,
larger spectroscopic samples are needed. Based on such spectroscopic studies,
a description of the distribution of star-formation rates as a function
of stellar mass for a {\it stellar-mass selected} sample 
will provide the ideal observational probe of this potentially meaningful
trend.

\section{THE INTERSTELLAR CONTENT OF HIGH-REDSHIFT GALAXIES}
\label{sec:ISM}

The stellar content of galaxies offers an incomplete version of the story of their
formation and evolution, which must be filled in by a characterization
of their interstellar environments.
Indeed, the multiwavelength study of the current and 
past history of star formation in
galaxies cannot be constructed without understanding the nature
of dust. In addition to attenuating and
reddening the radiation from stars, dust also reradiates the
absorbed emission in the IR with a characteristic overall spectral 
shape that depends on temperature. The detailed
emission spectrum in the mid-IR offers a further probe of the energy sources heating
dust grains (i.e., radiation from star formation or an AGN). 
While dust reprocesses the light from stars, it is the cool
gas content of galaxies that forms the very fuel for star formation.
The elevated rate of star formation in galaxies in the early Universe
is a direct result of the large cool gas fractions in these systems.
As stars evolve and die, they return gas and heavy elements 
to the ISM. The patterns of chemical enrichment
in both gas and stars therefore reflect the past history of star formation,
and of gas inflows and outflows in galaxies. In this section, we review
what is known about the interstellar contents of high-redshift galaxies.
We begin by discussing both dust extinction and reprocessed emission,
and then turn to observations of the molecular gas and metals,
all of which provide important insights into the nature of distant
galaxies.

\subsection{Dust Extinction}
\label{subsec:ISM-dustext}

Starburst galaxies in the local Universe follow a correlation between
attenuation and reddening. As these galaxies become more attenuated
in the rest-frame UV (as probed by the ratio of far-IR to UV luminosities),
their rest-frame UV continua become redder. Based on multiwavelength
observations spanning from the UV to far-IR, this correlation
has been quantified as a relationship between the rest-frame UV
slope, $\beta$ (where $f_\lambda \propto \lambda ^{\beta}$),
and $A_{1600}$ (the attenuation at 1600~\AA), such that
$A_{1600} = 4.43 + 1.99 \beta$ \citep{meurer1999}. The starburst obscuration curve
of \citet{calzetti2000} describes the wavelength dependence
of effective attenuation when dust is distributed in a patchy
foreground screen, relative to young stars. 
For reference, the \citet{calzetti2000}
curve lacks the 2175~\AA\ bump characteristic of the Milky
Way extinction curve, and has a different ratio
of total to selective extinction, with $R_V=A_V/E(B-V)=4.05$,
as opposed to the average Milky Way value of 3.1 
\citep{cardelli1989}. The \citet{calzetti2000}
law is also ``grayer" than the SMC law \citep{prevot1984},
rising less steeply in the near-UV, and therefore implying
more attenuation for a given reddening in the rest-frame UV.
This ``grayness" likely stems from a geometrical
configuration between gas and stars in which the dust
is distributed in a patchy foreground-like screen, 
and some stars suffer little or no extinction \citep{calzetti2001}.
This starburst attenuation curve predicts
an almost identical relation between $A_{1600}$ and $\beta$
to that of \citet{meurer1999},
with the assumption of fairly uniform intrinsic rest-frame
UV colors for starburst galaxies.

Soon after the discovery of LBGs at $z\sim 3$, 
their observed range of rest-frame UV colors
was interpreted in terms of a range of dust reddening
\citep{steidel1999,meurer1999,calzetti2000}, in analogy with the description
of local starbursts.  The \citeauthor{calzetti2000}
obscuration curve was used to translate between rest-frame UV color
and $E(B-V)$, and, by extension, $A_{1600}$, with the average
value of $E(B-V)=0.15$ corresponding to an attenuation factor of $\sim 4.7$
at $1500$~\AA. In \citet{steidel1999}, this factor was used to  correct the observed
UV-luminosity densities at $z\sim 3$ and $z\sim 4$ and show that 
only a small fraction of the intrinsic UV radiation typically escapes from 
even UV-selected galaxies at high redshift. Furthermore,
consistent with the trend observed among starbursts in the local 
Universe, \citet{adelberger2000}
find that objects with greater bolometric luminosities suffer
more extinction in the rest-frame UV. However, the trend at $z\sim 3$
is offset from the local one in the sense that, for a given
bolometric luminosity (i.e. star-formation rate), the extinction
is significantly smaller at higher redshift. \citet{bouwens2009}
have investigated the average reddening as a function of rest-frame
UV luminosity (uncorrected for dust), demonstrating that, 
at $z\sim 2.5$ and $z\sim 4.0$, $\beta$ becomes bluer
for fainter objects. Furthermore, the average $\beta$
value of the most UV-luminous galaxies is redder at $z\sim 2.5$
than at $z\sim 4.0$. The trend between $\beta$ and UV luminosity
was not apparent in the datasets of \citet{adelberger2000}
and \citet{reddy2008}, most likely due to the smaller dynamic range
of UV luminosities probed.

At this point, the \citet{calzetti2000} law is the most commonly adopted extinction
law for modeling the stellar populations of high-redshift
galaxies (Section~\ref{sec:stellarpop}). 
Therefore, it is important to consider the empirical support
for this choice, based on comparisons between UV-extinction-corrected and
extinction-free estimates of star-formation rates. While individual
star-forming galaxies are not detected in deep {\it Chandra} X-ray and 
and VLA 1.4 GHz radio imaging observations, stacking methods
have proven very powerful for comparing different multiwavelength 
star-formation rate indicators. \citet{reddy2004} measure the
stacked X-ray and radio fluxes for spectroscopically-confirmed,
UV-selected star-forming
galaxies at $1.5\leq z \leq 3.0$ in the GOODS-N field. On average,
the inferred X-ray and radio-derived star-formation rates
are, respectively, $42$ and $56\; M_{\odot}\mbox{ yr}^{-1}$.
The UV star-formation rate, extinction corrected 
based on rest-frame UV colors and assuming the validity
of the \citeauthor{calzetti2000} law, is
$50\; M_{\odot}\mbox{ yr}^{-1}$ -- consistent with 
the radio and X-ray estimates. Furthermore, the
ratio of star-formation rates derived from X-ray and
uncorrected UV fluxes implies a factor of $\sim 4.5-5.0$
attenuation in the UV, consistent with the attenuation
inferred from the UV colors. \citet{pannella2009} perform
a similar radio stacking analysis for star-forming
$sBzK$ galaxies with photometric redshifts at $z\sim 2$ in the COSMOS
field, finding that the UV attenuation, $A_{1500}$, derived
from the ratio between the extinction-free radio-derived
star-formation rate and the UV-uncorrected star-formation rate, agrees well with
that inferred on the basis of rest-frame UV colors alone, assuming
the \citet{calzetti2000} law applies.

{\it Spitzer}/MIPS $24 \mu$m observations have allowed
for a test of extinction on a per object
basis at $z\sim 2$. \citet{reddy2006,reddy2010} compute mid-IR
luminosities from observed $24 \mu$m fluxes for 
spectroscopically-confirmed UV-selected galaxies, and extrapolate
these to total IR luminosities using local templates
\citep{elbaz2002}. The relation between extinction ($L_{FIR}/L_{1600}$)
and reddening ($\beta$) for these UV-selected galaxies is compared
with the relation among local starbursts, revealing that the majority
of objects (though not all -- see below)  follow the local trend. Furthermore,
\citet{reddy2006} confirm the result of \citet{adelberger2000} of the positive
correlation between bolometric luminosity and extinction, but now
based on a {\it Spitzer}/MIPS estimate of the IR luminosity. The
evolution in extinction with redshift is also reproduced using
{\it Spitzer}, which suggests that extinction in the rest-frame UV
is a factor of $\sim 10$ smaller at $z\sim 2$ than at $z\sim 0$,
at fixed bolometric luminosity.
\citet{daddi2007} perform an analogous test of UV extinction laws, 
using star-formation rates for star-forming
$sBzK$ galaxies derived from both rest-frame UV and {\it Spitzer}
$24 \mu$m fluxes. \citet{daddi2007} also conclude that the 
\citeauthor{calzetti2000} law is valid for the the majority
of systems at $z\sim 2$.

\begin{figure}
\centerline{\psfig{figure=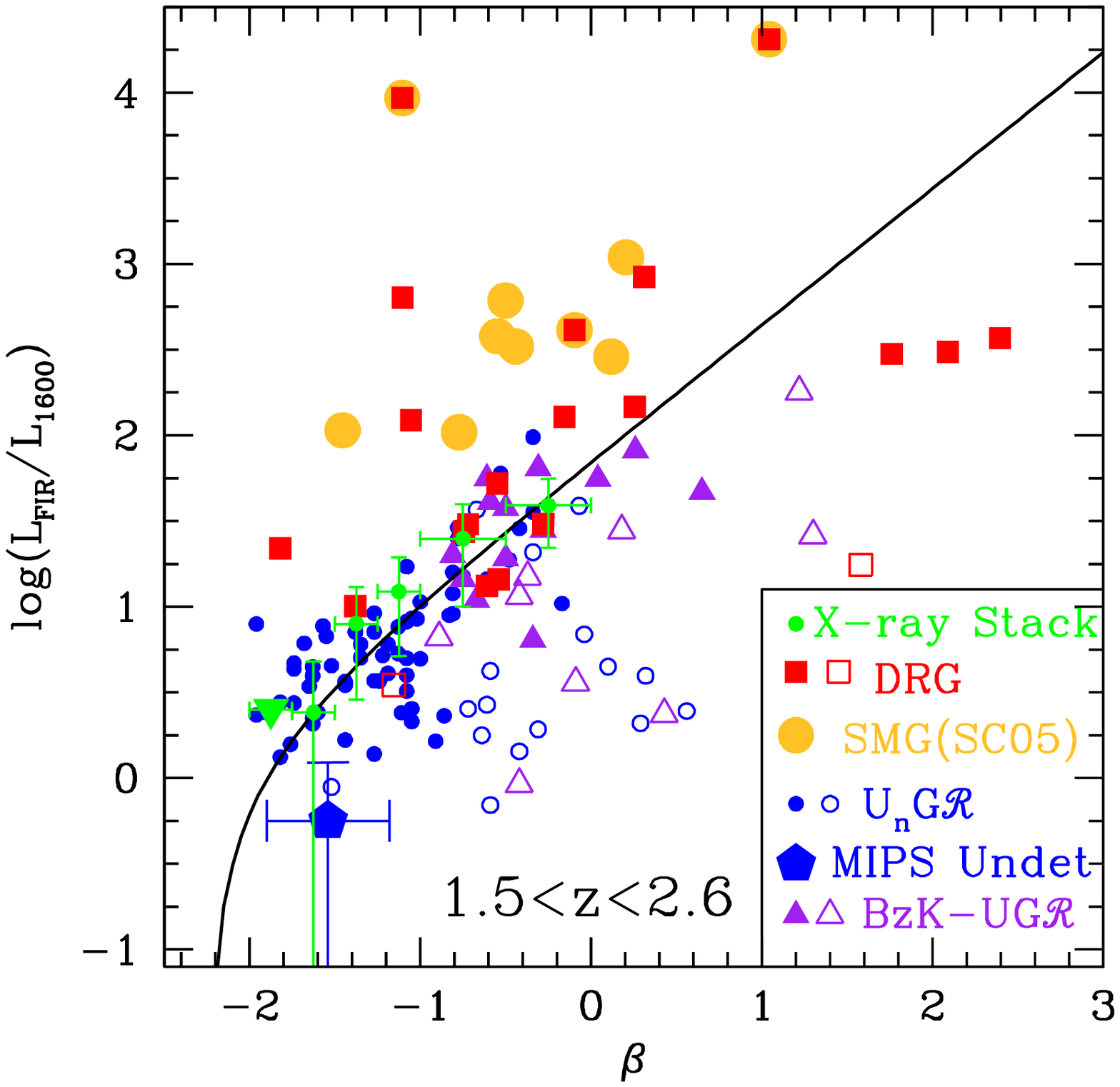,height=3in}
\psfig{figure=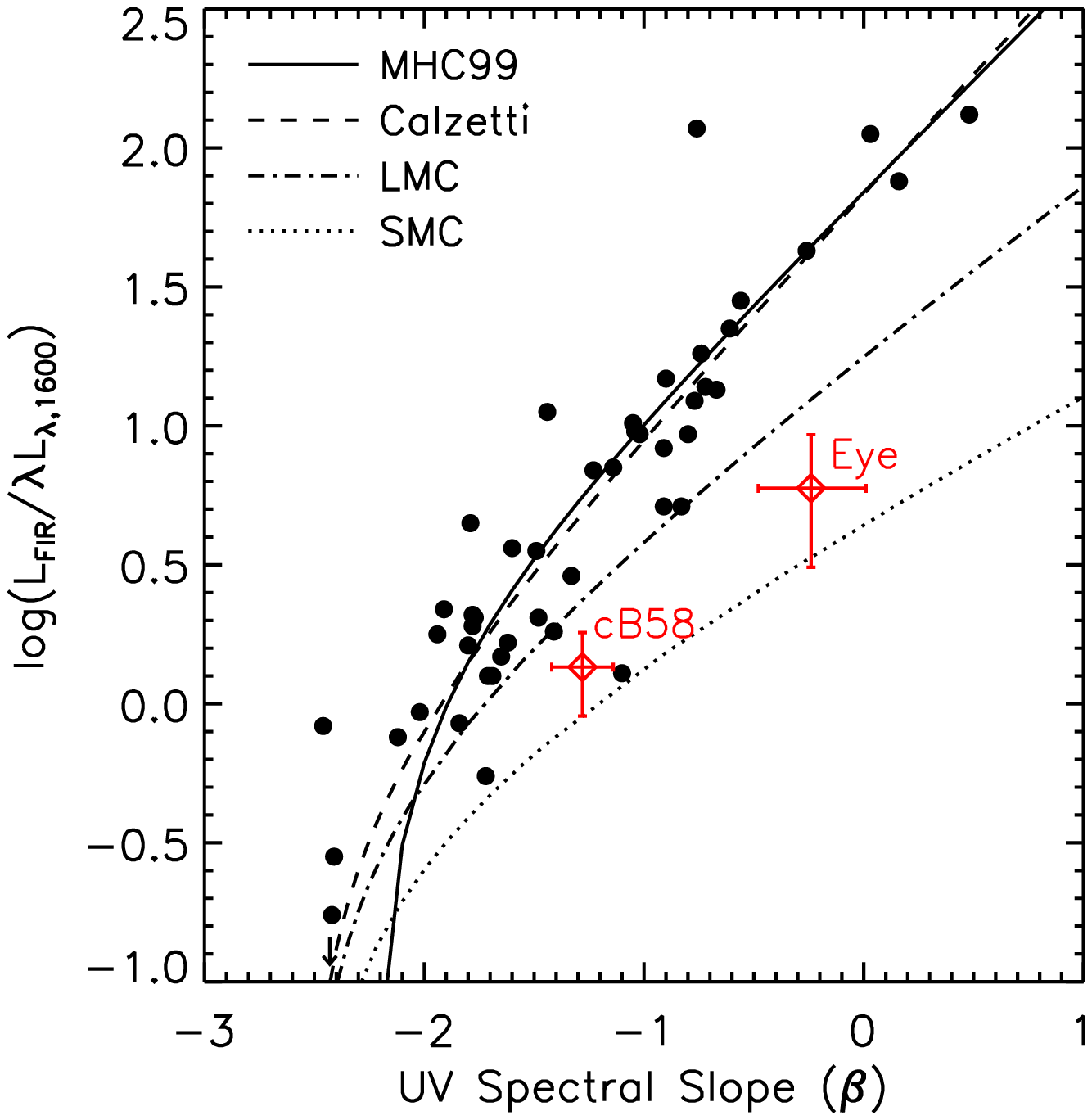,height=3.2in}}
\caption{Dust Extinction in High-redshift Galaxies.
{\bf Left:} \citep[From][]{reddy2006} Dust absorption,
parameterized by $F_{FIR}/F_{1600}$ vs. rest-frame UV spectral
slope, $\beta$, for galaxies at $1.5 <z < 2.6$. Filled and
open symbols, respectively, indicate galaxies with inferred
stellar population ages of $>100$ and $<100$~Myr, for UV-selected
galaxies (blue), $BzK$ galaxies (purple), and DRGs (red). Filled yellow circles
indicate SMGs. The large blue
pentagon shows the results for UV-selected galaxies
undetected at $24\mu$m, using $24\mu$m stacking results.
The green filled circles represent the results from 
X-ray stacking analysis. 
The solid line indicates the \citet{meurer1999} relation found for local UV-selected
starburst galaxies.  {\bf Right:} \citep[From][]{siana2009} 
Similar quantities, but for local starbursts (filled black circles), and
two $z\sim 3$ gravitationally-lensed objects, cB58 and The Cosmic Eye (empty red
diamonds). Also shown are the \citet{meurer1999} relation (solid curve),
and the predicted relation between $L_{FIR}/\lambda L_{\lambda,1600}$
and $\beta$ for different reddening curves (Calzetti, LMC, and SMC).}
\label{fig:ISM-dustext-reddy2006siana2009}
\end{figure}

While the success of the \citet{calzetti2000} law is impressive,
in terms of correcting
UV luminosities and matching various extinction-free tracers
of star-formation rates for large samples of high-redshift galaxies, it is also
important to highlight the cases in which it appears to fail.
For example, in ultraluminous SMGs at $z\sim 2$, the dust-corrected
UV luminosities (assuming the \citeauthor{calzetti2000} law)
underpredict the bolometric luminosities suggested by
submillimeter and radio luminosities by as much as a factor
of $\sim 10- 100$ \citep{chapman2005, reddy2006,daddi2007}. A
related yet rather non-intuitive fact is that $\geq 50$\%
of the SMGs in the \citet{chapman2005} sample have rest-frame
UV colors that actually satisfy the UV-selection criteria of \citet{steidel2003,steidel2004}.
The mismatch between predicted (based on rest-UV color) and observed IR luminosities,
shown in Figure~\ref{fig:ISM-dustext-reddy2006siana2009} (left),
may arise in systems where regions of massive star formation are
completely opaque to UV radiation, and the UV radiation that
does escape is from regions that are disjoint from the dusty ones dominating
the bolometric output \citep{daddi2007}. \citet{reddy2006}
find that the \citet{calzetti2000} law appears to break
down at $z\sim 2$ for systems more luminous than $\sim 2\times 10^{12} L_{\odot}$.
On the other hand, \citet{magdis2010a} demonstrate that, at $z\sim 3$,
the starburst attenuation curve yields consistent results for galaxies
with bolometric luminosities as large as $10^{13} L_{\odot}$.
In general, however, the \citet{calzetti2000} law does not
appear valid for describing dust extinction in the most luminous
sources at $z\geq 2$.

While the local starburst relation appears to underpredict
the UV attenuation for the most bolometrically luminous systems,
there is a discrepancy in the opposite sense for objects
that, based on their stellar population modeling, appear ``young,"
with best-fit ages (assuming constant star-formation histories)
less than 100 Myr. This second discrepancy was highlighted most
dramatically in the case of the strongly gravitationally-lensed objects
MS1512-cB58 (or ``cB58") \citep{pettini2000}  and the Cosmic Eye 
\citep{smail2007}. Both of these objects have rest-frame UV colors
and continuum slopes that suggest dust-corrected star-formation
rates several times larger than what is actually measured
using {\it Spitzer}/MIPS mid-IR photometry and InfraRed Spectrograph (IRS)
spectroscopy (a factor of $\sim 3-5$ for cB58, and $\sim 8$
for the Cosmic Eye) \citep{siana2008,siana2009}. This discrepancy,
shown in  Figure~\ref{fig:ISM-dustext-reddy2006siana2009} (right),
was previously noted for cB58 on the basis of millimeter and
submillimeter observations \citep{baker2001,sawicki2001}.

A similar discrepancy is observed by \citet{reddy2006,reddy2010}
among UV-selected galaxies with best-fit stellar population ages
of $t\leq 100$~Myr. As shown in  Figure~\ref{fig:ISM-dustext-reddy2006siana2009}
(left), the measured ratio of $L_{FIR}/L_{UV}$
for these young systems falls significantly below the local
starburst relation, given their observed rest-frame UV
slopes, $\beta$. Both cB58 and the Cosmic Eye are described
by similarly young ages as well. Therefore, it appears
that these young systems are described by a different
extinction law, one that is steeper (as in the 
case of the SMC curve), such that a given
observed amount of reddening corresponds to less attenuation
in the rest-frame UV. More work is needed to develop a 
self-consistent evolutionary scenario to explain the reddening
and attenuation properties of star-forming galaxies as a function
of the maturity of their stellar populations. Finally, 
we call attention to recent results based on {\it Herschel} PACS $160 \mu$m
measurements of star-forming galaxies at $z\sim 2$ \citep{nordon2010}. A comparison
of PACS and UV, dust-corrected star-formation rates
indicates that the UV-corrected values are, on average, 0.3 dex
higher (with a scatter of 0.35 dex). More extensive
comparisons with upcoming PACS far-IR measurements will be vital
for further testing of the \citeauthor{calzetti2000} law.

Related to the overall reliability of the \citeauthor{calzetti2000}
law, we must also consider the question of differential
extinction of the radiation from stars and ionized gas.
In local starbursts, emission-line tracers of ionized gas 
appear to be systematically more attenuated than
the stellar continuum at similar wavelengths. The relationship
derived between the stellar and nebular extinction is 
$E(B-V)_{stars} = 0.44 E(B-V)_{nebular}$ \citep{calzetti2000}.
Currently, there is conflicting evidence about the relative
extinction of stars and gas at $z\sim 2$. Comparing
star-formation rates inferred from rest-frame UV and
H$\alpha$ luminosities for a sample of $z\sim 2$ UV-selected galaxies
(see Section~\ref{subsec:ISM-metals}), \citet{erb2006c} conclude that
$E(B-V)_{stars} \approx E(B-V)_{nebular}$, and that
a \citeauthor{calzetti2000} law applied to correct both
UV-continuum and H$\alpha$ measurements
gives rise to the best agreement between
star-formation rate indicators.
On the other hand, \citet{forsterschreiber2009}
find for a set of star-forming galaxies with 
VLT/SINFONI integral-field unit (IFU) maps of H$\alpha$ emission at roughly
the same redshift as the \citeauthor{erb2006c} sample (see Section~\ref{subsubsec:structure-dynamics-ifu})
that the best agreement between H$\alpha$
and UV-derived star-formation rates
results when H$\alpha$ luminosities are corrected
by an additional factor of $\sim 2$, and with the
assumption $E(B-V)_{stars} = 0.44 E(B-V)_{nebular}$.
To settle the question of differential extinction, 
much larger samples of objects are required with both
measurements of multiple Balmer emission lines and
rest-frame UV estimates of star-formation rates. Assembling
these measurements will be possible with the next generation of
multi-object near-IR spectrographs presently coming on-line on $8-10$-meter
class telescopes.

\subsection{Dust Emission}
\label{subsec:ISM-dustem}

In addition to absorbing ultraviolet and optical radiation from
stars, dust re-emits at IR and submillimeter wavelengths. 
Direct observations of this re-radiated 
emission at long wavelengths have opened a window 
into the nature of dust in distant galaxies, in terms of its temperature, composition,
and the sources heating it. At far-IR and submillimeter wavelengths, direct
observations of dust emission are restricted to the most luminous sources,
with $L_{bol} > 10^{12} L_{\odot}$. While mid-IR imaging observations
have been obtained for lower-luminosity systems using {\it Spitzer}/MIPS, 
spectroscopy has been limited to the most luminous sources,
selected on the very basis of their bright submillimeter or mid-IR emission
[except in cases of strongly gravitationally-lensed systems \citep{siana2009}].

Until recently, direct measurements
of dust temperatures in $z>2$ ULIRGs only existed for small samples.
Indirect estimates of the dust temperature,
$T_d$, in SMGs were obtained by measuring the flux at individual
rest-frame far-IR (i.e., observed 850~$\mu$m)
and radio (i.e., 1.4~GHz) wavelengths, and assuming
that the local correlation between far-IR and radio
luminosities applies \citep{condon1992}. Based on the ratio between
850~$\mu$m flux and inferred total far-IR luminosity, 
$T_d$ can be inferred, assuming that the dust emission follows a single-temperature 
modified blackbody spectrum
of the form $S_{\nu}\propto \frac{\nu^{3+\beta}}{\exp(h\nu/kT_d)-1}$,
with the emissivity, $\beta$, set to a value of $1.5$.
\citet{chapman2005} use such a method to characterize their sample of 73 SMGs with
spectroscopic redshifts at a median redshift of $z=2.2$. 
For this sample, the median inferred dust temperature is $T_d=36\pm 7$~K,
$\sim 5$~K cooler than local ULIRGs with similar IR luminosities.
In order to constrain the dust temperature more directly,
photometric measurements at multiple rest-frame far-IR wavelengths
are required, and the observations between {\it Spitzer}/MIPS
and SCUBA/850~$\mu$m wavelengths have until recently been limited. These
include Caltech Submillimeter Observatory (CSO) Submillimeter
High Angular Resolution Camera (SHARC-2) observations at 350~$\mu$m
from \citet{kovacs2006} and \citet{coppin2008} for a total of $\sim 30$ SMGs
with previous $850 \mu$m detections.
In these studies the additional far-IR SED point
suggests a median $T_d=30-35$~K, consistent with the earlier,
indirect estimate of $T_d=36$~K for SMGs.
Furthermore, the total far-IR luminosities of SMGs
in these samples are better constrained and typically
$L_{FIR}\sim \mbox{ few} \times 10^{12} L_{\odot}$, with dust masses
$M_{dust}\sim 10^9 M_{\odot}$, significantly larger than
those observed in local starburst galaxies \citep{coppin2008}.

New {\it Herschel} observations at $100$ and $160 \mu$m with
PACS and at $250$, $350$, and $500 \mu$m with the 
Spectral and Photometric Imaging Receiver (SPIRE)
have recently provided significantly more refined estimates of dust temperatures
in high-redshift ULIRGs, and revealed a temperature diversity
hinted at by earlier, indirect studies
\citep{chapman2004,casey2009}. Combining data in the three SPIRE channels
with $850\mu$m SCUBA measurements, \citet{chapman2010}
find a median of $T_d=34\pm 5$~K for a sample of 31 SMGs
and a hotter median $T_d=41\pm 5$~K for 37 radio-selected
ULIRGs with fainter submillimeter fluxes (where the errors
represent the standard deviations of the samples, not of the median values).
These samples are characterized
by median far-IR luminosities of $L_{IR}=7.1\times 10^{12} L_{\odot}$
and $3.8\times 10^{12} L_{\odot}$, respectively. Such large
luminosities were inferred previously on the basis of much more
limited rest-frame far-IR data, with large uncertainties on the
total luminosity due to its strong dependence on assumed dust
temperature. The constraints on the far-IR SED shape now offer
a much more robust indication of the dust luminosity and temperature.
\citet{magdis2010b} focus on combined PACS and SPIRE observations
for a sample of star-forming MIPS $24\mu$m-selected ULIRGs at $z\sim 2$,
finding a broad range of dust temperatures spanning from  
$25 \leq T_d \leq 65$~K, with a median of $T_d=42$~K, and a range
of IR luminosities of 
$1.7\times 10^{12} L_{\odot} \leq L_{IR} \leq 8.7 \times 10^{12} L_{\odot}$. 
In particular, and as emphasized previously by \citet{chapman2004}, it is shown that
ULIRGs with hotter dust temperatures and fainter submillimeter fluxes at $850-1200\mu$m
would be missed in current ground-based
surveys of SMGs, given their sensitivity limits at $\sim 1$mm.
Therefore, ground-based surveys for $z>2$ ULIRGs selected on the basis
of $850\mu$m fluxes may be biased towards the coolest of the most luminous
galaxies. The origin of the diversity in dust temperatures among these most
luminous systems is an open question.

While the far-IR SED offers constraints on
the thermal properties and total luminosities of the most
luminous galaxies at $z>2$, the mid-IR spectral range reveals
the nature of their smaller dust grains and PAH molecules.
Furthermore, a fundamental issue regarding ULIRGs at 
high redshift concerns the nature of the sources heating
the dust that emits such copious amounts of
far-IR radiation. The two basic alternatives are star-formation
or AGN activity. While rest-frame UV and optical 
spectra \citep{chapman2005,swinbank2004}
and X-ray observations \citep{alexander2005}
of SMGs indicate evidence for AGN activity in these systems,
it is challenging to assess the bolometric importance of the AGNs from these data.
The rest-frame mid-IR spectral region, from $\sim 5-15 \mu$m,
offers strong discriminatory power between star-formation and AGN activity,
and constraints on the relative contributions of each to the 
bolometric luminosity of high-redshift ULIRGs. The energetics
of the radiation field are determined by the prominence
of PAH emission features at $6.2$, $7.7$, $8.6$, and $11.3 \mu$m,
relative to the strength of an underlying power-law continuum
from hot dust and silicate dust absorption at $9.7\mu$m. 
The PAH emission is primarily tied to star formation, whereas
the hot dust continuum is associated with AGN activity.
The strength of silicate absorption potentially
indicates the importance of a buried nuclear component \citep{sajina2007,
pope2008}.

The IRS instrument onboard {\it Spitzer} has proven critical for untangling
the processes powering the dust emission in ULIRGs at $z\sim 2$.
Strikingly, the mid-IR spectra of the majority of SMGs exhibit strong 
PAH emission features \citep{menendez2009,pope2008}, indicating
the dominance of star-formation over AGN activity in these systems.
While many SMGs have X-ray properties suggesting
the presence of an AGN, this component does not
appear to be energetically dominant in terms of the
bolometric luminosity, typically contributing $\leq 30$\% of the luminosity
in the mid-IR \citep{menendez2009,pope2008}. Only 4 out of the 24 sources
presented in \citet{menendez2009}, and 2 out of 13 of those described
in \citet{pope2008}, have mid-IR spectra dominated by a power-law continuum,
and the composite spectra for both samples exhibit pronounced
PAH features. On the other hand, the $z\sim 2$
$24\mu$m-selected ULIRGs analyzed by \citet{sajina2007}
are more heterogeneous in the mid-IR. The majority ($\sim 75$\%) of these
sources have mid-IR spectra dominated by a power-law continuum
and therefore AGN activity. At the same time, more than half of these power-law
sources have PAH emission features as well, indicating contributions
from both AGN and star-formation processes. The minority ($\sim 25$\%) of 
PAH-dominated sources also indicate evidence for AGN
activity in the form of hot dust continuum. Furthermore, the 
significant strength of $9.7\mu$m silicate absorption in $\sim 25$\% of the
sample indicates the presence of an obscured, compact nuclear component.

\begin{figure}
\centerline{\psfig{figure=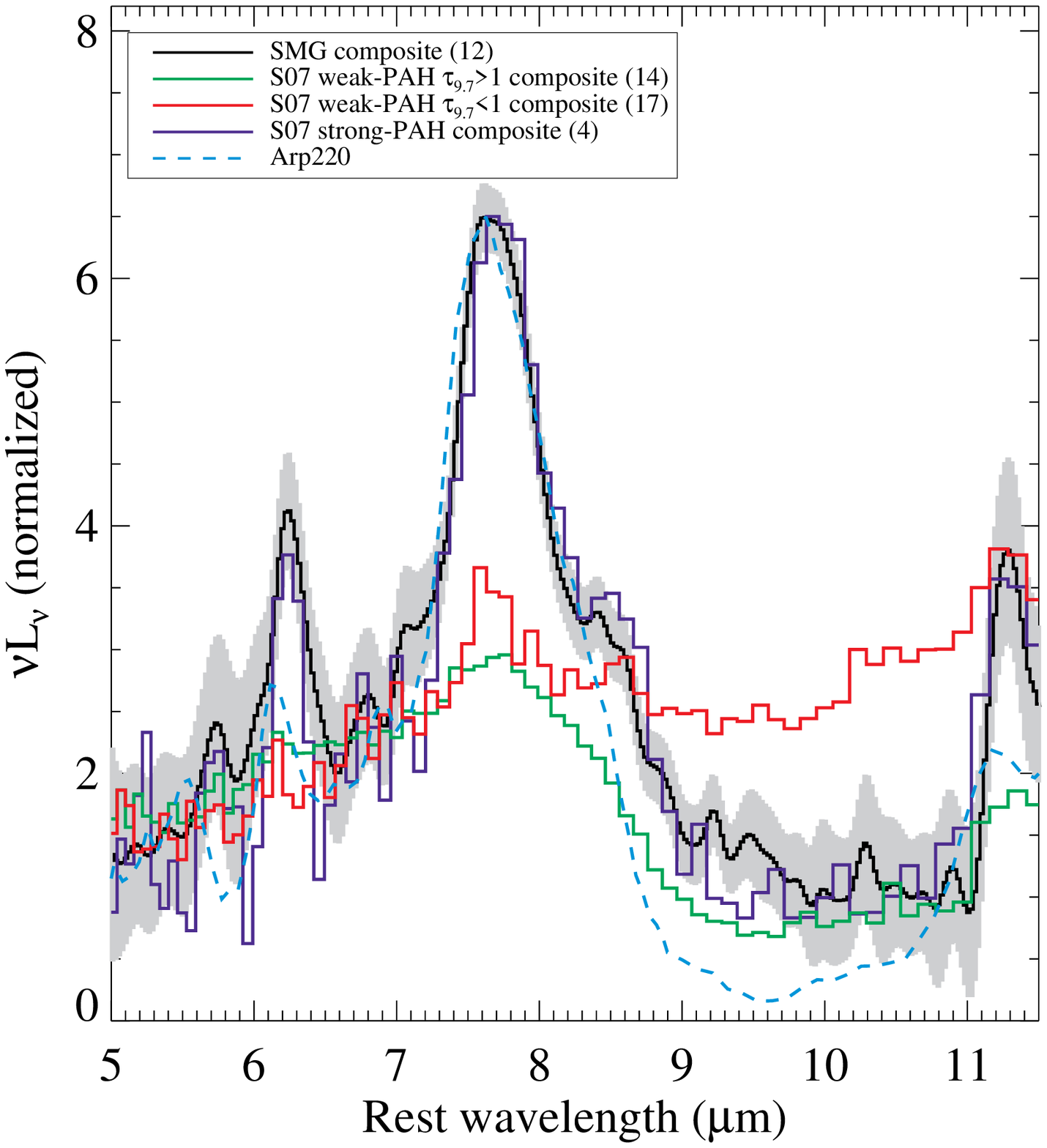,height=3in}}
\caption{\citep[From][]{pope2008} Mid-IR Spectrum of ULIRGs at high redshift.
The composite {\it Spitzer}/IRS spectrum of 12 SMGs from \citet{pope2008} is shown
as the solid black line. The purple, red, and green solid histograms
are composites from the sample of {\it Spitzer} $24 \mu$m-selected
ULIRGs from \citet{sajina2007}, with purple indicating the
composite of 4 strong-PAH
sources, green indicating the composite of 14 weak-PAH, power-law-dominated sources with
significant 9.7-$\mu$m silicate absorption ($\tau_{9.7}>1$), 
and red indicating the composite of 17 weak-PAH, 
power-law-dominated sources with weaker 9.7-$\mu$m silicate absorption ($\tau_{9.7}<1$).
The light-blue dashed curve is the mid-IR spectrum of the local
ULIRG, Arp 220. All curves have been normalized at $\sim 7 \mu$m.
The numbers in the legend indicate the number of sources in each composite
spectrum. The prominence of PAH emission in the
SMG composite indicates that star formation is the main
energy source powering the mid-IR spectrum. On the other hand,
the $24 \mu$m-selected ULIRGs indicate a range of spectral
types, the majority of which appear to be dominated
by AGN emission.
}
\label{fig:ISM-dustem-pope2008}
\end{figure}

Figure ~\ref{fig:ISM-dustem-pope2008}, from \citet{pope2008},
illustrates the diversity among high-redshift ULIRG mid-IR spectra,
including the PAH-dominated flavor common among SMGs and in
a minority of $24 \mu$m-selected ULIRGs, and the power-law
dominated spectra common in $24 \mu$m-selected ULIRGs, with
different amounts of silicate absorption. While SMGs and
$24 \mu$m-selected ULIRGs appear to have similar bolometric luminosities,
AGNs appear to play a more significant role in the $24 \mu$m-selected ULIRGs,
which also have hotter dust temperatures. 
These differences have been interpreted in terms of an
evolutionary scenario, in which
SMGs and $24 \mu$m-selected ULIRGs
represent, respectively, earlier and later stages of a 
gas-rich, major merger event \citep{pope2008,yan2010}.
Through the progression of these stages, obscured nuclear AGN activity grows
in importance as the system evolves into an unobscured QSO,
and, eventually, a massive elliptical galaxy. While this proposed scenario
is intriguing, additional
constraints on the number densities, stellar mass and gas content
of these various high-redshift samples is necessary
to establish robust connections between them \citep{yan2010}.

\subsection{Molecular Gas Content}
\label{subsec:ISM-gas}

For a complete characterization of the process of star 
formation at high redshift, observations of the molecular 
phase of the ISM are critical. Stars form directly from this 
dense interstellar phase, which dominates the cool gas content 
of the most actively star-forming galaxies \citep{blitz2006}. 
In the local Universe, the star-formation rate per unit area,
$\Sigma_{SFR}$, is directly related to the total gas surface 
density, $\Sigma_{gas}$, according to the empirical Schmidt 
Law, $\Sigma_{SFR} \propto \Sigma_{gas}^N$, with observational
determinations of $N$ ranging from $0.9-1.7$ 
\citep{kennicutt1998}. Tracing this relationship at high 
redshift is key to understanding how gas is converted into 
stars during the epoch of peak star formation.
Furthermore, the star-formation rate 
and gas content can be related to characterize the 
efficiency of star formation, as well as the timescale on 
which gas will be depleted. With short (compared to the Hubble 
time) gas depletion timescales, star formation can only be 
sustained by the ongoing accretion of gas from the IGM. The 
relationship between star formation and the available gas 
reservoir, in terms of its mass, baryonic mass fraction, and 
spatial extent, therefore offers crucial inputs into models of 
galaxy formation, which must include a proper description of 
the balance between gas accretion, the conversion of gas into 
stars, and the energetic feedback related to the process of 
star formation.


In the local Universe, rotational transitions of carbon 
monoxide (CO) serve as excellent tracers of molecular hydrogen 
gas \citep{young1991}. These same transitions are used to 
trace the molecular gas content of high-redshift galaxies, 
using millimeter-wave and radio telescopes. The first such 
observations were of extreme sources with known ultraluminous 
far-IR luminosities, such as QSOs and SMGs, and were suggestive of 
large amounts of dust and gas \citep{omont1996,genzel2003}. 
Recently, however, CO detections have been achieved for less 
extreme systems, well into the LIRG regime and falling on the 
correlation between $SFR$ and $M_{star}$ that appears to 
describe more quiescently star-forming galaxies over the range 
$10 M_{\odot}\mbox{yr}^{-1} \leq SFR \leq \sim \mbox{few} 
\times 100 M_{\odot}\mbox{yr}^{-1}$ 
\citep{daddi2007,tacconi2008}. For observations of CO at high 
redshift, the IRAM Plateau de Bure Interferometer (PdBI) has 
played a dominant role, tuned to detecting various upper-level 
transitions [e.g. CO(2-1), CO(3-2), CO(4-3), and so on] in the 
millimeter range of the spectrum. Very recently, 
longer-wavelength observations carried out at the Expanded 
Very Large Array (EVLA) and Green Bank Telescope (GBT) have 
been tuned to the ground-state CO(1-0) transition, which is 
more directly tied to the molecular gas mass 
\citep{ivison2011,harris2010}.


In order to infer the mass of molecular gas, $M_{gas}$, 
associated with detected CO emission, there are two major 
sources of systematic uncertainty. One is in the conversion 
factor between CO(1-0) emission luminosity and $M_{gas}$. The 
second results from the fact that, for the most part, 
high-redshift CO observations have been of upper-level $J$ 
transitions, whereas the CO-to-$H_2$ conversion factor is 
calibrated for the CO(1-0) transition. The ratios between 
upper-level and ground-state $J$ transitions depend on the 
excitation and physical conditions in the molecular gas, and 
incorrect assumptions about these conditions will result in a bias 
in the inferred CO(1-0) luminosity. In some cases, higher-$J$ 
transitions, which are sensitive to warmer and denser regions, 
may even offer an incorrect representation of the spatial 
distribution of the full molecular gas reservoir 
\citep{ivison2011}.

As for the CO-to-$H_2$ conversion factor, $\alpha$, it has been 
calibrated in the Milky Way with a value of $\alpha_{MW}\sim 
4-5$, in units of $M_{\odot} (K \mbox{ km s}^{-1}\mbox{ 
pc}^2)^{-1}$. This value is very close to the theoretical 
expectation based on the assumption that CO line emission is 
produced in discrete, virialized clouds obeying scaling 
relations among their masses, sizes, and linewidths 
\citep{young1991}. On the other hand, in galactic nuclei and 
local starburst galaxies, different dynamical and geometrical 
conditions apply to the molecular gas, which may reside in a 
smoother, diskier configuration whose motions are additionally 
affected by the gravitational potential from stars. These 
differences result in a lower conversion factor for starbursts 
of $\alpha=0.8-1.6$ \citep{tacconi2008}. For observations of 
high-redshift SMGs, the starburst conversion factor has been 
adopted \citep{tacconi2006,tacconi2008}. In fact, using the 
higher, Milky Way conversion factor would result in baryonic 
(gas plus stellar) masses in excess of the measured dynamical 
masses. On the other hand, for the lower-luminosity LIRGs 
(both UV-selected and $BzK$ sources), a Milky Way type 
conversion factor is favored on the basis of similar dynamical 
arguments \citep{daddi2010} and the idea that CO emission 
in these LIRGs arises in virialized clouds with densities 
similar to those observed in local quiescent disk galaxies 
\citep{tacconi2008}.

We now consider the excitation of the molecular gas, which will 
affect the conversion between higher-$J$ transitions and 
CO(1-0). While the CO levels in SMGs appear to be thermally 
populated up to $J\geq 3$, simultaneous observations of 
CO(1-0) (VLA), and CO(2-1), and CO(3-2) (PdBI) transitions in 
a $BzK$ LIRG at $z\sim 1.5$ suggests that the CO(3-2) level is 
significantly subthermally excited, similar to what is 
observed in the Milky Way and other local disk galaxies 
\citep{dannerbauer2009}. These low-excitation physical 
conditions require a larger conversion factor from CO(3-2) to 
CO(1-0), and therefore, a larger inferred molecular gas mass 
for a given CO(3-2) line luminosity. 

\begin{figure}
\centerline{\psfig{figure=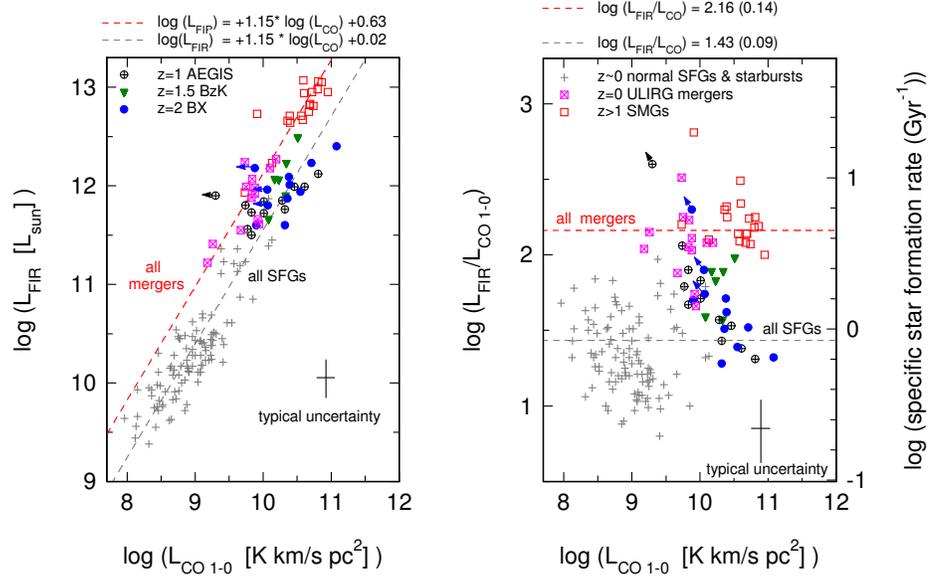,height=3in,angle=270}}
\caption{\citep[Adapted from][]{genzel2010a} Relationship between FIR
luminosity and CO(1-0) luminosity in star-forming galaxies at
low and high redshift. Gray crosses indicate isolated star-forming
galaxies at $z\sim 0$. Magenta crossed squares are merging ULIRG galaxies
at $z\sim 0$. Black crossed circles are $z\sim 1$ star-forming galaxies
from the AEGIS field \citep{davis2007}. Filled blue circles are $z \sim 2$ UV-selected
``BX" galaxies \citep{steidel2004}. 
Filled green triangles are $z\sim 1.5$ $BzK$ galaxies \citep{daddi2010}. 
Red empty squares are $z\sim 1-3.5$ SMGs \citep[SMG references in][]{genzel2010a}.
In cases where upper-$J$ CO transitions were observed (i.e.,  most $z\geq 1$ systems), empirically-calibrated
correction factors were applied to infer the CO(1-0) luminosity plotted on the horizontal axis.
{\bf Left:} $L_{FIR}$ plotted vs. $L_{CO,1-0}$. 
Dashed gray and red lines indicate, respectively, the fits to the
data for non-mergers (i.e. isolated star-forming galaxies at $z\sim 0$,
$z\sim 1$ star-forming galaxies, and UV-selected and $BzK$ galaxies
at higher redshift) and mergers (i.e. $z\sim 0$ merging and
interacting galaxies, and SMGs at higher redshift). {\bf Right:}
$L_{FIR}/L_{CO,1-0}$ vs. $L_{CO,1-0}$. The left-hand vertical
axis, $L_{FIR}/L_{CO,1-0}$, can be re-expressed in terms of
the specific star-formation rate (in units of Gyr$^{-1}$), 
shown as the right-hand vertical axis.
}
\label{fig:ISM-gas-genzel2010}
\end{figure}

CO observations of both ULIRGs and LIRGs at $z\sim 2$ have 
yielded many important insights into the nature of star 
formation in different regimes of total galaxy luminosity. 
Before even considering inferred molecular gas masses, it is 
worth emphasizing the relationship between far-IR luminosity, 
$L_{FIR}$, and CO luminosity, $L_{CO}$. As shown in 
Figure~\ref{fig:ISM-gas-genzel2010}, \citet{genzel2010a} have
assembled measurements of local star-forming galaxies (both 
quiescent and merging systems), as well as $z\sim 1.5-2.0$ 
UV-selected (``BX") and $BzK$ systems and (more IR-luminous) SMGs. 
Most strikingly (and also pointed out by \citet{daddi2010}), 
at a fixed $L_{CO}$, high-redshift SMGs appear to produce 
$4-10$ times more far-IR luminosity than their UV-selected and 
$BzK$ counterparts, which cannot simply be attributed to 
enhanced AGN activity in SMGs. This trend mirrors the one observed
between local mergers and more quiescently star-forming galaxies.

In terms of star-formation rate and molecular gas surface 
densities, SMGs are characterized by significantly higher 
$\Sigma_{SFR}$ at a given $\Sigma_{gas}$. The UV-selected and 
$BzK$ galaxies follow relations between $\Sigma_{SFR}$ and 
$\Sigma_{gas}$ that are similar to the one observed among 
quiescently star-forming galaxies in the local Universe, with 
a slope of $N=1.1-1.2$ \citep{genzel2010a}. SMGs follow an 
analogous relation but with a higher overall normalization, 
perhaps reflective of their shorter dynamical timescales 
\citep{genzel2010a}, and also of processes that increase the 
efficiency of star formation in the turbulent, merger-driven 
environment that may be common in SMGs \citep{engel2010}. More 
speculatively, a top-heavy IMF in SMGs may produce more far-IR 
luminosity for a given mass of stars formed. However, the 
evidence for IMF variations at high-redshift is only indirect 
at this point, and not specific to SMGs 
\citep{vandokkum2008a,dave2008}. At the same time, 
\citet{ivison2011} offer a potential caveat regarding the 
inferred bimodality of star-formation efficiencies among SMGs 
and other, more quiescent $z\sim 2$ systems. In so far as the 
$L_{CO}$ values are based on a range of CO transitions, the 
conversion of these to CO(1-0) relies on a proper 
characterization of the gas excitation conditions. Larger 
samples of $z\sim 2$ star-forming galaxies with uniform 
CO(1-0) measurements will be crucial for characterizing the 
diversity (and actual bimodality) among star-formation 
efficiencies in high-redshift star-forming systems.

The inferred molecular gas reservoirs in UV-selected and $BzK$ 
galaxies are extended on scales of several kpc, with typical 
masses of $0.5-1.0\times 10^{11} M_{\odot}$ 
\citep{tacconi2010,daddi2010}. These correspond to median gas 
fractions of 0.44 and 0.57, respectively, for the 10 
UV-selected and 6 $BzK$ galaxies, and typical gas depletion 
timescales of $\sim 0.5$~Gyr \citep{genzel2010a}. The typical 
size of the molecular gas distribution is smaller (half-light 
radii of $\sim 2-3$~kpc) in SMGs, and, while the molecular gas 
masses and fractions are similar on average, the apparent gas 
depletion scales are a factor of several shorter than for the 
UV-selected and $BzK$ galaxies, due to the higher 
star-formation rates. With the Atacama Large Millimeter Array 
(ALMA) it will be possible to extend these studies down to 
fainter luminosities and characterize the gas reservoirs of 
more typical systems at $z>2$ that make up the bulk of the 
star formation at those early epochs.

\subsection{Metal Content}
\label{subsec:ISM-metals}

While molecular gas provides the material out of which stars 
form, heavy elements constitute an important product of star 
formation, returned to the ISM by supernova explosions and 
stellar winds. As such, the metal content of galaxies reflects 
the past integral of star formation, modified by the effects 
of gas inflow (i.e. gas accretion) and outflow (i.e. feedback 
from star formation or black-hole accretion). The relative 
abundances of different chemical elements also provides clues 
about the past history of star formation, in so far as 
so-called $\alpha$ elements (e.g., O, S, Si, Mg) are produced 
primarily in Type~II supernovae events, on short timescales 
($\sim 10$~Myr), while Fe-peak elements (e.g., Fe, Mn, Ni)
are produced primarily in Type~Ia supernovae over longer 
timescales ($\sim 1$~Gyr).  The metal content of galaxies is 
especially meaningful when 
considered in concert with their stellar and gas masses, since 
the relationships among these quantities -- and deviations 
from ``closed-box" expectations provide constraints on the 
nature of large-scale gas flows (in both directions). 

There are many different methods for measuring the metallicities of 
galaxies at high redshift, probing both their stellar and 
gaseous components using rest-frame UV and optical 
spectroscopic features. Stellar metallicity is measured from 
absorption lines, while the metal content of the interstellar 
gas can be gauged from either absorption lines arising in the 
neutral and ionized ISM, or emission features originating in 
H~II regions.

In general, the continuum S/N and spectral resolution obtained 
for the rest-frame UV and optical spectra of typical $z\geq 2$ 
galaxies are not 
sufficient for robust absorption-line metallicity measurements 
in individual objects (even using $8-10$-meter class telescopes).
The rest-frame UV spectra of 
star-forming galaxies include a host of interstellar features 
arising from neutral hydrogen and both neutral and ionized 
metal species, but only the strongest, highly-saturated 
interstellar absorption lines are detected on an 
individual-object basis within high-redshift samples 
\citep{shapley2003}. These saturated features are not useful 
for metallicity estimates. In exceptional cases of the spectra 
of strongly gravitationally-lensed objects, for which both the 
continuum S/N and resolution are at least an order of 
magnitude better than average, weak, unsaturated interstellar 
metal absorption features are detected and can be used for 
interstellar metallicity estimates. \citet{pettini2002} 
measure weak features from $\alpha$, Fe-peak, and 
intermediate elements (i.e. nitrogen) in the spectrum of the 
gravitationally-lensed $z=2.73$ galaxy, cB58 
(see Section~\ref{subsec:ISM-dustext}), 
to infer an actual interstellar abundance pattern. Based on the
relative enhancement of $\alpha$ to both Fe-peak elements and 
nitrogen, \citet{pettini2002} estimate a ``young" age for cB58, 
of less than $\sim 300$~Myr, the timescale for nitrogen 
enrichment -- an unusual case of a {\it chemical} constraint
on the past history of star formation.

Rest-frame UV stellar absorption features from both the 
photospheric and wind features of hot stars can in principle 
be used to infer stellar metallicity. \citet{rix2004} develop 
calibrations for metal-sensitive stellar photospheric 
absorption indices at 1370, 1425, and 1978~\AA. These have 
been applied to a small number of individual lensed and 
unlensed star-forming galaxies at $z\sim 2-3$ 
\citep{steidel2004,quider2009,dessauges2010}, yielding results 
consistent for the most part with other metallicity indicators. 
On the other hand, the 1978~\AA\ 
Fe~III index was measured in a composite spectrum of 75 
star-forming galaxies drawn from the GMASS survey 
\citep{halliday2008}, in fact suggesting a systematic 
enhancement of $\alpha$ relative to Fe, when compared
with the expected oxygen abundance for objects of the same
stellar mass (see below). The shape of the 
C~IV$\lambda 1549$ P-Cygni wind feature from O and B stars is 
also sensitive to metallicity (as well as the form of the 
IMF), and has also been used to estimate stellar metallicity 
in a few (mainly gravitationally-lensed) $z\sim 2-3$ galaxies 
with adequate S/N and spectral resolution 
\citep{pettini2000,quider2009,quider2010}. These interstellar 
and stellar absorption metallicities offer intriguing and detailed probes 
of small numbers of special high-redshift galaxies, but await 
the power of future 30-meter-class ground-based telescopes for 
application to large samples of individual, unlensed objects.

Most results about the metal content of high-redshift galaxies 
are based on measurements of rest-frame optical emission lines 
from H~II regions. These include combinations of hydrogen recombination lines 
(H$\alpha$, H$\beta$), and collisionally excited forbidden 
lines from heavy elements such as oxygen ([OIII], [OII]), 
nitrogen ([NII]), and neon ([NeIII]). At $z\geq 2$, rest-frame 
optical features shift out of the observed optical range and 
require near-IR spectroscopic observations. The study of 
rest-frame optical emission from H~II regions at high redshift 
to date has relied on both long-slit and IFU
spectroscopy, using instruments such as NIRSPEC and OSIRIS at 
the Keck Observatory, ISAAC and SINFONI at the VLT, MOIRCS on 
Subaru, and GNIRS on Gemini-South. Emission lines are 
primarily used to infer the gas-phase abundance of oxygen 
\citep[expressed as $12+\log(\mbox{O/H})$, 
where the solar value in these units is 8.66;][]{asplund2004},
and are based on the 
relations between particular sets of emission-line ratios and 
metallicity. These relations have been both empirically 
calibrated in the local Universe 
\citep[e.g.,][]{pettinipagel2004,nagao2006} and theoretically 
modelled using photoionization codes 
\citep[e.g.,][]{kewley2002,tremonti2004}. Two commonly-used 
indicators of $12+\log(\mbox{O/H})$ at high redshift are the 
so-called $N2$ index, defined as $\log(\mbox{[NII]}\lambda 
6584 / \mbox{H}\alpha)$, and $R_{23} \equiv 
(\mbox{[OIII]}+\mbox{[OII]})/\mbox{H}\beta$. From the ground, 
$R_{23}$ can be measured within near-IR windows of atmospheric 
transmision for various redshift intervals between $z\sim 2$ 
and $z\sim 4$, while $N2$ is only measurable up to $z\sim 
2.6$, at which point the thermal background becomes preventively high. 

In addition to the significant scatter among different
emission-line indicators (differences as large
as 0.7 dex in metallicity for the same galaxy) \citep{kewley2008},
another potential source of bias when using locally-calibrated 
metallicity indicators to interpret the emission-line ratios 
of high-redshift galaxies is the assumption that the physical 
conditions in high-redshift galaxy H~II regions are similar to 
those in local galaxies. Small samples of UV-selected galaxies 
at $z\sim 2$ with measurements of both [OIII]/H$\beta$ and 
[NII]/H$\alpha$, and/or [OIII]/[OII] and $R_{23}$, indicate 
systematic offsets from the excitation sequence of 
low-redshift galaxies \citep{erb2006a,hainline2009}. These 
differences may be indicative of a systematically higher 
ionization parameter and/or electron density, or harder 
ionizing radiation spectrum, with correspondingly different
translations between empirical line ratios and physical
metallicities. Larger samples of objects at 
$z\sim 2$ with such measurements will be required to determine 
the origin of these apparent offsets, relative to local 
star-forming galaxies.

At this point, $N2$ measurements have been obtained for a wide 
variety of $z\sim 2$ sources, numbering $\sim 200$ and ranging 
from UV- and near-IR selected galaxies
\citep[e.g.,][]{shapley2004,erb2006a,hayashi2009,yoshikawa2010,kriek2007} to SMGs 
\citep{swinbank2004}. The sample of objects with 
emission-line metallicity measurements
at $z\geq 3$ is much smaller \citep{pettini2001,maiolino2008,mannucci2009}, 
with metallicities based on $R_{23}$ or other combinations of 
oxygen, [NeIII], and Balmer lines for $\sim 30$ objects.

The measurement of galaxy metallicities along with stellar masses allows
for the construction of the galaxy mass-metallicity (or $M_{star}$-Z)
relation. This relationship has been studied in the local
Universe for $>50,000$ galaxies with SDSS \citep{tremonti2004}, and used
as a tool to constrain the importance of inflows and outflows as a
function of galaxy mass. Another necessary component for this 
type of analysis is an estimate of the gas mass, which enables a calculation
of the metallicity as a function of gas fraction, or, equivalently,
the effective yield, $y_{eff}$.  Given the small sample of galaxies
with direct atomic and/or molecular gas measurements both
locally and at high redshift,
gas masses have typically been estimated indirectly
on the basis of $\Sigma_{SFR}$ values and galaxy sizes,
and an assumption that the locally-calibrated relation 
between $\Sigma_{SFR}$ and $\Sigma_{gas}$ applies
\citep{tremonti2004,erb2006a}. We note here that
the average gas fraction of $\sim 50$\% inferred by
\citet{erb2006b} for UV-selected star-forming galaxies
at $z\sim 2$ agrees well with the CO-based estimate of $44$\%
from \citet{tacconi2010}.

\begin{figure}
\centerline{\psfig{figure=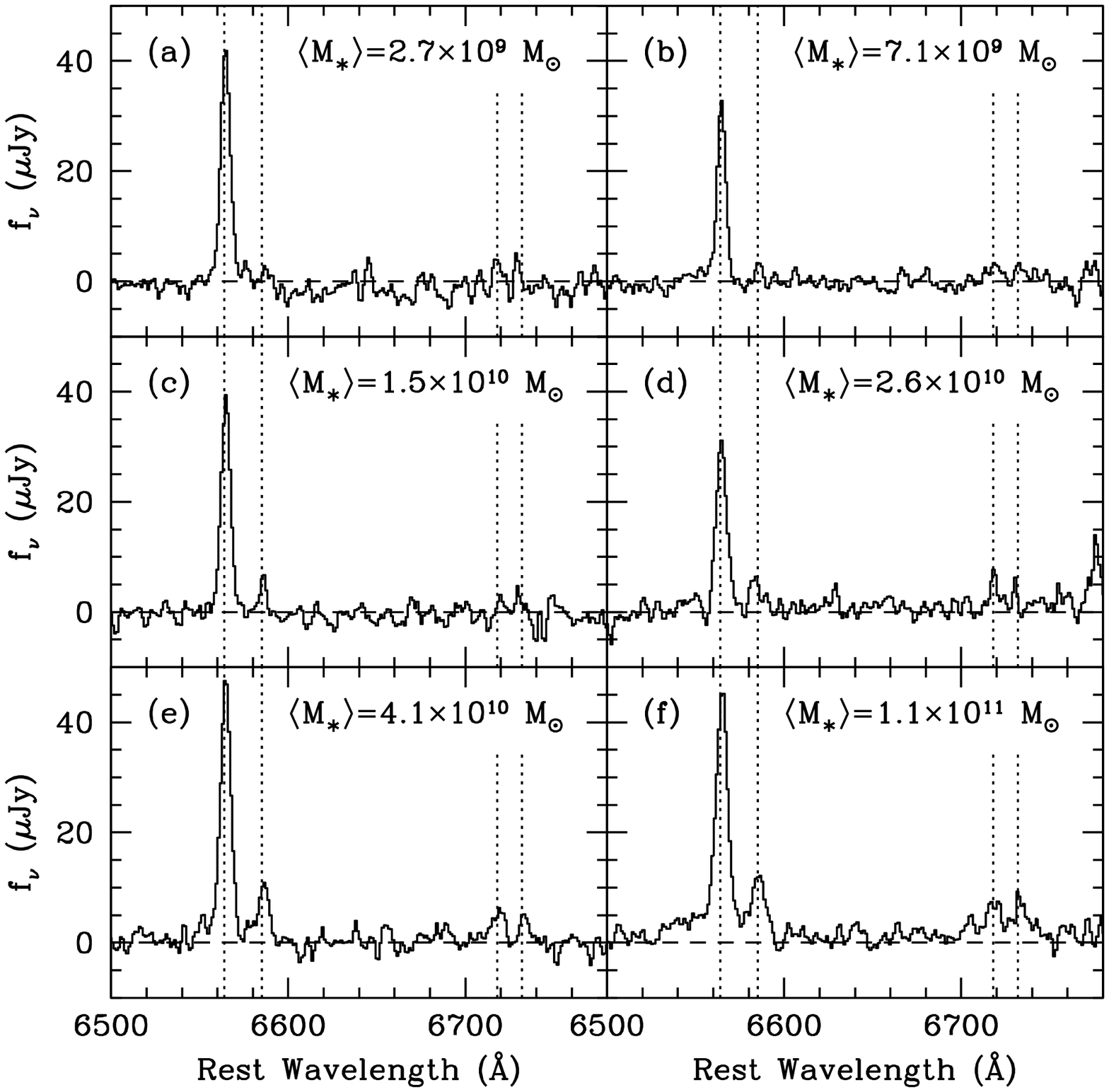,height=2.0in}\psfig{figure=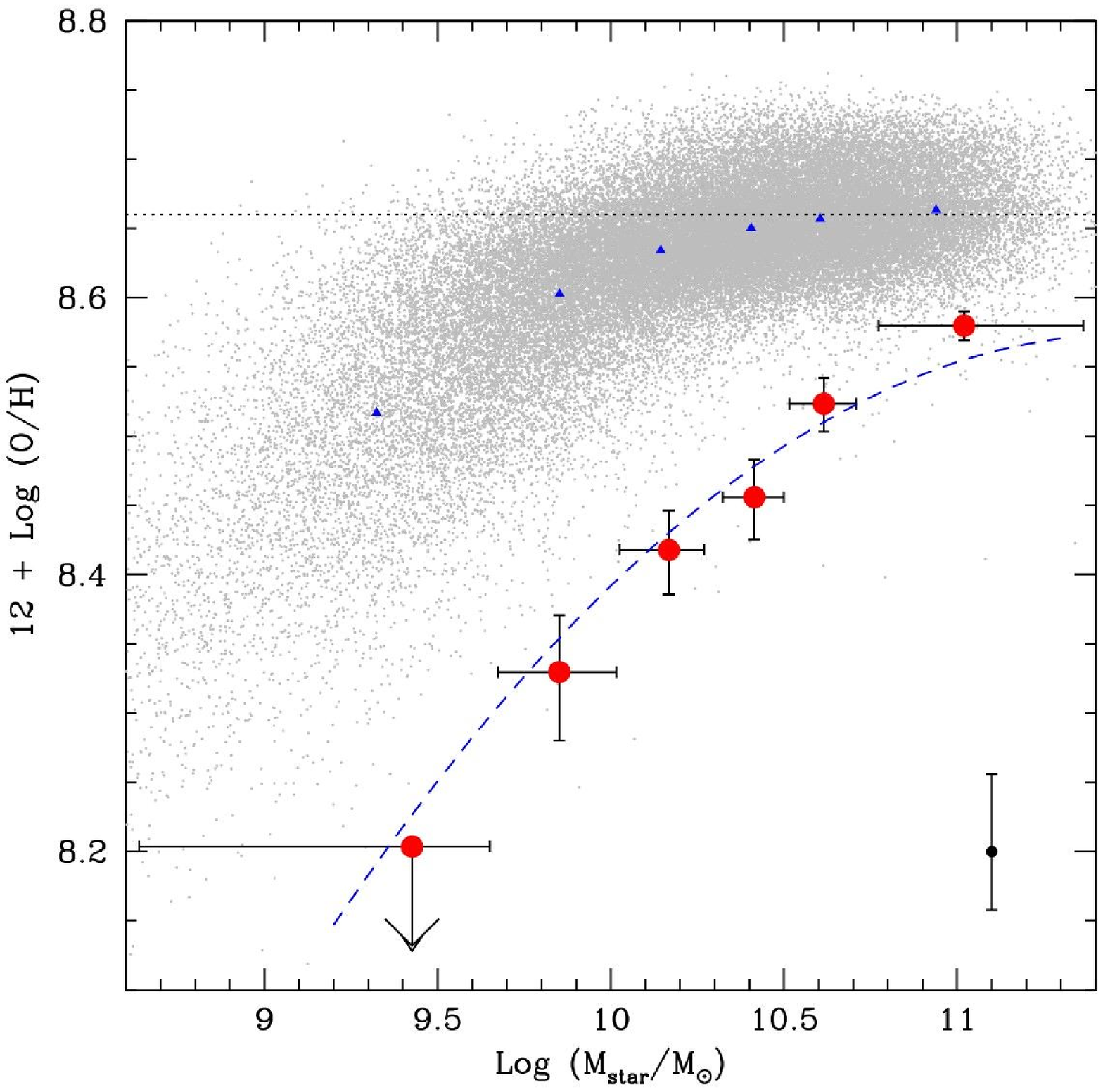,height=2.0in}}
\caption{\citep[From][]{erb2006a} The $M_{star}-Z$ relation at $z\sim 2$.
{\bf Left:} Relationship between [NII]/H$\alpha$
and stellar mass. Shown here are composite Keck/NIRSPEC spectra of the H$\alpha$
and [NII] region for
the 87 galaxies in the sample of \citet{erb2006a}, grouped into
six roughly equal bins of increasing stellar mass. In each panel, the mean
stellar mass in the bin is indicated, and the H$\alpha$, [NII], and [SII]
lines are marked by dotted lines (left to right, respectively).
As stellar mass increases, the ratio between [NII] and H$\alpha$ increases
as well.
{\bf Right:} The corresponding relationship between 12+$\log(\mbox{O/H})$
and stellar mass, based on the empirical trend shown in the left-hand panel.
Large gray circles represent the averages in each $z\sim 2$ stellar mass
bin, with metallicity estimated from the observed [NII]/H$\alpha$
using the $N2$ calibration of \citet{pettinipagel2004}. Vertical
error bars show the uncertainty in the [NII]/H$\alpha$ ratio,
while the additional error bar in the lower right corner shows the 
additional uncertainty in the $N2$ calibration
itself. The dashed line is the best-fit mass-metallicity relation
of \citet{tremonti2004}, shifted downwards by $0.56$~dex.
In order to compare with the metallicities of $\sim 53,000$ SDSS
galaxies in \citet{tremonti2004}, the $N2$ calibration was
applied to the low-redshift sample, shown as small gray dots. The
filled triangles indicate the mean metallicity of the SDSS galaxies
in the same mass bins used for the $z\sim 2$ sample. While the SDSS
sample clearly shows the saturation of the $N2$ indicator, 
the more reliable, lower-metallicity bins indicate that $z\sim 2$
galaxies are $\sim 0.3$~dex lower in metalllicity at a given stellar
mass.
}
\label{fig:ISM-metals-gas-erb2006a}
\end{figure}

The largest survey of rest-frame optical emission lines at $z\geq 2$
is presented in \citet{erb2006a}, where spectra covering
the H$\alpha$ region were obtained for 87 UV-selected galaxies.  As shown in 
Figure~\ref{fig:ISM-metals-gas-erb2006a}, the empirically-measured
$N2$ indicator increases monotonically in bins of increasing
stellar mass, corresponding
to an  increase in $12+\log(\mbox{O/H})$. At a given
stellar mass, \citet{erb2006a} find that $z\sim 2$ galaxies
are $0.3$~dex lower in metallicity than the local galaxies
studied in \citet{tremonti2004}. Based on
indirect estimates of gas masses and gas fractions, these authors
compare their observations with simple chemical evolution models. 
The shallow slope of the $z\sim 2$ $M_{star}-Z$ and gas-fraction$-Z$ 
relations are then used to demonstrate the importance of galaxy-scale winds
with mass-outflow rates roughly equal to the star-formation rate \citep{erb2008}.
The observed $M_{star}-Z$ slope at $z\sim 2$  has been viewed as 
evidence in favor of a 
``momentum-driven" feedback recipe in the simulations of \citet{finlator2008},
however we caution against placing too much significance on the precise
observed value of this slope, given the known limitations of the $N2$ 
indicator (which saturates at roughly solar metallicity).

\citet{maiolino2008} construct the $M_{star}-Z$ relation for
9 galaxies at $z\sim 3.5$, showing that, at fixed stellar
mass, galaxies at $z\sim 3.5$ are 0.4 dex lower in $12+\log(\mbox{O/H})$
than at $z\sim 2$.
\citet{mannucci2010} explain the observed evolution of the $M_{star}-Z$
relation up to $z\sim 2.5$ in terms of galaxies populating
a more general $M_{star}-Z-SFR$ relation, with no redshift evolution. According
to this ``Fundamental Metallicity Relation" \citep[FMR;][]{mannucci2010}, galaxies at fixed
stellar mass but higher star-formation rate will have lower
metallicity, resulting from the interplay between infalling
IGM gas and outflowing enriched material. 
The rapid apparent evolution in the $M_{star}-Z$ relation beyond
$z\sim 2.5$ requires some evolution in the FMR, yet the sample
of galaxies at these redshifts with metallicity measurements
is too small to draw definitive conclusions.

In order to use the $M_{star}-Z$ relation to constrain
models of gas inflow and outflow in star-forming galaxies
at $z\sim 2-4$, significantly larger (i.e. at least an order of magnitude)
samples of galaxies with
metallicity measurements are required, which are complete down
to a given stellar mass or star-formation rate. Furthermore, constructing
the relations between metal and baryonic (stellar and gas) content
for galaxies with direct estimates of molecular gas content
will provide more robust estimates of the metallicity as 
a function of gas fraction. Spatially-resolved estimates of
chemical abundance gradients will offer additional
constraints on the inflow and outflow of gas, and the build-up
of the galaxy stellar population. These have been now reported for
a handful of objects, with conflicting results about the sign
of the radial gradient \citep{cresci2010,jones2010}.
Deeper data for larger samples will be required to settle this discrepancy.
The next generation of near-IR multi-slit spectrographs and 
IFUs should easily enable these observations.

\section{STRUCTURAL AND DYNAMICAL PROPERTIES}
\label{sec:structure}

The diversity among galaxy structural and dynamical parameters in the local Universe
reflects a range of mass assembly histories, and the relative
importance of mergers and smoother mass accretion.
Furthermore, properties
such as size, concentration, mass-surface density,  and velocity dispersion
are correlated with indicators of the nature of stellar populations, such as
luminosity, stellar mass, and specific star-formation rate
\citep{kauffmann2003b,shen2003}. Reproducing the
observed connections among these galaxy structural and stellar
properties and their redshift evolution within a unified framework
constitutes a crucial test for models of galaxy formation.
In this section we review recent results about the structural and 
dynamical properties of high-redshift galaxies. 
Due to the small apparent sizes of most
$z\geq 2$ galaxies ($\lesssim 1$"), {\it HST} imaging and ground-based IFU maps assisted
by adaptive optics (AO) are both critical for resolving overall
shapes and $\sim$kpc-scale fine structure. Interferometric
observations of CO line emission have also provided valuable
insight into the dynamical properties of the most extreme
sources at high-redshift, such as SMGs.
While high-resolution imaging and dynamical maps are each powerful
probes individually, combining these two methods of observations
\citep{forsterschreiber2011} will provide the deepest insights into the 
nature of the galaxy assembly process at high redshift.

\subsection{Structural Properties}
\label{subsec:structure-structure}

In general, the morphologies of galaxies at $z\geq 2$
are characterized by two important differences with respect to
those of local galaxies. First, the traditional Hubble sequence of 
regular spirals and elliptical galaxies has not settled into place by
$z\sim 2$, and
a much higher frequency of clumpy, irregular morphologies
is observed among star-forming systems \citep{ravindranath2006,lotz2006}.
Second, both star-forming and quiescent galaxies are more compact
at fixed stellar mass or rest-frame optical luminosity \citep{buitrago2008}.
In order to characterize the structural properties of high-redshift
galaxies, both parametric and non-parametric methods have been employed.
In the former, model S\'ersic profiles are fit to the data, 
yielding best-fit values for $n$, the S\'ersic power-law index,
and $r_{e}$, the effective radius, within which half the galaxy luminosity
is emitted. S\'ersic profile fits work best for regular, axially-symmetric
morphologies with well-defined centers, so, in order to address
the clumpy, irregular structures observed among high-redshift
systems non-parametric statistics such as the $Gini$, $M_{20}$,
$Multiplicity$, and $CAS$ coefficients have been developed 
\citep{abraham2003,lotz2004,lotz2006,law2007a,conselice2003}
to characterize their concentrations, asymmetries, and clumpiness
(i.e. deviations from maximally compact configurations).

The first high-redshift galaxies to be observed with
{\it HST} were UV-selected galaxies at $z\sim 3$
\citep{giavalisco1996,lowenthal1997}.
The irregular morphologies of these systems,
often composed of multiple compact components and/or
irregular nebulosity, were first attributed
to the fact that the Wide Field Planetary Camera 2 (WFPC2)
and ACS are optical imagers, probing rest-frame UV
wavelengths at $z\sim 3$ -- a so-called
``morphological $k$-correction." In principle, measurements 
at these wavelengths might be sensitive to
only the most active regions of star formation rather than the bulk
of the stellar mass, and potentially affected by patchy dust extinction.
However, in practice, rest-frame optical imaging with the 
Near Infrared Camera and Multi-Object Spectrometer
(NICMOS), using both NIC2 and NIC3 cameras 
\citep{dickinson2000,papovich2005,forsterschreiber2011,kriek2009} and
the Wide Field Camera 3 (WFC3) \citep{overzier2010} has demonstrated that
actively star-forming galaxies at $z\sim 2-3$
have very similar rest-frame UV and optical morphologies.
On the other hand, among the most massive, evolved
systems at $z\sim 2-3$, there is evidence for more centrally-concentrated,
regular morphologies at rest-frame optical wavelengths, compared
with the structure observed in the rest-frame UV
\citep{toft2005,cameron2010}. With the recent installation of WFC3,
it will now be possible to measure rest-frame optical morphologies
for statistical samples of $z\sim 2-3$ objects at $\sim$kpc-scale
resolution. The performance of the IR channel of WFC3 is
superior to the NICMOS NIC3 camera in terms of
resolution, sensitivity, and area.

\begin{figure}
\centerline{\psfig{figure=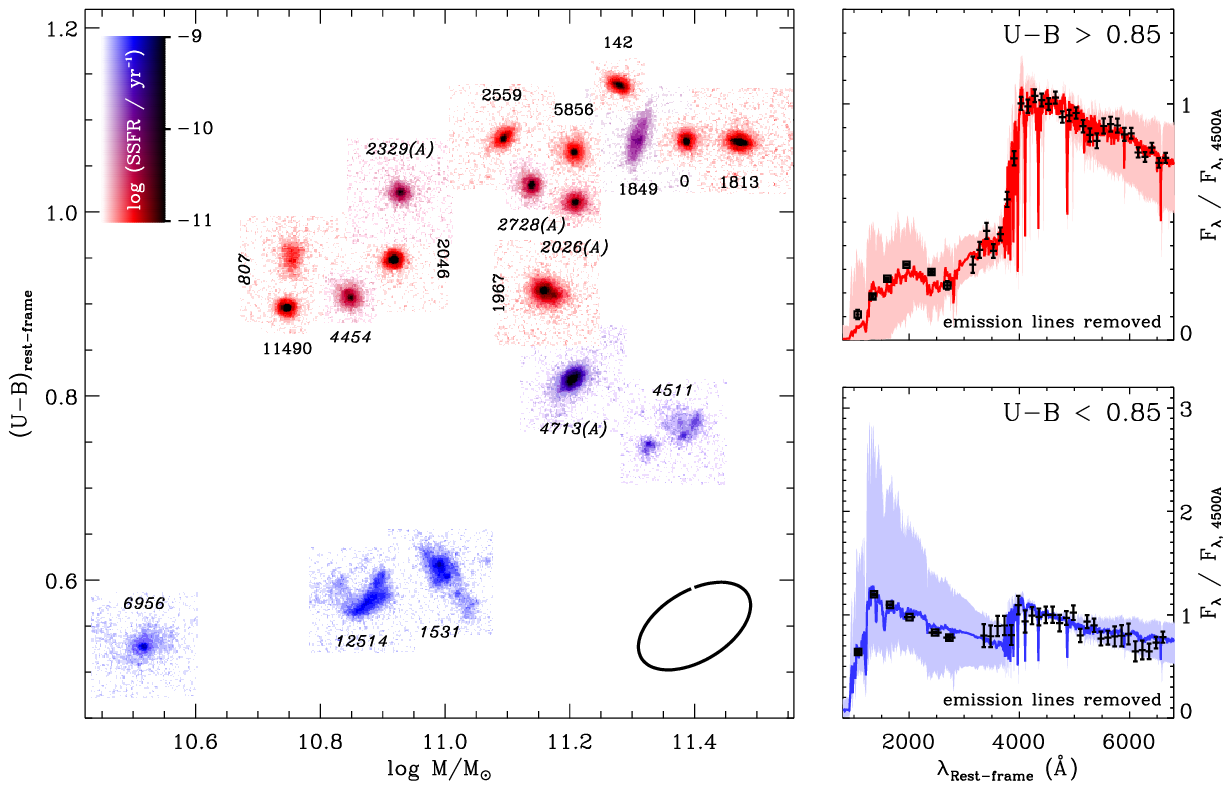,height=3in}}
\caption{\citep[From][]{kriek2009} Morphologies
and SEDs of massive galaxies. 
{\bf Left:} $U-B$ color vs. stellar mass, for a massive galaxy
sample at $z\sim 2.3$ from \citet{kriek2009} with rest-frame
optical spectroscopy. The actual NIC2 image of each galaxy
is used as a symbol, in order to indicate the trends between
morphology and stellar population parameters. Color coding
reflects the specific star-formation rate of the galaxies. Emission-line
galaxies are indicated with italic ID numbers, and AGNs are indicated
additionally with ``(A)." Large, irregular galaxies reside mainly
in the ``blue cloud", while compact, quiescent galaxies lie
on a red sequence. The ellipse represents the average $1\sigma$ confidence
interval. {\bf Right:} Stacked SEDs for blue (bottom panel) and
red (top panel) galaxies in the $2\leq z \leq 3$ spectroscopic sample of
\citet{kriek2008a}, based on rest-frame
UV photometry and rest-frame  optical spectra.}
\label{fig:structure-structure-kriek2009}
\end{figure}

While the traditional Hubble sequence of disk and elliptical
is not in place by $z\sim 2$, {\it HST} imaging has
revealed a large diversity of structural parameters
among galaxies at $z\sim 2-3$, which correlate with
their stellar populations in analogy with the trends
observed in the local Universe. Based on NICMOS/NIC3
F160W imaging, \citet{zirm2007} and \citet{toft2007} show
that quiescent, high-redshift $z\sim 2$ galaxies
have systematically smaller sizes and higher
stellar-mass surface densities than actively
star-forming systems of the same stellar mass. \citet{franx2008}, \citet{toft2009},
and \citet{williams2010} use deep, ground-based
$K$-band images to demonstrate the same trend.
Based on higher-resolution NICMOS/NIC2 F160W imaging
for a sample of spectroscopically-confirmed
$K$-selected massive galaxies at $z\sim 2$, \citet{kriek2009}
presents a comparison of the S\'ersic profile fits
for the quiescent and emission-line objects. As shown
in Figure~\ref{fig:structure-structure-kriek2009},
galaxies with redder rest-frame $U-B$ colors, lower
specific star-formation rates, and SEDs indicative of more
evolved stellar populations, are characterized by 
smaller $r_e$ and more concentrated
surface-brightness profiles (larger $n$). This figure
also shows the clumpy rest-frame optical morphologies
of the massive, blue galaxies. Objects whose emission-line
ratios actually suggest AGN activity show structural
properties and SEDs that are similar to those of the
quiescent galaxies. \citet{forsterschreiber2011} find
even larger sizes and shallower profiles in NIC2 images of the six clumpy
UV-selected galaxies in their sample, which also have higher specific
star-formation rates. Also similar to the trends observed
in the local Universe, \citet{franx2008} and \citet{forsterschreiber2011}
find a correlation between stellar mass surface density
and specific star-formation rate, such that galaxies
with higher stellar mass densities have lower 
specific star-formation rates \citep{kauffmann2003b}.

We now highlight two important observed morphological
phenomena, which have sparked much interest from the theoretical
community. One is the nature of the clumpy morphology
and ``clumps" in actively star-forming galaxies at $z\geq 2$.
The other is the incredibly compact nature of massive,
quiescent galaxies at high redshift, and their connection to today's 
massive, early-type galaxies.

\subsubsection{CLUMPY MORPHOLOGIES IN STAR-FORMING SYSTEMS}
\label{subsubsec:structure-structure-clump}

While it is possible to model the rest-frame
UV and optical morphologies of $z\geq 2$ galaxies using smooth S\'ersic
profiles, significant residuals result due to the presence
of non-axisymmetric surface-brightness fluctuations.
These ``clumps" have been fairly ubiquitously
observed among the rest-frame
UV and optical morphologies of distant galaxies
\citep{cowie1995,elmegreen2005,law2007a,lotz2004},
and are visible among the actively star-forming
galaxies in Figure~\ref{fig:structure-structure-kriek2009}.
Clumps are characterized by typical sizes of $\sim 1$~kpc,
and make up as much as $\sim 40$\% of the rest-frame UV
light distribution \citep{elmegreen2009}.
While clumpy, irregular structure was first explained
as the natural outcome of the increased prevalence
of major merger events at high redshift \citep[e.g.,][]{conselice2003},
various lines of evidence suggest that other factors
may contribute to this phenomenon. 

First, \citet{law2007a}
demonstrated a decoupling between
the $Gini$ coefficient (and any other rest-frame UV non-parametric
morphological statistic) and star-formation rate. If
clumpy morphology was connected with the incidence of a major
merger, the $Gini$ coefficient, sensitive to non-uniformities
in the light distribution, should display some correlation
with the parameters describing galaxy stellar populations.
\citet{swinbank2010} also highlight the fact that the distribution of rest-frame
UV and optical morphologies of SMGs \citep[plausibly the sites
of major merger events;][]{engel2010,swinbank2006}
were statistically indistinguishable
from those of UV-selected star-forming galaxies, which
are significantly more quiescent in terms of
bolometric luminosity and star-formation rate. Furthermore,
the clumpy phenomenon is far too common among
star-forming galaxies at $z\sim 2-3$ to be accounted
for quantitatively by the predicted rate of major merger events
in, e.g., the Millennium Simulation \citep{genel2008,conroy2008}.
Finally, as described below in Section~\ref{subsubsec:structure-dynamics-ifu},
clumpy structures are commonly observed in galaxies with ordered velocity
fields indicative of rotating disks -- demonstrating that a major
merger is not necessarily the cause for clumps. Minor-merger interactions
and instabilities in gas-rich, turbulent disks are more
likely alternatives for the formation of clumps
\citep{genzel2008,bournaud2009,dekel2009}.

\subsubsection{THE COMPACTNESS OF QUIESCENT GALAXIES}
\label{subsubsec:structure-structure-compact}

At fixed stellar mass, both star-forming and quiescent
galaxies have smaller radii at higher redshift.
The evolution in the $M_{star}-r_e$ relationship
is especially dramatic for quiescent systems.
As noted by many authors 
\citep[e.g.,][]{daddi2005,trujillo2007,toft2007,cimatti2008,vandokkum2008b,
buitrago2008,damjanov2009} using rest-frame UV and optical {\it HST}
imaging and ground-based data, $z\sim 1.5-2$ quiescent galaxies have radii that are
$\sim 2-5$ times smaller than those of local early-type galaxies of 
the same stellar mass. For example, $z\sim 2$ quiescent galaxies
with $M_{star}\sim 10^{11} M_{\odot}$
are measured to have $r_e\sim 1$~kpc \citep{vandokkum2008b},
compared with $\sim 3$~kpc in local early-type galaxies of
the same mass \citep{damjanov2009}.
The observed differences in size correspond to 
factors of $\sim 10-100$ in physical density! 
Furthermore, \citet{taylor2010} demonstrate the extreme scarcity
of such compact, early-type  systems in the local Universe.
The resulting challenge from these discoveries consists of
relating high-redshift massive, compact quiescent galaxies to 
their low-redshift counterparts -- since structural evolution
is required for the distant objects to resemble the current
ones. Scenarios considered to explain the observed evolution in size
include the effects of major and minor dissipationless mergers \citep{naab2009,
hopkins2009,hopkins2010}
as well as energetic feedback from an accreting black hole \citep{fan2008}.

Many important caveats have been raised about the observations
of apparent compactness among quiescent galaxies. These
include $(a)$ the possibility of not accounting for extended low-surface brightness
emission which would tend to increase the inferred radius;  $(b)$
using in some cases rest-frame UV rather than rest-frame optical imaging;
$(c)$ radially-dependent stellar $M/L$ ratios that increase outwards,
causing an underestimate of the mass at larger radii; $(d)$ a reliance
on photometric rather than spectroscopic redshifts; 
and $(e)$ uncertainties in stellar population
models used to estimate stellar masses \citep{vandokkum2008b,vandokkum2010}.
Most of these concerns have been addressed, using samples with 
spectroscopic redshifts and deeper surface-brightness limits.
Additionally, crude dynamical information has been obtained
from stacked and individual spectra of compact, quiescent
galaxies, based on an estimate of velocity dispersion
from stellar absorption lines \citep{vandokkum2009_nature,
cenarro2009,cappellari2009}. The estimated velocity dispersions
appear to confirm the large masses inferred from stellar
population modeling. Furthermore, results at $z\sim 1$ based on
a comparison of radii and {\it dynamical} mass estimates for 50 galaxies
suggest a factor of $2$ evolution in size at fixed dynamical mass, 
consistent with the evolutionary trend observed to $z\sim 2$
based on stellar masses \citep{vanderwel2008}.  

Employing a complementary technique, \citet{vandokkum2010} stack
the ground-based near-IR surface-brightness profiles for
rare, massive galaxies of constant number density in five redshift bins
between $z=2.0$ and $z=0.1$, such that the higher-redshift objects
can plausibly represent the progenitors of the objects
at lower redshift. The observed evolution in these
stacked surface brightness profiles shows that the 
mass within an inner region of $r=5$~kpc remains roughly
constant, while the mass at larger radii builds up smoothly
as a function of decreasing redshift. This result suggests
the importance of minor merger events
in the growth of massive, elliptical galaxies
\citep{naab2009,hopkins2009}, and 
that the structure of elliptical galaxies is not self-similar
as a function of redshift. Based on deep near-IR spectroscopy
with upcoming instruments on the ground and in space,
statistical samples of dynamical mass estimates for passive galaxies
at $z\geq 2$ will be crucial for even more robustly
distinguishing among evolutionary scenarios.

\subsection{Dynamical Properties}
\label{subsec:structure-dynamics}

Dynamical studies of high-redshift galaxies
have mainly been limited to star-forming objects,
using rest-frame optical emission lines from ionized
gas or millimeter-wave emission from CO tracing
molecular gas. While the sample of existing rest-frame
UV spectra for high-redshift galaxies
is much larger than either of these other types
of dataset \citep{steidel2003,steidel2004}, the emission
and absorption features typically detected in the
rest-frame UV range are broadened by non-virial
motions from outflowing gas and radiative transfer effects, and 
unfortunately do not permit useful dynamical measurements
\citep{pettini2001,shapley2003,steidel2010}.
The first dynamical probes of high-redshift
star-forming galaxies consisted of
long-slit near-IR spectroscopy of $\sim 100$ UV-selected galaxies
\citep{pettini2001,erb2003,erb2004,erb2006b}
and CO maps of very small samples of SMGs
\citep{genzel2003,neri2003}. Indeed,
CO observations can be used not only for
estimating molecular gas masses but also 
for probing the dynamical properties of a galaxy. These early observations
revealed a typical H$\alpha$ or [OIII] emission-line velocity dispersion 
of $\sigma\sim 100\mbox{ km s}^{-1}$
for UV-selected galaxies. Measured linewidths were $\sim 2-3$ times
larger in the SMG CO maps, which also showed
evidence for complex, multi-component emission-line morphology.
A particularly striking result from the long-slit H$\alpha$ studies
of \citet{erb2006b} consists of the measurement of spatially-resolved,
tilted emission-lines, indicative of velocity shear, and,
perhaps, rotation -- i.e., preliminary evidence for disks at high redshift. 
Below we summarize some of the latest results
from both near-IR IFU and CO dynamical studies of
high-redshift galaxies, highlighting many open questions
in this rapidly evolving field.

\subsubsection{INTEGRAL-FIELD UNIT OBSERVATIONS OF TURBULENT STAR-FORMING GALAXIES}
\label{subsubsec:structure-dynamics-ifu}

With the advent of both seeing-limited and AO-assisted near-IR
IFU spectrographs on Keck, the VLT and Gemini telescopes several
years ago, the study of high-redshift galaxy dynamics has advanced considerably.
Some striking trends have emerged from 
a total sample of $\sim 100$ objects with IFU kinematic observations,
which are no longer subject to the spatial-sampling limitations
of long-slit spectroscopy taken at a fixed position angle. 
We focus here on the VLT/SINFONI survey of
\citet{forsterschreiber2009} and the Keck/OSIRIS survey of \citet{law2009},
which represent two of the largest campaigns at $z\geq 2$
using this new observational capability. The ``SINS H$\alpha$
Survey" presented in \citet{forsterschreiber2009} includes
62 star-forming galaxies at $1.3 \leq z \leq 2.6$
selected in the rest-frame UV, optical,
near-IR, and submillimeter. The sample is 
representative of massive, $M_{star}>10^{10} M_{\odot}$, star-forming
galaxies in this redshift range, with a median stellar mass and star-formation
rate of $M_{star}=3\times 10^{10} M_{\odot}$ and 
$SFR=70 \; M_{\odot}\mbox{yr}^{-1}$, respectively. The sample of \citet{law2009}
includes 12 UV-selected star-forming galaxies at $2.0\leq z \leq 2.5$
with a median stellar mass and star-formation rate
of  $M_{star}=10^{10} M_{\odot}$ and $SFR=40\; M_{\odot}\mbox{yr}^{-1}$,
respectively.
In addition to having lower stellar masses on average, the objects
in the \cite{law2009} sample are also characterized by a smaller
median H$\alpha$ half-light radius than the SINS H$\alpha$ sample
(1.3 versus 3.1~kpc). All OSIRIS observations in
\citet{law2009} are AO-assisted, with an effective PSF
of $\sim 0.15$", corresponding to $\sim 1.2$~kpc. 
The majority ($\sim 85$\%) of the SINFONI observations
discussed by \citet{forsterschreiber2009} are seeing-limited,
with a typical resolution of $\sim 0.6$", corresponding
to $\sim 4.9$~kpc; the remainder were observed
using either laser- or natural-guide-star AO, with
resolutions ranging from $0.17"-0.41"$, corresponding
to $\sim 1.6-3.4$~kpc.

\begin{figure}
\centerline{\psfig{figure=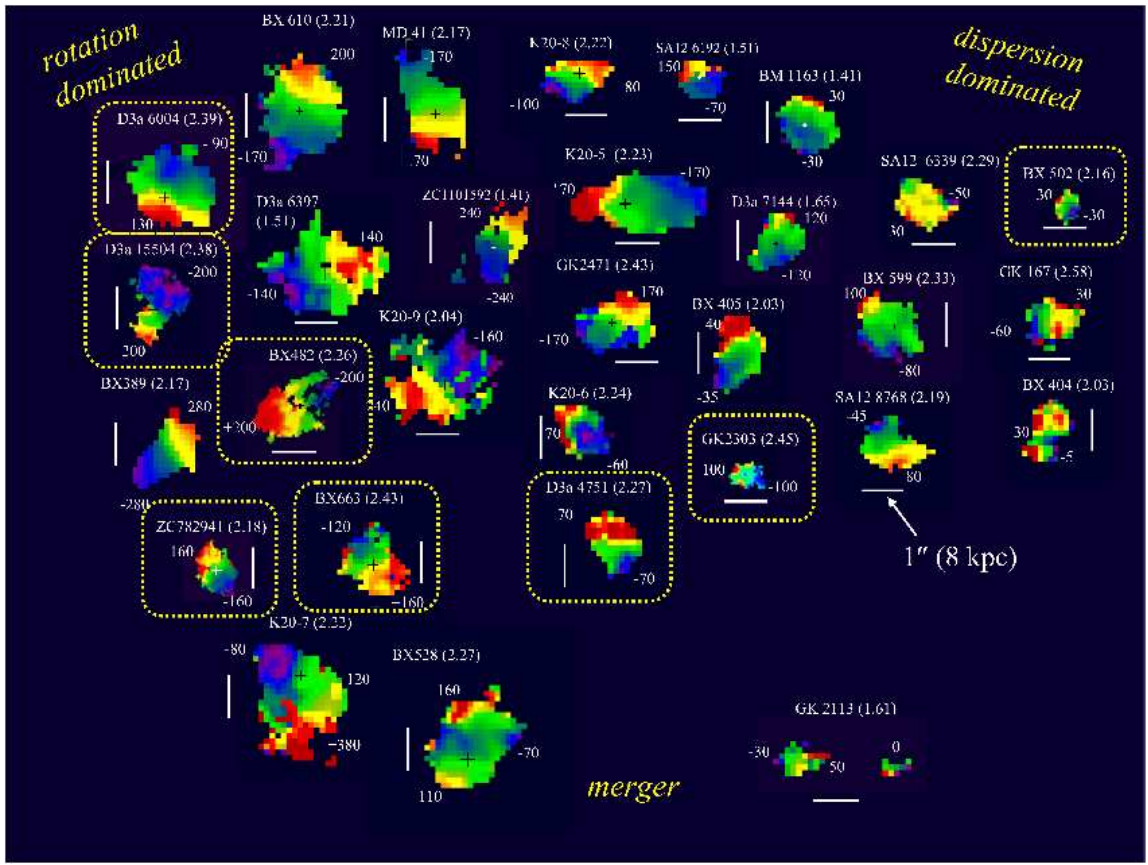,height=4.5in}}
\caption{\citep[From][]{forsterschreiber2009} Velocity fields from the
SINS H$\alpha$ sample, obtained using SINFONI on the VLT.
Shown here are velocity fields for 30 of the 62 galaxies in the SINS
H$\alpha$ survey. Color coding is such that blue to 
red colors correspond to the range of blueshifted to redshifted
line emission with respect to the systemic velocity. The minimum
and maximum relative velocities are labeled for each galaxy
(in $\mbox{km s}^{-1}$). All sources are shown on the same
angular scale. The white bars correspond to 1", or $\sim 8$~kpc
at $z=2$. The galaxies are approximately sorted from left to right
according to whether their kinematics are rotation-dominated or 
dispersion-dominated, and from top to bottom according to whether 
they are disk-like or merger-like as quantified by kinemetry
analysis \citep{shapiro2009}. Galaxies observed with AO
are indicated by the yellow dashed lines.
}
\label{fig:structure-dynamics-disks-forsterschreiber2009}
\end{figure}

These IFU surveys have revealed a wealth of diversity
among the dynamical properties of distant star-forming galaxies.
The key observables are maps of velocity, velocity dispersion,
and line-intensity, which can then be used as inputs to detailed
dynamical models, or analyzed in a more empirical sense.
Figure~\ref{fig:structure-dynamics-disks-forsterschreiber2009}
demonstrates the range of velocity fields observed in
the SINS H$\alpha$ survey. Roughly 1/3 of
these velocity maps show clear gradients
and are classified as ``rotation-dominated,"
with corresponding maps of velocity dispersion showing central
maxima; roughly another 1/3 qualify as ``mergers" as evidenced
by kinemetric modeling \citep{shapiro2009} or distinct multiple
components; finally, 1/3 of the sample are ``dispersion-dominated",
with no evidence for rotation or shear, but characterized by a significant
velocity dispersion. Half of the \citet{law2009} sample belong
to this last category, with no evidence for rotation.

Another basic result consists of the high degree of turbulence
observed in the ISM of high-redshift galaxies, along with
a small ratio of rotational to random motions.
The spatially-resolved velocity dispersion maps indicate
local velocity dispersion values ranging from 
$\sigma_{local}=30-90 \mbox{ km s}^{-1}$ in the SINS H$\alpha$
sample and $\sigma_{local}=60-100 \mbox{ km s}^{-1}$ in the
sample of \citeauthor{law2009}. Even in rotation-dominated
systems, the velocity dispersion is significant. The 
median value of $v_{rot}/\sigma$ is $4.5$, which is
considerably lower than the values of $\sim 10-20$ observed in local 
spiral galaxies \citep{forsterschreiber2009}. Of course,
in the dispersion-dominated systems, $v_{rot}/\sigma$ is
even smaller (i.e., $\leq 1$). The prevalence of rotation and rotation-dominated
systems appears to be correlated with stellar mass, with a higher
fraction of rotating systems at higher stellar masses. At the
same time, galaxies with small stellar masses and large
gas fractions tend to have negligible rotation 
\citep{law2009,forsterschreiber2009}.

The star-forming galaxies with IFU observations also appear
to follow scaling relations between their basic dynamical
and stellar properties. Both \citet{bouche2007b} and
\citet{forsterschreiber2009} present a strong correlation
between galaxy circular velocity, $v_{rot}$,  and size for galaxies
selected in the rest-frame UV, optical, and near-IR.
This correlation is indistinguishable from the one
describing local disk galaxies (despite the significant
differences in $v_{rot}/\sigma$), and both UV-selected and $BzK$ 
galaxies follow it. On the other 
hand, SMGs occupy a distinct region of $v_{rot}-$size
parameter space, with significantly larger
velocities and smaller sizes. This difference
potentially reflects the lower angular momenta
and higher matter densities present in SMGs
relative to more quiescently star-forming UV-selected
and $BzK$ systems \citep{bouche2007b}. Limiting
their  analysis to 18 rotation-dominated systems
from the SINS H$\alpha$ sample, \citet{cresci2009}
discover a ``Tully-Fisher" correlation between stellar mass and $v_{rot}$,
which has the same slope as the correlation
among local disk galaxies, but offset by $\sim 0.4$ dex
towards lower stellar mass at fixed $v_{rot}$. Furthermore,
a comparison of dynamical and stellar masses indicates
a strong correlation, but suggests that galaxies
must contain significant gas fractions to explain
the differences between the two mass estimates. In
particular, galaxies with young stellar ages appear
to contain larger gas fractions, consistent
with the previous, long-slit results of \citet{erb2006b}.

The spatially-resolved H$\alpha$ maps indicate that clumpy
morphologies are a feature of not only continuum surface brightness
\citep{elmegreen2005}, but also the emission from ionized gas 
\citep{forsterschreiber2009,genzel2008}. 
Given that star-forming
clumps are present even in galaxies with ordered rotation fields
and no evidence of major merging, and that the clumps appear
to follow the general velocity field of the larger systems
in which they are embedded, it is unlikely that clumps
constitute distinct, accreted systems. An alternative
explanation for observed clump properties is based
on predicting the size and mass-scale on which gas
should fragment in a turbulent, high-surface-density
disk.  Specifically, the high turbulent velocities observed
(as described above) lead to an expected Jeans length
for fragmentation of $\sim 2.5$~kpc \citep{genzel2008}.
Folding in the typical observed disk surface densities
($\sim 10^2 M_{\odot}\mbox{ pc}^{-2}$), \citet{genzel2008}
estimate clump masses of  $\sim 10^9 M_{\odot}$, similar
to what is observed. Numerical simulations of turbulent
gas-rich disks also appear to produce $\sim$kpc-scale
clumps \citep{bournaud2009}. Theoretical arguments (both analytical
and numerical) suggest that these clumps may migrate
inwards and coalesce to form a bulge on $\sim0.5-1.0$~Gyr
timescales \citep{dekel2009,elmegreen2008}. 
On the other hand, feedback from star formation
may disrupt the clumps before they reach the center,
and they may simply contribute to the overall growth of the 
disk. The role of clumpy structures in the growth of massive
galaxies is an extremely active area of research;
new observational results will help to elucidate the
origin and fate of the clumps.

Local velocity dispersions of $\sim 50-100 \mbox{ km s}^{-1}$
appear to be a generic property of high-redshift star-forming galaxies
at $z\sim 2$. Determining the cause of this large apparent
turbulence is an important goal for models of galaxy formation.
Proposed scenarios include the conversion of the gravitational
potential energy from infalling accreting matter into random 
kinetic energy or collisions between large clumps 
as they migrate inwards \citep[see, e.g.,][for a review
of these possibilities]{genzel2010b}. On the other hand,
\citet{lehnert2009} argue that mechanical energy feedback
associated with star formation -- in the form of stellar winds
and supernovae explosions -- is the driving force behind
the large observed interstellar turbulence. Various
authors have compared the surface-density of star-formation
$\Sigma_{SFR}$ with the local velocity dispersion (presumably
an indication of the connection between star-formation feedback
and interstellar turbulence) and have arrived at different
conclusions about what the weak observed correlation (if any) signifies
\citep{lehnert2009,forsterschreiber2009,genzel2010b}. 
The cause of large turbulent velocities in high-redshift
star-forming galaxies clearly remains an open question.

\subsubsection{THE NATURE OF SMGS}
\label{subsubsec:structure-dynamics-smgs}

Dynamical information can also be potentially used
to investigate the origin of the extreme luminosities
among SMGs. Both near-IR IFU maps and CO
observations of SMGs have been used in this 
endeavor. The observed CO profiles are broad (FWHM typically
several hundred $\mbox{km s}^{-1}$) and often
double-peaked \citep{greve2005}. A broad, double-peaked
profile can be indicative of either a massive, rotating
disk or, alternatively, a merger event. 
\citet{swinbank2006} present near-IR IFU
observations of 8 SMGs at $1.3 \leq z\leq 2.6$, probing rest-frame
optical line emission. At least five of these
systems show evidence for two or more distinct
dynamical components, suggestive of merging.
Based on subarcsecond-resolution CO maps
for 12 SMGs at $1.2\leq z \leq 3.4$, \citet{engel2010}
find evidence in 5 systems for two distinct spatial
components, with mass ratios (when possible to determine)
closer than $1:3$ -- the standard threshold for being
considered a major merger. In the remaining systems,
morphologies are either disturbed or compact,
and \citet{engel2010} argue that these represent
later-stage, coalesced merger events. 
A significant fraction of SMGs clearly show dynamical
evidence for major merging. Whether an alternative
scenario of smooth gas infall and minor mergers
\citep[e.g.,][]{dave2010} can explain the luminosities of the remaining sources
will require more robust observations of the space
densities, and stellar and dynamical masses of SMGs. The
extreme matter densities inferred for SMGs \citep{bouche2007b},
their high apparent star-formation efficiencies \citep{daddi2010,genzel2010a},
and the deviation of these systems from global scaling relations
between star-formation rate and stellar mass \citep{daddi2007}, all suggest
that they constitute very unusual events. 
It remains an open challenge to relate SMGs to other 
ULIRGs at the same epochs \citep[e.g., the $24\mu$m-selected
objects of][]{yan2007}, as well plausible
descendants among the lower-redshift massive-galaxy population.

\section{CONCLUDING REMARKS}
\label{sec:future}

Our path has traveled through
many different wavelength ranges and observational
techniques along the way to characterizing the stars,
dust, gas, heavy elements, structure and dynamics
of high-redshift galaxies at $2\leq z \leq 4$. 
By design, this review has focused primarily on the translation
of observed quantities to physical ones,
instead of making systematic comparisons with particular
theories (although at times, when appropriate,
a connection was made). At the same time, the rapid development
of observations and their physical interpretation 
over the previous decade has yielded an important body of data for input into
models of galaxy formation (i.e., the diversity of
star-formation histories and structures among the highest-mass systems;
the connection between stellar mass and star-formation
rate; the nature of dust extinction
as a function of luminosity;  the estimate
of star-formation efficiency for LIRGs and ULIRGs;
the mass-metallicity relation as a function of redshift;
the compactness of massive, quiescent systems;
the high degree of turbulence in the ISM of star-forming galaxies; and so on).

There are some definite limitations in the nature
of this review. Indeed, the description of physical
properties was often still couched in terms
of how the results applied to a specific galaxy sample,
assembled using a specific selection technique
from among the ones described in Section~\ref{sec:technique}.
In order for measurements of galaxy properties to
have true discriminating power among galaxy formation
models, it is crucial to use samples that are complete
with respect to a well-defined property, such
as rest-frame optical or near-IR luminosity, stellar mass,
dynamical mass, or star-formation rate. The majority
of the samples described in this review slice the underlying 
high-redshift galaxy population in particular ways that add an extra
layer of complexity if a comparison with a simulated
galaxy population is desired. We advocate the design
of future surveys with the type of physical completeness described above
in mind. The NEWFIRM Medium-Band Survey \citep{vandokkum2009_pasp} represents
an important step in this direction, but an even deeper survey would be desirable,
with both rest-UV and optical spectroscopic follow-up.

Furthermore, we point out that many of the results
reported here were (by necessity) based on small samples -- 
these include mid-IR spectra of the most luminous sources,
measurements of molecular gas content and
dynamics, individual emission and absorption-line metallicity 
measurements at $z\geq 2$, high-resolution rest-frame
optical morphological measurements, and AO-assisted emission-line maps. 
While these results highlight truly compelling questions,
they also await confirmation from samples an order of magnitude
larger for a robust comparison with models. 

Also, by necessity,
some of the most intriguing results regarding gas, dust,
and structural properties are limited to the luminous
extreme of the galaxy population. With the steep luminosity
functions described in Section~\ref{subsec:empirical-LF},
galaxies with $L\leq 10^{12} L_{\odot}$ comprise the bulk ($\sim 75$\%) of the luminosity
and star-formation rate density of the Universe at $z\sim 2$. Even galaxies
with $L\leq 10^{11} L_{\odot}$ may contribute $\sim 30-35$\% 
of the bolometric luminosity density at this redshift \citep{reddy2008}. We
must find ways of extending our physical studies towards fainter luminosities --
either by using gravitationally-lensed objects, or the next generation of instruments
and telescopes. These faint objects are important, and perhaps
more analogous to the galaxies playing a crucial role in
the reionization process at $z\geq 6$.

We close by highlighting two important observational
challenges, and looking towards the future.
First, the direct measurement of gas inflow (accretion) and outflow
(star-formation and AGN feedback) received only passing reference
during the course of this review.
The ubiquity of galaxy-scale outflows among $z\geq 2$ UV-selected
galaxies has long been known on the basis
of rest-frame UV and optical spectra \citep{pettini2001,adelberger2003,
shapley2003,steidel2004,adelberger2005,steidel2010},
yet obtaining robust constraints on the physical properties associated with
these outflows (e.g., mass outflow rates) remains a challenge
\citep[but see, e.g.,][]{steidel2010}. At the same time,
there is no obvious connection between these observations and
the extremely popular theoretical models of cold gas accretion
\citep[e.g.,][]{keres2005,keres2009,dekel2009_nature}.
An open challenge is to observe a ``smoking gun" of gas accretion
in high-redshift galaxies. Second, the study of galaxy
environments at $z\geq 2$ is a field in its infancy.
While overdensities of LAEs have been identified near
luminous radio galaxies \citep{venemans2007}, and a small number of apparent
protoclusters have been discovered serendipitously
during the course of high-redshift galaxy spectroscopic
surveys \citep{steidel1998,steidel2005}, the study of the dependence
of galaxy properties on environment at $z\geq 2$ remains largely
untapped.  Environmental studies will require
extensive spectroscopy of mass-complete samples
in both overdense and ``field" environments,
and will have potentially fundamental implications
for understanding the origin of local
galaxy environmental trends. Finally, we look forward
to the truly exciting insights that will be made possible
by upcoming facilities, including sensitive multi-object
near-IR spectrographs on $8-10$-meter class telescopes,
ALMA, JWST, and the extremely large ground-based telescopes
of the next decade.

\bigskip

\section*{Acknowledgments}
I would like to thank Katherine Kornei and Gwen Rudie for careful readings
of the manuscript, and Naveen Reddy for many stimulating 
discussions and helpful comments. I would also like to thank the 
numerous authors who so generously shared
their figures for this review, and note that figures previously appearing
in {\it The Astrophysical Journal} have been reproduced by permission of the AAS. 
Very importantly, my thanks go to Sandy Faber, for 
giving me the opportunity to review the field of high-redshift galaxy
properties, and for her patience and constructive feedback
as editor.  Finally, I would like to thank my collaborators
for useful dialogs over the past decade and more, which have shaped my
understanding of this rapidly developing field. AES is a Packard Fellow; financial
support from the Packard Foundation is gratefully acknowledged.

\bibliographystyle{Astronomy}
\bibliography{apj-jour,araa}

\begin{thebibliography}{}
\expandafter\ifx\csname natexlab\endcsname\relax\def\natexlab#1{#1}\fi

\bibitem[{{Abazajian} et~al.(2009){Abazajian}, {Adelman-McCarthy},
  {Ag{\"u}eros}, {Allam}, {Allende Prieto} et~al.}]{abazajian2009}
{Abazajian} KN, {Adelman-McCarthy} JK, {Ag{\"u}eros} MA, {Allam} SS, {Allende
  Prieto} C, et~al. 2009.
\newblock \textit{\apjs} 182:543--558

\bibitem[{{Abraham} et~al.(2004){Abraham}, {Glazebrook}, {McCarthy},
  {Crampton}, {Murowinski} et~al.}]{abraham2004}
{Abraham} RG, {Glazebrook} K, {McCarthy} PJ, {Crampton} D, {Murowinski} R,
  et~al. 2004.
\newblock \textit{\aj} 127:2455--2483

\bibitem[{{Abraham}, {van den Bergh} \& {Nair}(2003)}]{abraham2003}
{Abraham} RG, {van den Bergh} S, {Nair} P. 2003.
\newblock \textit{\apj} 588:218--229

\bibitem[{{Adelberger} et~al.(2005){Adelberger}, {Shapley}, {Steidel},
  {Pettini}, {Erb} \& {Reddy}}]{adelberger2005}
{Adelberger} KL, {Shapley} AE, {Steidel} CC, {Pettini} M, {Erb} DK, {Reddy} NA.
  2005.
\newblock \textit{\apj} 629:636--653

\bibitem[{{Adelberger} \& {Steidel}(2000)}]{adelberger2000}
{Adelberger} KL, {Steidel} CC. 2000.
\newblock \textit{\apj} 544:218--241

\bibitem[{{Adelberger} et~al.(2004){Adelberger}, {Steidel}, {Shapley}, {Hunt},
  {Erb} et~al.}]{adelberger2004}
{Adelberger} KL, {Steidel} CC, {Shapley} AE, {Hunt} MP, {Erb} DK, et~al. 2004.
\newblock \textit{\apj} 607:226--240

\bibitem[{{Adelberger} et~al.(2003){Adelberger}, {Steidel}, {Shapley} \&
  {Pettini}}]{adelberger2003}
{Adelberger} KL, {Steidel} CC, {Shapley} AE, {Pettini} M. 2003.
\newblock \textit{\apj} 584:45--75

\bibitem[{{Alexander} et~al.(2005){Alexander}, {Bauer}, {Chapman}, {Smail},
  {Blain} et~al.}]{alexander2005}
{Alexander} DM, {Bauer} FE, {Chapman} SC, {Smail} I, {Blain} AW, et~al. 2005.
\newblock \textit{\apj} 632:736--750

\bibitem[{{Asplund} et~al.(2004){Asplund}, {Grevesse}, {Sauval}, {Allende
  Prieto} \& {Kiselman}}]{asplund2004}
{Asplund} M, {Grevesse} N, {Sauval} AJ, {Allende Prieto} C, {Kiselman} D. 2004.
\newblock \textit{\aap} 417:751--768

\bibitem[{{Baker} et~al.(2001){Baker}, {Lutz}, {Genzel}, {Tacconi} \&
  {Lehnert}}]{baker2001}
{Baker} AJ, {Lutz} D, {Genzel} R, {Tacconi} LJ, {Lehnert} MD. 2001.
\newblock \textit{\aap} 372:L37--L40

\bibitem[{{Baldry} et~al.(2004){Baldry}, {Glazebrook}, {Brinkmann},
  {Ivezi{\'c}}, {Lupton} et~al.}]{baldry2004}
{Baldry} IK, {Glazebrook} K, {Brinkmann} J, {Ivezi{\'c}} {\v Z}, {Lupton} RH,
  et~al. 2004.
\newblock \textit{\apj} 600:681--694

\bibitem[{{Baldry}, {Glazebrook} \& {Driver}(2008)}]{baldry2008}
{Baldry} IK, {Glazebrook} K, {Driver} SP. 2008.
\newblock \textit{\mnras} 388:945--959

\bibitem[{{Bastian}, {Covey} \& {Meyer}(2010)}]{bastian2010}
{Bastian} N, {Covey} KR, {Meyer} MR. 2010.
\newblock \textit{\araa} 48:339--389

\bibitem[{{Bell} et~al.(2004){Bell}, {Wolf}, {Meisenheimer}, {Rix}, {Borch}
  et~al.}]{bell2004}
{Bell} EF, {Wolf} C, {Meisenheimer} K, {Rix} H, {Borch} A, et~al. 2004.
\newblock \textit{\apj} 608:752--767

\bibitem[{{Blanton} et~al.(2003){Blanton}, {Hogg}, {Bahcall}, {Brinkmann},
  {Britton} et~al.}]{blanton2003}
{Blanton} MR, {Hogg} DW, {Bahcall} NA, {Brinkmann} J, {Britton} M, et~al. 2003.
\newblock \textit{\apj} 592:819--838

\bibitem[{{Blanton} \& {Moustakas}(2009)}]{blanton2009}
{Blanton} MR, {Moustakas} J. 2009.
\newblock \textit{\araa} 47:159--210

\bibitem[{{Blitz} \& {Rosolowsky}(2006)}]{blitz2006}
{Blitz} L, {Rosolowsky} E. 2006.
\newblock \textit{\apj} 650:933--944

\bibitem[{{Borys} et~al.(2005){Borys}, {Smail}, {Chapman}, {Blain}, {Alexander}
  \& {Ivison}}]{borys2005}
{Borys} C, {Smail} I, {Chapman} SC, {Blain} AW, {Alexander} DM, {Ivison} RJ.
  2005.
\newblock \textit{\apj} 635:853--863

\bibitem[{{Bouch{\'e}} et~al.(2007){Bouch{\'e}}, {Cresci}, {Davies},
  {Eisenhauer}, {F{\"o}rster Schreiber} et~al.}]{bouche2007b}
{Bouch{\'e}} N, {Cresci} G, {Davies} R, {Eisenhauer} F, {F{\"o}rster Schreiber}
  NM, et~al. 2007.
\newblock \textit{\apj} 671:303--309

\bibitem[{{Bournaud} \& {Elmegreen}(2009)}]{bournaud2009}
{Bournaud} F, {Elmegreen} BG. 2009.
\newblock \textit{\apjl} 694:L158--L161

\bibitem[{{Bouwens} et~al.(2009){Bouwens}, {Illingworth}, {Franx}, {Chary},
  {Meurer} et~al.}]{bouwens2009}
{Bouwens} RJ, {Illingworth} GD, {Franx} M, {Chary} R, {Meurer} GR, et~al. 2009.
\newblock \textit{\apj} 705:936--961

\bibitem[{{Bouwens} et~al.(2007){Bouwens}, {Illingworth}, {Franx} \&
  {Ford}}]{bouwens2007}
{Bouwens} RJ, {Illingworth} GD, {Franx} M, {Ford} H. 2007.
\newblock \textit{\apj} 670:928--958

\bibitem[{{Bouwens} et~al.(2010){Bouwens}, {Illingworth}, {Oesch}, {Stiavelli},
  {van Dokkum} et~al.}]{bouwens2010}
{Bouwens} RJ, {Illingworth} GD, {Oesch} PA, {Stiavelli} M, {van Dokkum} P,
  et~al. 2010.
\newblock \textit{\apjl} 709:L133--L137

\bibitem[{{Boylan-Kolchin} et~al.(2009){Boylan-Kolchin}, {Springel}, {White},
  {Jenkins} \& {Lemson}}]{boylankolchin2009}
{Boylan-Kolchin} M, {Springel} V, {White} SDM, {Jenkins} A, {Lemson} G. 2009.
\newblock \textit{\mnras} 398:1150--1164

\bibitem[{{Brammer} et~al.(2009){Brammer}, {Whitaker}, {van Dokkum},
  {Marchesini}, {Labb{\'e}} et~al.}]{brammer2009}
{Brammer} GB, {Whitaker} KE, {van Dokkum} PG, {Marchesini} D, {Labb{\'e}} I,
  et~al. 2009.
\newblock \textit{\apjl} 706:L173--L177

\bibitem[{{Brinchmann} et~al.(2004){Brinchmann}, {Charlot}, {White},
  {Tremonti}, {Kauffmann} et~al.}]{brinchmann2004}
{Brinchmann} J, {Charlot} S, {White} SDM, {Tremonti} C, {Kauffmann} G, et~al.
  2004.
\newblock \textit{\mnras} 351:1151--1179

\bibitem[{{Bruzual} \& {Charlot}(2003)}]{bruzualcharlot2003}
{Bruzual} G, {Charlot} S. 2003.
\newblock \textit{\mnras} 344:1000--1028

\bibitem[{{Buitrago} et~al.(2008){Buitrago}, {Trujillo}, {Conselice},
  {Bouwens}, {Dickinson} \& {Yan}}]{buitrago2008}
{Buitrago} F, {Trujillo} I, {Conselice} CJ, {Bouwens} RJ, {Dickinson} M, {Yan}
  H. 2008.
\newblock \textit{\apjl} 687:L61--L64

\bibitem[{{Calzetti}(2001)}]{calzetti2001}
{Calzetti} D. 2001.
\newblock \textit{\pasp} 113:1449--1485

\bibitem[{{Calzetti} et~al.(2000){Calzetti}, {Armus}, {Bohlin}, {Kinney},
  {Koornneef} \& {Storchi-Bergmann}}]{calzetti2000}
{Calzetti} D, {Armus} L, {Bohlin} RC, {Kinney} AL, {Koornneef} J,
  {Storchi-Bergmann} T. 2000.
\newblock \textit{\apj} 533:682--695

\bibitem[{{Cameron} et~al.(2010){Cameron}, {Carollo}, {Oesch}, {Bouwens},
  {Illingworth} et~al.}]{cameron2010}
{Cameron} E, {Carollo} CM, {Oesch} PA, {Bouwens} RJ, {Illingworth} GD, et~al.
  2010.
\newblock \textit{ArXiv e-prints, (astro-ph/1007.2422)}

\bibitem[{{Cappellari} et~al.(2009){Cappellari}, {di Serego Alighieri},
  {Cimatti}, {Daddi}, {Renzini} et~al.}]{cappellari2009}
{Cappellari} M, {di Serego Alighieri} S, {Cimatti} A, {Daddi} E, {Renzini} A,
  et~al. 2009.
\newblock \textit{\apjl} 704:L34--L39

\bibitem[{{Caputi} et~al.(2007){Caputi}, {Lagache}, {Yan}, {Dole}, {Bavouzet}
  et~al.}]{caputi2007}
{Caputi} KI, {Lagache} G, {Yan} L, {Dole} H, {Bavouzet} N, et~al. 2007.
\newblock \textit{\apj} 660:97--116

\bibitem[{{Cardelli}, {Clayton} \& {Mathis}(1989)}]{cardelli1989}
{Cardelli} JA, {Clayton} GC, {Mathis} JS. 1989.
\newblock \textit{\apj} 345:245--256

\bibitem[{{Casey} et~al.(2009){Casey}, {Chapman}, {Beswick}, {Biggs}, {Blain}
  et~al.}]{casey2009}
{Casey} CM, {Chapman} SC, {Beswick} RJ, {Biggs} AD, {Blain} AW, et~al. 2009.
\newblock \textit{\mnras} 399:121--128

\bibitem[{{Cassata} et~al.(2008){Cassata}, {Cimatti}, {Kurk}, {Rodighiero},
  {Pozzetti} et~al.}]{cassata2008}
{Cassata} P, {Cimatti} A, {Kurk} J, {Rodighiero} G, {Pozzetti} L, et~al. 2008.
\newblock \textit{\aap} 483:L39--L42

\bibitem[{{Cenarro} \& {Trujillo}(2009)}]{cenarro2009}
{Cenarro} AJ, {Trujillo} I. 2009.
\newblock \textit{\apjl} 696:L43--L47

\bibitem[{{Ceverino}, {Dekel} \& {Bournaud}(2010)}]{ceverino2010}
{Ceverino} D, {Dekel} A, {Bournaud} F. 2010.
\newblock \textit{\mnras} 404:2151--2169

\bibitem[{{Chabrier}(2003)}]{chabrier2003}
{Chabrier} G. 2003.
\newblock \textit{\pasp} 115:763--795

\bibitem[{{Chapman} et~al.(2003){Chapman}, {Blain}, {Ivison} \&
  {Smail}}]{chapman2003}
{Chapman} SC, {Blain} AW, {Ivison} RJ, {Smail} IR. 2003.
\newblock \textit{\nat} 422:695--698

\bibitem[{{Chapman} et~al.(2005){Chapman}, {Blain}, {Smail} \&
  {Ivison}}]{chapman2005}
{Chapman} SC, {Blain} AW, {Smail} I, {Ivison} RJ. 2005.
\newblock \textit{\apj} 622:772--796

\bibitem[{{Chapman} et~al.(2010){Chapman}, {Ivison}, {Roseboom}, {Auld}, {Bock}
  et~al.}]{chapman2010}
{Chapman} SC, {Ivison} RJ, {Roseboom} IG, {Auld} R, {Bock} J, et~al. 2010.
\newblock \textit{\mnras} 409:L13--L18

\bibitem[{{Chapman} et~al.(2004){Chapman}, {Smail}, {Blain} \&
  {Ivison}}]{chapman2004}
{Chapman} SC, {Smail} I, {Blain} AW, {Ivison} RJ. 2004.
\newblock \textit{\apj} 614:671--678

\bibitem[{{Charlot} \& {Fall}(2000)}]{charlotfall2000}
{Charlot} S, {Fall} SM. 2000.
\newblock \textit{\apj} 539:718--731

\bibitem[{{Chary} \& {Elbaz}(2001)}]{chary2001}
{Chary} R, {Elbaz} D. 2001.
\newblock \textit{\apj} 556:562--581

\bibitem[{{Cimatti} et~al.(2008){Cimatti}, {Cassata}, {Pozzetti}, {Kurk},
  {Mignoli} et~al.}]{cimatti2008}
{Cimatti} A, {Cassata} P, {Pozzetti} L, {Kurk} J, {Mignoli} M, et~al. 2008.
\newblock \textit{\aap} 482:21--42

\bibitem[{{Cirasuolo} et~al.(2010){Cirasuolo}, {McLure}, {Dunlop}, {Almaini},
  {Foucaud} \& {Simpson}}]{cirasuolo2010}
{Cirasuolo} M, {McLure} RJ, {Dunlop} JS, {Almaini} O, {Foucaud} S, {Simpson} C.
  2010.
\newblock \textit{\mnras} 401:1166--1176

\bibitem[{{Cirasuolo} et~al.(2007){Cirasuolo}, {McLure}, {Dunlop}, {Almaini},
  {Foucaud} et~al.}]{cirasuolo2007}
{Cirasuolo} M, {McLure} RJ, {Dunlop} JS, {Almaini} O, {Foucaud} S, et~al. 2007.
\newblock \textit{\mnras} 380:585--595

\bibitem[{{Cole} et~al.(2001){Cole}, {Norberg}, {Baugh}, {Frenk},
  {Bland-Hawthorn} et~al.}]{cole2001}
{Cole} S, {Norberg} P, {Baugh} CM, {Frenk} CS, {Bland-Hawthorn} J, et~al. 2001.
\newblock \textit{\mnras} 326:255--273

\bibitem[{{Colless} et~al.(2001){Colless}, {Dalton}, {Maddox}, {Sutherland},
  {Norberg} et~al.}]{colless2001}
{Colless} M, {Dalton} G, {Maddox} S, {Sutherland} W, {Norberg} P, et~al. 2001.
\newblock \textit{\mnras} 328:1039--1063

\bibitem[{{Condon}(1992)}]{condon1992}
{Condon} JJ. 1992.
\newblock \textit{\araa} 30:575--611

\bibitem[{{Conroy}, {Gunn} \& {White}(2009)}]{conroy2009}
{Conroy} C, {Gunn} JE, {White} M. 2009.
\newblock \textit{\apj} 699:486--506

\bibitem[{{Conroy} et~al.(2008){Conroy}, {Shapley}, {Tinker}, {Santos} \&
  {Lemson}}]{conroy2008}
{Conroy} C, {Shapley} AE, {Tinker} JL, {Santos} MR, {Lemson} G. 2008.
\newblock \textit{\apj} 679:1192--1203

\bibitem[{{Conselice}, {Chapman} \& {Windhorst}(2003)}]{conselice2003}
{Conselice} CJ, {Chapman} SC, {Windhorst} RA. 2003.
\newblock \textit{\apjl} 596:L5--L8

\bibitem[{{Coppin} et~al.(2008){Coppin}, {Halpern}, {Scott}, {Borys}, {Dunlop}
  et~al.}]{coppin2008}
{Coppin} K, {Halpern} M, {Scott} D, {Borys} C, {Dunlop} J, et~al. 2008.
\newblock \textit{\mnras} 384:1597--1610

\bibitem[{{Cowie} \& {Hu}(1998)}]{cowie1998}
{Cowie} LL, {Hu} EM. 1998.
\newblock \textit{\aj} 115:1319--1328

\bibitem[{{Cowie}, {Hu} \& {Songaila}(1995)}]{cowie1995}
{Cowie} LL, {Hu} EM, {Songaila} A. 1995.
\newblock \textit{\aj} 110:1576--+

\bibitem[{{Cresci} et~al.(2009){Cresci}, {Hicks}, {Genzel}, {Schreiber},
  {Davies} et~al.}]{cresci2009}
{Cresci} G, {Hicks} EKS, {Genzel} R, {Schreiber} NMF, {Davies} R, et~al. 2009.
\newblock \textit{\apj} 697:115--132

\bibitem[{{Cresci} et~al.(2010){Cresci}, {Mannucci}, {Maiolino}, {Marconi},
  {Gnerucci} \& {Magrini}}]{cresci2010}
{Cresci} G, {Mannucci} F, {Maiolino} R, {Marconi} A, {Gnerucci} A, {Magrini} L.
  2010.
\newblock \textit{\nat} 467:811--813

\bibitem[{{Daddi} et~al.(2010){Daddi}, {Bournaud}, {Walter}, {Dannerbauer},
  {Carilli} et~al.}]{daddi2010}
{Daddi} E, {Bournaud} F, {Walter} F, {Dannerbauer} H, {Carilli} CL, et~al.
  2010.
\newblock \textit{\apj} 713:686--707

\bibitem[{{Daddi} et~al.(2004){Daddi}, {Cimatti}, {Renzini}, {Fontana},
  {Mignoli} et~al.}]{daddi2004}
{Daddi} E, {Cimatti} A, {Renzini} A, {Fontana} A, {Mignoli} M, et~al. 2004.
\newblock \textit{\apj} 617:746--764

\bibitem[{{Daddi} et~al.(2007){Daddi}, {Dickinson}, {Morrison}, {Chary},
  {Cimatti} et~al.}]{daddi2007}
{Daddi} E, {Dickinson} M, {Morrison} G, {Chary} R, {Cimatti} A, et~al. 2007.
\newblock \textit{\apj} 670:156--172

\bibitem[{{Daddi} et~al.(2005){Daddi}, {Renzini}, {Pirzkal}, {Cimatti},
  {Malhotra} et~al.}]{daddi2005}
{Daddi} E, {Renzini} A, {Pirzkal} N, {Cimatti} A, {Malhotra} S, et~al. 2005.
\newblock \textit{\apj} 626:680--697

\bibitem[{{Dale} \& {Helou}(2002)}]{dale2002}
{Dale} DA, {Helou} G. 2002.
\newblock \textit{\apj} 576:159--168

\bibitem[{{Damjanov} et~al.(2009){Damjanov}, {McCarthy}, {Abraham},
  {Glazebrook}, {Yan} et~al.}]{damjanov2009}
{Damjanov} I, {McCarthy} PJ, {Abraham} RG, {Glazebrook} K, {Yan} H, et~al.
  2009.
\newblock \textit{\apj} 695:101--115

\bibitem[{{Dannerbauer} et~al.(2009){Dannerbauer}, {Daddi}, {Riechers},
  {Walter}, {Carilli} et~al.}]{dannerbauer2009}
{Dannerbauer} H, {Daddi} E, {Riechers} DA, {Walter} F, {Carilli} CL, et~al.
  2009.
\newblock \textit{\apjl} 698:L178--L182

\bibitem[{{Dav{\'e}}(2008)}]{dave2008}
{Dav{\'e}} R. 2008.
\newblock \textit{\mnras} 385:147--160

\bibitem[{{Dav{\'e}} et~al.(2010){Dav{\'e}}, {Finlator}, {Oppenheimer},
  {Fardal}, {Katz} et~al.}]{dave2010}
{Dav{\'e}} R, {Finlator} K, {Oppenheimer} BD, {Fardal} M, {Katz} N, et~al.
  2010.
\newblock \textit{\mnras} 404:1355--1368

\bibitem[{{Davis} et~al.(2007){Davis}, {Guhathakurta}, {Konidaris}, {Newman},
  {Ashby} et~al.}]{davis2007}
{Davis} M, {Guhathakurta} P, {Konidaris} NP, {Newman} JA, {Ashby} MLN, et~al.
  2007.
\newblock \textit{\apjl} 660:L1--L6

\bibitem[{{Dekel} et~al.(2009){Dekel}, {Birnboim}, {Engel}, {Freundlich},
  {Goerdt} et~al.}]{dekel2009_nature}
{Dekel} A, {Birnboim} Y, {Engel} G, {Freundlich} J, {Goerdt} T, et~al. 2009.
\newblock \textit{\nat} 457:451--454

\bibitem[{{Dekel}, {Sari} \& {Ceverino}(2009)}]{dekel2009}
{Dekel} A, {Sari} R, {Ceverino} D. 2009.
\newblock \textit{\apj} 703:785--801

\bibitem[{{Dessauges-Zavadsky} et~al.(2010){Dessauges-Zavadsky}, {D'Odorico},
  {Schaerer}, {Modigliani}, {Tapken} \& {Vernet}}]{dessauges2010}
{Dessauges-Zavadsky} M, {D'Odorico} S, {Schaerer} D, {Modigliani} A, {Tapken}
  C, {Vernet} J. 2010.
\newblock \textit{\aap} 510:A26+

\bibitem[{{Dey} et~al.(2008){Dey}, {Soifer}, {Desai}, {Brand}, {Le Floc'h}
  et~al.}]{dey2008}
{Dey} A, {Soifer} BT, {Desai} V, {Brand} K, {Le Floc'h} E, et~al. 2008.
\newblock \textit{\apj} 677:943--956

\bibitem[{{Dickinson}(2000)}]{dickinson2000}
{Dickinson} M. 2000.
\newblock In \textit{Astronomy, physics and chemistry of $H^{+}_{3}$}, vol. 358
  of \textit{Royal Society of London Philosophical Transactions Series A}

\bibitem[{{Dickinson} et~al.(2003){Dickinson}, {Papovich}, {Ferguson} \&
  {Budav{\'a}ri}}]{dickinson2003}
{Dickinson} M, {Papovich} C, {Ferguson} HC, {Budav{\'a}ri} T. 2003.
\newblock \textit{\apj} 587:25--40

\bibitem[{{Drory} et~al.(2005){Drory}, {Salvato}, {Gabasch}, {Bender}, {Hopp}
  et~al.}]{drory2005}
{Drory} N, {Salvato} M, {Gabasch} A, {Bender} R, {Hopp} U, et~al. 2005.
\newblock \textit{\apjl} 619:L131--L134

\bibitem[{{Elbaz} et~al.(2002){Elbaz}, {Cesarsky}, {Chanial}, {Aussel},
  {Franceschini} et~al.}]{elbaz2002}
{Elbaz} D, {Cesarsky} CJ, {Chanial} P, {Aussel} H, {Franceschini} A, et~al.
  2002.
\newblock \textit{\aap} 384:848--865

\bibitem[{{Elbaz} et~al.(2007){Elbaz}, {Daddi}, {Le Borgne}, {Dickinson},
  {Alexander} et~al.}]{elbaz2007}
{Elbaz} D, {Daddi} E, {Le Borgne} D, {Dickinson} M, {Alexander} DM, et~al.
  2007.
\newblock \textit{\aap} 468:33--48

\bibitem[{{Elmegreen}, {Bournaud} \& {Elmegreen}(2008)}]{elmegreen2008}
{Elmegreen} BG, {Bournaud} F, {Elmegreen} DM. 2008.
\newblock \textit{\apj} 688:67--77

\bibitem[{{Elmegreen} \& {Elmegreen}(2005)}]{elmegreen2005}
{Elmegreen} BG, {Elmegreen} DM. 2005.
\newblock \textit{\apj} 627:632--646

\bibitem[{{Elmegreen} et~al.(2009){Elmegreen}, {Elmegreen}, {Fernandez} \&
  {Lemonias}}]{elmegreen2009}
{Elmegreen} BG, {Elmegreen} DM, {Fernandez} MX, {Lemonias} JJ. 2009.
\newblock \textit{\apj} 692:12--31

\bibitem[{{Elsner}, {Feulner} \& {Hopp}(2008)}]{elsner2008}
{Elsner} F, {Feulner} G, {Hopp} U. 2008.
\newblock \textit{\aap} 477:503--512

\bibitem[{{Engel} et~al.(2010){Engel}, {Tacconi}, {Davies}, {Neri}, {Smail}
  et~al.}]{engel2010}
{Engel} H, {Tacconi} LJ, {Davies} RI, {Neri} R, {Smail} I, et~al. 2010.
\newblock \textit{\apj} 724:233--243

\bibitem[{{Erb}(2008)}]{erb2008}
{Erb} DK. 2008.
\newblock \textit{\apj} 674:151--156

\bibitem[{{Erb} et~al.(2006{\natexlab{a}}){Erb}, {Shapley}, {Pettini},
  {Steidel}, {Reddy} \& {Adelberger}}]{erb2006a}
{Erb} DK, {Shapley} AE, {Pettini} M, {Steidel} CC, {Reddy} NA, {Adelberger} KL.
  2006{\natexlab{a}}.
\newblock \textit{\apj} 644:813--828

\bibitem[{{Erb} et~al.(2003){Erb}, {Shapley}, {Steidel}, {Pettini},
  {Adelberger} et~al.}]{erb2003}
{Erb} DK, {Shapley} AE, {Steidel} CC, {Pettini} M, {Adelberger} KL, et~al.
  2003.
\newblock \textit{\apj} 591:101--118

\bibitem[{{Erb} et~al.(2004){Erb}, {Steidel}, {Shapley}, {Pettini} \&
  {Adelberger}}]{erb2004}
{Erb} DK, {Steidel} CC, {Shapley} AE, {Pettini} M, {Adelberger} KL. 2004.
\newblock \textit{\apj} 612:122--130

\bibitem[{{Erb} et~al.(2006{\natexlab{b}}){Erb}, {Steidel}, {Shapley},
  {Pettini}, {Reddy} \& {Adelberger}}]{erb2006c}
{Erb} DK, {Steidel} CC, {Shapley} AE, {Pettini} M, {Reddy} NA, {Adelberger} KL.
  2006{\natexlab{b}}.
\newblock \textit{\apj} 647:128--139

\bibitem[{{Erb} et~al.(2006{\natexlab{c}}){Erb}, {Steidel}, {Shapley},
  {Pettini}, {Reddy} \& {Adelberger}}]{erb2006b}
{Erb} DK, {Steidel} CC, {Shapley} AE, {Pettini} M, {Reddy} NA, {Adelberger} KL.
  2006{\natexlab{c}}.
\newblock \textit{\apj} 646:107--132

\bibitem[{{Fan} et~al.(2008){Fan}, {Lapi}, {De Zotti} \& {Danese}}]{fan2008}
{Fan} L, {Lapi} A, {De Zotti} G, {Danese} L. 2008.
\newblock \textit{\apjl} 689:L101--L104

\bibitem[{{Finlator} \& {Dav{\'e}}(2008)}]{finlator2008}
{Finlator} K, {Dav{\'e}} R. 2008.
\newblock \textit{\mnras} 385:2181--2204

\bibitem[{{Fioc} \& {Rocca-Volmerange}(1997)}]{fioc1997}
{Fioc} M, {Rocca-Volmerange} B. 1997.
\newblock \textit{\aap} 326:950--962

\bibitem[{{Fontana} et~al.(2004){Fontana}, {Pozzetti}, {Donnarumma}, {Renzini},
  {Cimatti} et~al.}]{fontana2004}
{Fontana} A, {Pozzetti} L, {Donnarumma} I, {Renzini} A, {Cimatti} A, et~al.
  2004.
\newblock \textit{\aap} 424:23--42

\bibitem[{{Fontana} et~al.(2006){Fontana}, {Salimbeni}, {Grazian}, {Giallongo},
  {Pentericci} et~al.}]{fontana2006}
{Fontana} A, {Salimbeni} S, {Grazian} A, {Giallongo} E, {Pentericci} L, et~al.
  2006.
\newblock \textit{\aap} 459:745--757

\bibitem[{{F{\"o}rster Schreiber} et~al.(2009){F{\"o}rster Schreiber},
  {Genzel}, {Bouch{\'e}}, {Cresci}, {Davies} et~al.}]{forsterschreiber2009}
{F{\"o}rster Schreiber} NM, {Genzel} R, {Bouch{\'e}} N, {Cresci} G, {Davies} R,
  et~al. 2009.
\newblock \textit{\apj} 706:1364--1428

\bibitem[{{F{\"o}rster Schreiber} et~al.(2011){F{\"o}rster Schreiber},
  {Shapley}, {Erb}, {Genzel}, {Steidel} et~al.}]{forsterschreiber2011}
{F{\"o}rster Schreiber} NM, {Shapley} AE, {Erb} DK, {Genzel} R, {Steidel} CC,
  et~al. 2011.
\newblock \textit{\apj} 731:65--+

\bibitem[{{F{\"o}rster Schreiber} et~al.(2004){F{\"o}rster Schreiber}, {van
  Dokkum}, {Franx}, {Labb{\'e}}, {Rudnick} et~al.}]{forsterschreiber2004}
{F{\"o}rster Schreiber} NM, {van Dokkum} PG, {Franx} M, {Labb{\'e}} I,
  {Rudnick} G, et~al. 2004.
\newblock \textit{\apj} 616:40--62

\bibitem[{{Franx} et~al.(2003){Franx}, {Labb{\' e}}, {Rudnick}, {van Dokkum},
  {Daddi} et~al.}]{franx2003}
{Franx} M, {Labb{\' e}} I, {Rudnick} G, {van Dokkum} PG, {Daddi} E, et~al.
  2003.
\newblock \textit{\apjl} 587:L79--L82

\bibitem[{{Franx} et~al.(2008){Franx}, {van Dokkum}, {Schreiber}, {Wuyts},
  {Labb{\'e}} \& {Toft}}]{franx2008}
{Franx} M, {van Dokkum} PG, {Schreiber} NMF, {Wuyts} S, {Labb{\'e}} I, {Toft}
  S. 2008.
\newblock \textit{\apj} 688:770--788

\bibitem[{{Gabasch} et~al.(2004){Gabasch}, {Bender}, {Seitz}, {Hopp}, {Saglia}
  et~al.}]{gabasch2004}
{Gabasch} A, {Bender} R, {Seitz} S, {Hopp} U, {Saglia} RP, et~al. 2004.
\newblock \textit{\aap} 421:41--58

\bibitem[{{Gawiser} et~al.(2007){Gawiser}, {Francke}, {Lai}, {Schawinski},
  {Gronwall} et~al.}]{gawiser2007}
{Gawiser} E, {Francke} H, {Lai} K, {Schawinski} K, {Gronwall} C, et~al. 2007.
\newblock \textit{\apj} 671:278--284

\bibitem[{{Genel} et~al.(2008){Genel}, {Genzel}, {Bouch{\'e}}, {Sternberg},
  {Naab} et~al.}]{genel2008}
{Genel} S, {Genzel} R, {Bouch{\'e}} N, {Sternberg} A, {Naab} T, et~al. 2008.
\newblock \textit{\apj} 688:789--793

\bibitem[{{Genzel} et~al.(2003){Genzel}, {Baker}, {Tacconi}, {Lutz}, {Cox}
  et~al.}]{genzel2003}
{Genzel} R, {Baker} AJ, {Tacconi} LJ, {Lutz} D, {Cox} P, et~al. 2003.
\newblock \textit{\apj} 584:633--642

\bibitem[{{Genzel} et~al.(2008){Genzel}, {Burkert}, {Bouch{\'e}}, {Cresci},
  {F{\"o}rster Schreiber} et~al.}]{genzel2008}
{Genzel} R, {Burkert} A, {Bouch{\'e}} N, {Cresci} G, {F{\"o}rster Schreiber}
  NM, et~al. 2008.
\newblock \textit{\apj} 687:59--77

\bibitem[{{Genzel} et~al.(2011){Genzel}, {Newman}, {Jones}, {F{\"o}rster
  Schreiber}, {Shapiro} et~al.}]{genzel2010b}
{Genzel} R, {Newman} S, {Jones} T, {F{\"o}rster Schreiber} NM, {Shapiro} K,
  et~al. 2011.
\newblock \textit{\apj} 733:101--+

\bibitem[{{Genzel} et~al.(2010){Genzel}, {Tacconi}, {Gracia-Carpio},
  {Sternberg}, {Cooper} et~al.}]{genzel2010a}
{Genzel} R, {Tacconi} LJ, {Gracia-Carpio} J, {Sternberg} A, {Cooper} MC, et~al.
  2010.
\newblock \textit{\mnras} 407:2091--2108

\bibitem[{{Giavalisco}, {Steidel} \& {Macchetto}(1996)}]{giavalisco1996}
{Giavalisco} M, {Steidel} CC, {Macchetto} FD. 1996.
\newblock \textit{\apj} 470:189--+

\bibitem[{{Graves} \& {Faber}(2010)}]{graves2010}
{Graves} GJ, {Faber} SM. 2010.
\newblock \textit{\apj} 717:803--824

\bibitem[{{Greve} et~al.(2005){Greve}, {Bertoldi}, {Smail}, {Neri}, {Chapman}
  et~al.}]{greve2005}
{Greve} TR, {Bertoldi} F, {Smail} I, {Neri} R, {Chapman} SC, et~al. 2005.
\newblock \textit{\mnras} 359:1165--1183

\bibitem[{{Gronwall} et~al.(2007){Gronwall}, {Ciardullo}, {Hickey}, {Gawiser},
  {Feldmeier} et~al.}]{gronwall2007}
{Gronwall} C, {Ciardullo} R, {Hickey} T, {Gawiser} E, {Feldmeier} JJ, et~al.
  2007.
\newblock \textit{\apj} 667:79--91

\bibitem[{{Hainline} et~al.(2009){Hainline}, {Shapley}, {Kornei}, {Pettini},
  {Buckley-Geer} et~al.}]{hainline2009}
{Hainline} KN, {Shapley} AE, {Kornei} KA, {Pettini} M, {Buckley-Geer} E, et~al.
  2009.
\newblock \textit{\apj} 701:52--65

\bibitem[{{Halliday} et~al.(2008){Halliday}, {Daddi}, {Cimatti}, {Kurk},
  {Renzini} et~al.}]{halliday2008}
{Halliday} C, {Daddi} E, {Cimatti} A, {Kurk} J, {Renzini} A, et~al. 2008.
\newblock \textit{\aap} 479:417--425

\bibitem[{{Harris} et~al.(2010){Harris}, {Baker}, {Zonak}, {Sharon}, {Genzel}
  et~al.}]{harris2010}
{Harris} AI, {Baker} AJ, {Zonak} SG, {Sharon} CE, {Genzel} R, et~al. 2010.
\newblock \textit{\apj} 723:1139--1149

\bibitem[{{Hayashi} et~al.(2009){Hayashi}, {Motohara}, {Shimasaku}, {Onodera},
  {Uchimoto} et~al.}]{hayashi2009}
{Hayashi} M, {Motohara} K, {Shimasaku} K, {Onodera} M, {Uchimoto} YK, et~al.
  2009.
\newblock \textit{\apj} 691:140--151

\bibitem[{{Hayes} et~al.(2011){Hayes}, {Schaerer}, {{\"O}stlin}, {Mas-Hesse},
  {Atek} \& {Kunth}}]{hayes2011}
{Hayes} M, {Schaerer} D, {{\"O}stlin} G, {Mas-Hesse} JM, {Atek} H, {Kunth} D.
  2011.
\newblock \textit{\apj} 730:8--+

\bibitem[{{Hopkins}(2004)}]{hopkins2004}
{Hopkins} AM. 2004.
\newblock \textit{\apj} 615:209--221

\bibitem[{{Hopkins} \& {Beacom}(2006)}]{hopkinsbeacom2006}
{Hopkins} AM, {Beacom} JF. 2006.
\newblock \textit{\apj} 651:142--154

\bibitem[{{Hopkins} et~al.(2010){Hopkins}, {Bundy}, {Hernquist}, {Wuyts} \&
  {Cox}}]{hopkins2010}
{Hopkins} PF, {Bundy} K, {Hernquist} L, {Wuyts} S, {Cox} TJ. 2010.
\newblock \textit{\mnras} 401:1099--1117

\bibitem[{{Hopkins} et~al.(2009){Hopkins}, {Bundy}, {Murray}, {Quataert},
  {Lauer} \& {Ma}}]{hopkins2009}
{Hopkins} PF, {Bundy} K, {Murray} N, {Quataert} E, {Lauer} TR, {Ma} C. 2009.
\newblock \textit{\mnras} 398:898--910

\bibitem[{{Ivison} et~al.(2011){Ivison}, {Papadopoulos}, {Smail}, {Greve},
  {Thomson} et~al.}]{ivison2011}
{Ivison} RJ, {Papadopoulos} PP, {Smail} I, {Greve} TR, {Thomson} AP, et~al.
  2011.
\newblock \textit{\mnras} 412:1913--1925

\bibitem[{{Jones} et~al.(2010){Jones}, {Swinbank}, {Ellis}, {Richard} \&
  {Stark}}]{jones2010}
{Jones} TA, {Swinbank} AM, {Ellis} RS, {Richard} J, {Stark} DP. 2010.
\newblock \textit{\mnras} 404:1247--1262

\bibitem[{{Kajisawa} et~al.(2009){Kajisawa}, {Ichikawa}, {Tanaka}, {Konishi},
  {Yamada} et~al.}]{kajisawa2009}
{Kajisawa} M, {Ichikawa} T, {Tanaka} I, {Konishi} M, {Yamada} T, et~al. 2009.
\newblock \textit{\apj} 702:1393--1412

\bibitem[{{Kajisawa} et~al.(2010){Kajisawa}, {Ichikawa}, {Yamada}, {Uchimoto},
  {Yoshikawa} et~al.}]{kajisawa2010}
{Kajisawa} M, {Ichikawa} T, {Yamada} T, {Uchimoto} YK, {Yoshikawa} T, et~al.
  2010.
\newblock \textit{\apj} 723:129--145

\bibitem[{{Kauffmann} et~al.(2003{\natexlab{a}}){Kauffmann}, {Heckman},
  {White}, {Charlot}, {Tremonti} et~al.}]{kauffmann2003a}
{Kauffmann} G, {Heckman} TM, {White} SDM, {Charlot} S, {Tremonti} C, et~al.
  2003{\natexlab{a}}.
\newblock \textit{\mnras} 341:33--53

\bibitem[{{Kauffmann} et~al.(2003{\natexlab{b}}){Kauffmann}, {Heckman},
  {White}, {Charlot}, {Tremonti} et~al.}]{kauffmann2003b}
{Kauffmann} G, {Heckman} TM, {White} SDM, {Charlot} S, {Tremonti} C, et~al.
  2003{\natexlab{b}}.
\newblock \textit{\mnras} 341:54--69

\bibitem[{{Kennicutt}(1998)}]{kennicutt1998}
{Kennicutt} Jr. RC. 1998.
\newblock \textit{\araa} 36:189--232

\bibitem[{{Kere{\v s}} et~al.(2009){Kere{\v s}}, {Katz}, {Fardal}, {Dav{\'e}}
  \& {Weinberg}}]{keres2009}
{Kere{\v s}} D, {Katz} N, {Fardal} M, {Dav{\'e}} R, {Weinberg} DH. 2009.
\newblock \textit{\mnras} 395:160--179

\bibitem[{{Kere{\v s}} et~al.(2005){Kere{\v s}}, {Katz}, {Weinberg} \&
  {Dav{\'e}}}]{keres2005}
{Kere{\v s}} D, {Katz} N, {Weinberg} DH, {Dav{\'e}} R. 2005.
\newblock \textit{\mnras} 363:2--28

\bibitem[{{Kewley} \& {Dopita}(2002)}]{kewley2002}
{Kewley} LJ, {Dopita} MA. 2002.
\newblock \textit{\apjs} 142:35--52

\bibitem[{{Kewley} \& {Ellison}(2008)}]{kewley2008}
{Kewley} LJ, {Ellison} SL. 2008.
\newblock \textit{\apj} 681:1183--1204

\bibitem[{{Komatsu} et~al.(2011){Komatsu}, {Smith}, {Dunkley}, {Bennett},
  {Gold} et~al.}]{komatsu2011}
{Komatsu} E, {Smith} KM, {Dunkley} J, {Bennett} CL, {Gold} B, et~al. 2011.
\newblock \textit{\apjs} 192:18--+

\bibitem[{{Kornei} et~al.(2010){Kornei}, {Shapley}, {Erb}, {Steidel}, {Reddy}
  et~al.}]{kornei2010}
{Kornei} KA, {Shapley} AE, {Erb} DK, {Steidel} CC, {Reddy} NA, et~al. 2010.
\newblock \textit{\apj} 711:693--710

\bibitem[{{Kov{\'a}cs} et~al.(2006){Kov{\'a}cs}, {Chapman}, {Dowell}, {Blain},
  {Ivison} et~al.}]{kovacs2006}
{Kov{\'a}cs} A, {Chapman} SC, {Dowell} CD, {Blain} AW, {Ivison} RJ, et~al.
  2006.
\newblock \textit{\apj} 650:592--603

\bibitem[{{Kriek} et~al.(2010){Kriek}, {Labb{\'e}}, {Conroy}, {Whitaker}, {van
  Dokkum} et~al.}]{kriek2010}
{Kriek} M, {Labb{\'e}} I, {Conroy} C, {Whitaker} KE, {van Dokkum} PG, et~al.
  2010.
\newblock \textit{\apjl} 722:L64--L69

\bibitem[{{Kriek} et~al.(2008{\natexlab{a}}){Kriek}, {van der Wel}, {van
  Dokkum}, {Franx} \& {Illingworth}}]{kriek2008b}
{Kriek} M, {van der Wel} A, {van Dokkum} PG, {Franx} M, {Illingworth} GD.
  2008{\natexlab{a}}.
\newblock \textit{\apj} 682:896--906

\bibitem[{{Kriek} et~al.(2007){Kriek}, {van Dokkum}, {Franx}, {Illingworth},
  {Coppi} et~al.}]{kriek2007}
{Kriek} M, {van Dokkum} PG, {Franx} M, {Illingworth} GD, {Coppi} P, et~al.
  2007.
\newblock \textit{\apj} 669:776--790

\bibitem[{{Kriek} et~al.(2009){Kriek}, {van Dokkum}, {Franx}, {Illingworth} \&
  {Magee}}]{kriek2009}
{Kriek} M, {van Dokkum} PG, {Franx} M, {Illingworth} GD, {Magee} DK. 2009.
\newblock \textit{\apjl} 705:L71--L75

\bibitem[{{Kriek} et~al.(2008{\natexlab{b}}){Kriek}, {van Dokkum}, {Franx},
  {Illingworth}, {Marchesini} et~al.}]{kriek2008a}
{Kriek} M, {van Dokkum} PG, {Franx} M, {Illingworth} GD, {Marchesini} D, et~al.
  2008{\natexlab{b}}.
\newblock \textit{\apj} 677:219--237

\bibitem[{{Law} et~al.(2009){Law}, {Steidel}, {Erb}, {Larkin}, {Pettini}
  et~al.}]{law2009}
{Law} DR, {Steidel} CC, {Erb} DK, {Larkin} JE, {Pettini} M, et~al. 2009.
\newblock \textit{\apj} 697:2057--2082

\bibitem[{{Law} et~al.(2007){Law}, {Steidel}, {Erb}, {Pettini}, {Reddy}
  et~al.}]{law2007a}
{Law} DR, {Steidel} CC, {Erb} DK, {Pettini} M, {Reddy} NA, et~al. 2007.
\newblock \textit{\apj} 656:1--26

\bibitem[{{Le F{\`e}vre} et~al.(2005){Le F{\`e}vre}, {Vettolani}, {Garilli},
  {Tresse}, {Bottini} et~al.}]{lefevre2005}
{Le F{\`e}vre} O, {Vettolani} G, {Garilli} B, {Tresse} L, {Bottini} D, et~al.
  2005.
\newblock \textit{\aap} 439:845--862

\bibitem[{{Le Floc'h} et~al.(2005){Le Floc'h}, {Papovich}, {Dole}, {Bell},
  {Lagache} et~al.}]{lefloch2005}
{Le Floc'h} E, {Papovich} C, {Dole} H, {Bell} EF, {Lagache} G, et~al. 2005.
\newblock \textit{\apj} 632:169--190

\bibitem[{{Lehnert} et~al.(2009){Lehnert}, {Nesvadba}, {Tiran}, {Matteo}, {van
  Driel} et~al.}]{lehnert2009}
{Lehnert} MD, {Nesvadba} NPH, {Tiran} LL, {Matteo} PD, {van Driel} W, et~al.
  2009.
\newblock \textit{\apj} 699:1660--1678

\bibitem[{{Leitherer} et~al.(1999){Leitherer}, {Schaerer}, {Goldader},
  {Gonz{\'a}lez Delgado}, {Robert} et~al.}]{leitherer1999}
{Leitherer} C, {Schaerer} D, {Goldader} JD, {Gonz{\'a}lez Delgado} RM, {Robert}
  C, et~al. 1999.
\newblock \textit{\apjs} 123:3--40

\bibitem[{{Lilly} et~al.(1996){Lilly}, {Le Fevre}, {Hammer} \&
  {Crampton}}]{lilly1996}
{Lilly} SJ, {Le Fevre} O, {Hammer} F, {Crampton} D. 1996.
\newblock \textit{\apjl} 460:L1+

\bibitem[{{Lotz} et~al.(2006){Lotz}, {Madau}, {Giavalisco}, {Primack} \&
  {Ferguson}}]{lotz2006}
{Lotz} JM, {Madau} P, {Giavalisco} M, {Primack} J, {Ferguson} HC. 2006.
\newblock \textit{\apj} 636:592--609

\bibitem[{{Lotz}, {Primack} \& {Madau}(2004)}]{lotz2004}
{Lotz} JM, {Primack} J, {Madau} P. 2004.
\newblock \textit{\aj} 128:163--182

\bibitem[{{Lowenthal} et~al.(1997){Lowenthal}, {Koo}, {Guzman}, {Gallego},
  {Phillips} et~al.}]{lowenthal1997}
{Lowenthal} JD, {Koo} DC, {Guzman} R, {Gallego} J, {Phillips} AC, et~al. 1997.
\newblock \textit{\apj} 481:673--+

\bibitem[{{Madau} et~al.(1996){Madau}, {Ferguson}, {Dickinson}, {Giavalisco},
  {Steidel} \& {Fruchter}}]{madau1996}
{Madau} P, {Ferguson} HC, {Dickinson} ME, {Giavalisco} M, {Steidel} CC,
  {Fruchter} A. 1996.
\newblock \textit{\mnras} 283:1388--1404

\bibitem[{{Magdis} et~al.(2010{\natexlab{a}}){Magdis}, {Elbaz}, {Daddi},
  {Morrison}, {Dickinson} et~al.}]{magdis2010a}
{Magdis} GE, {Elbaz} D, {Daddi} E, {Morrison} GE, {Dickinson} M, et~al.
  2010{\natexlab{a}}.
\newblock \textit{\apj} 714:1740--1745

\bibitem[{{Magdis} et~al.(2010{\natexlab{b}}){Magdis}, {Elbaz}, {Hwang},
  {Amblard}, {Arumugam} et~al.}]{magdis2010b}
{Magdis} GE, {Elbaz} D, {Hwang} HS, {Amblard} A, {Arumugam} V, et~al.
  2010{\natexlab{b}}.
\newblock \textit{\mnras} 409:22--28

\bibitem[{{Maiolino} et~al.(2008){Maiolino}, {Nagao}, {Grazian}, {Cocchia},
  {Marconi} et~al.}]{maiolino2008}
{Maiolino} R, {Nagao} T, {Grazian} A, {Cocchia} F, {Marconi} A, et~al. 2008.
\newblock \textit{\aap} 488:463--479

\bibitem[{{Mannucci} et~al.(2010){Mannucci}, {Cresci}, {Maiolino}, {Marconi} \&
  {Gnerucci}}]{mannucci2010}
{Mannucci} F, {Cresci} G, {Maiolino} R, {Marconi} A, {Gnerucci} A. 2010.
\newblock \textit{\mnras} 408:2115--2127

\bibitem[{{Mannucci} et~al.(2009){Mannucci}, {Cresci}, {Maiolino}, {Marconi},
  {Pastorini} et~al.}]{mannucci2009}
{Mannucci} F, {Cresci} G, {Maiolino} R, {Marconi} A, {Pastorini} G, et~al.
  2009.
\newblock \textit{\mnras} 398:1915--1931

\bibitem[{{Maraston}(2005)}]{maraston2005}
{Maraston} C. 2005.
\newblock \textit{\mnras} 362:799--825

\bibitem[{{Maraston} et~al.(2006){Maraston}, {Daddi}, {Renzini}, {Cimatti},
  {Dickinson} et~al.}]{maraston2006}
{Maraston} C, {Daddi} E, {Renzini} A, {Cimatti} A, {Dickinson} M, et~al. 2006.
\newblock \textit{\apj} 652:85--96

\bibitem[{{Maraston} et~al.(2010){Maraston}, {Pforr}, {Renzini}, {Daddi},
  {Dickinson} et~al.}]{maraston2010}
{Maraston} C, {Pforr} J, {Renzini} A, {Daddi} E, {Dickinson} M, et~al. 2010.
\newblock \textit{\mnras} 407:830--845

\bibitem[{{Marchesini} et~al.(2007){Marchesini}, {van Dokkum}, {Quadri},
  {Rudnick}, {Franx} et~al.}]{marchesini2007}
{Marchesini} D, {van Dokkum} P, {Quadri} R, {Rudnick} G, {Franx} M, et~al.
  2007.
\newblock \textit{\apj} 656:42--65

\bibitem[{{Marchesini} et~al.(2009){Marchesini}, {van Dokkum}, {F{\"o}rster
  Schreiber}, {Franx}, {Labb{\'e}} \& {Wuyts}}]{marchesini2009}
{Marchesini} D, {van Dokkum} PG, {F{\"o}rster Schreiber} NM, {Franx} M,
  {Labb{\'e}} I, {Wuyts} S. 2009.
\newblock \textit{\apj} 701:1765--1796

\bibitem[{{Men{\'e}ndez-Delmestre} et~al.(2009){Men{\'e}ndez-Delmestre},
  {Blain}, {Smail}, {Alexander}, {Chapman} et~al.}]{menendez2009}
{Men{\'e}ndez-Delmestre} K, {Blain} AW, {Smail} I, {Alexander} DM, {Chapman}
  SC, et~al. 2009.
\newblock \textit{\apj} 699:667--685

\bibitem[{{Meurer}, {Heckman} \& {Calzetti}(1999)}]{meurer1999}
{Meurer} GR, {Heckman} TM, {Calzetti} D. 1999.
\newblock \textit{\apj} 521:64--80

\bibitem[{{Micha{\l}owski}, {Hjorth} \& {Watson}(2010)}]{michalowski2010}
{Micha{\l}owski} M, {Hjorth} J, {Watson} D. 2010.
\newblock \textit{\aap} 514:A67+

\bibitem[{{Muzzin} et~al.(2009){Muzzin}, {Marchesini}, {van Dokkum},
  {Labb{\'e}}, {Kriek} \& {Franx}}]{muzzin2009}
{Muzzin} A, {Marchesini} D, {van Dokkum} PG, {Labb{\'e}} I, {Kriek} M, {Franx}
  M. 2009.
\newblock \textit{\apj} 701:1839--1864

\bibitem[{{Naab}, {Johansson} \& {Ostriker}(2009)}]{naab2009}
{Naab} T, {Johansson} PH, {Ostriker} JP. 2009.
\newblock \textit{\apjl} 699:L178--L182

\bibitem[{{Nagao}, {Maiolino} \& {Marconi}(2006)}]{nagao2006}
{Nagao} T, {Maiolino} R, {Marconi} A. 2006.
\newblock \textit{\aap} 459:85--101

\bibitem[{{Neri} et~al.(2003){Neri}, {Genzel}, {Ivison}, {Bertoldi}, {Blain}
  et~al.}]{neri2003}
{Neri} R, {Genzel} R, {Ivison} RJ, {Bertoldi} F, {Blain} AW, et~al. 2003.
\newblock \textit{\apjl} 597:L113--L116

\bibitem[{{Nilsson} et~al.(2011){Nilsson}, {{\"O}stlin}, {M{\o}ller},
  {M{\"o}ller-Nilsson}, {Tapken} et~al.}]{nilsson2011}
{Nilsson} KK, {{\"O}stlin} G, {M{\o}ller} P, {M{\"o}ller-Nilsson} O, {Tapken}
  C, et~al. 2011.
\newblock \textit{\aap} 529:A9+

\bibitem[{{Noeske} et~al.(2007){Noeske}, {Weiner}, {Faber}, {Papovich}, {Koo}
  et~al.}]{noeske2007}
{Noeske} KG, {Weiner} BJ, {Faber} SM, {Papovich} C, {Koo} DC, et~al. 2007.
\newblock \textit{\apjl} 660:L43--L46

\bibitem[{{Nordon} et~al.(2010){Nordon}, {Lutz}, {Shao}, {Magnelli}, {Berta}
  et~al.}]{nordon2010}
{Nordon} R, {Lutz} D, {Shao} L, {Magnelli} B, {Berta} S, et~al. 2010.
\newblock \textit{\aap} 518:L24+

\bibitem[{{Oesch} et~al.(2010){Oesch}, {Bouwens}, {Illingworth}, {Carollo},
  {Franx} et~al.}]{oesch2010}
{Oesch} PA, {Bouwens} RJ, {Illingworth} GD, {Carollo} CM, {Franx} M, et~al.
  2010.
\newblock \textit{\apjl} 709:L16--L20

\bibitem[{{Omont} et~al.(1996){Omont}, {Petitjean}, {Guilloteau}, {McMahon},
  {Solomon} \& {P{\'e}contal}}]{omont1996}
{Omont} A, {Petitjean} P, {Guilloteau} S, {McMahon} RG, {Solomon} PM,
  {P{\'e}contal} E. 1996.
\newblock \textit{\nat} 382:428--431

\bibitem[{{Ouchi} et~al.(2008){Ouchi}, {Shimasaku}, {Akiyama}, {Simpson},
  {Saito} et~al.}]{ouchi2008}
{Ouchi} M, {Shimasaku} K, {Akiyama} M, {Simpson} C, {Saito} T, et~al. 2008.
\newblock \textit{\apjs} 176:301--330

\bibitem[{{Ouchi} et~al.(2004){Ouchi}, {Shimasaku}, {Okamura}, {Furusawa},
  {Kashikawa} et~al.}]{ouchi2004a}
{Ouchi} M, {Shimasaku} K, {Okamura} S, {Furusawa} H, {Kashikawa} N, et~al.
  2004.
\newblock \textit{\apj} 611:660--684

\bibitem[{{Overzier} et~al.(2010){Overzier}, {Heckman}, {Schiminovich},
  {Basu-Zych}, {Gon{\c c}alves} et~al.}]{overzier2010}
{Overzier} RA, {Heckman} TM, {Schiminovich} D, {Basu-Zych} A, {Gon{\c c}alves}
  T, et~al. 2010.
\newblock \textit{\apj} 710:979--991

\bibitem[{{Paltani} et~al.(2007){Paltani}, {Le F{\`e}vre}, {Ilbert}, {Arnouts},
  {Bardelli} et~al.}]{paltani2007}
{Paltani} S, {Le F{\`e}vre} O, {Ilbert} O, {Arnouts} S, {Bardelli} S, et~al.
  2007.
\newblock \textit{\aap} 463:873--882

\bibitem[{{Pannella} et~al.(2009){Pannella}, {Carilli}, {Daddi}, {McCracken},
  {Owen} et~al.}]{pannella2009}
{Pannella} M, {Carilli} CL, {Daddi} E, {McCracken} HJ, {Owen} FN, et~al. 2009.
\newblock \textit{\apjl} 698:L116--L120

\bibitem[{{Papovich}, {Dickinson} \& {Ferguson}(2001)}]{papovich2001}
{Papovich} C, {Dickinson} M, {Ferguson} HC. 2001.
\newblock \textit{\apj} 559:620--653

\bibitem[{{Papovich} et~al.(2005){Papovich}, {Dickinson}, {Giavalisco},
  {Conselice} \& {Ferguson}}]{papovich2005}
{Papovich} C, {Dickinson} M, {Giavalisco} M, {Conselice} CJ, {Ferguson} HC.
  2005.
\newblock \textit{\apj} 631:101--120

\bibitem[{{Papovich} et~al.(2011){Papovich}, {Finkelstein}, {Ferguson}, {Lotz}
  \& {Giavalisco}}]{papovich2011}
{Papovich} C, {Finkelstein} SL, {Ferguson} HC, {Lotz} JM, {Giavalisco} M. 2011.
\newblock \textit{\mnras} 412:1123--1136

\bibitem[{{Papovich} et~al.(2006){Papovich}, {Moustakas}, {Dickinson}, {Le
  Floc'h}, {Rieke} et~al.}]{papovich2006}
{Papovich} C, {Moustakas} LA, {Dickinson} M, {Le Floc'h} E, {Rieke} GH, et~al.
  2006.
\newblock \textit{\apj} 640:92--113

\bibitem[{{P{\'e}rez-Gonz{\'a}lez} et~al.(2005){P{\'e}rez-Gonz{\'a}lez},
  {Rieke}, {Egami}, {Alonso-Herrero}, {Dole} et~al.}]{perezgonzalez2005}
{P{\'e}rez-Gonz{\'a}lez} PG, {Rieke} GH, {Egami} E, {Alonso-Herrero} A, {Dole}
  H, et~al. 2005.
\newblock \textit{\apj} 630:82--107

\bibitem[{{P{\'e}rez-Gonz{\'a}lez} et~al.(2008){P{\'e}rez-Gonz{\'a}lez},
  {Rieke}, {Villar}, {Barro}, {Blaylock} et~al.}]{perezgonzalez2008}
{P{\'e}rez-Gonz{\'a}lez} PG, {Rieke} GH, {Villar} V, {Barro} G, {Blaylock} M,
  et~al. 2008.
\newblock \textit{\apj} 675:234--261

\bibitem[{{Pettini} et~al.(2000){Pettini}, {Ellison}, {Steidel}, {Shapley} \&
  {Bowen}}]{pettini2000}
{Pettini} M, {Ellison} SL, {Steidel} CC, {Shapley} AE, {Bowen} DV. 2000.
\newblock \textit{\apj} 532:65--76

\bibitem[{{Pettini} \& {Pagel}(2004)}]{pettinipagel2004}
{Pettini} M, {Pagel} BEJ. 2004.
\newblock \textit{\mnras} 348:L59--L63

\bibitem[{{Pettini} et~al.(2002){Pettini}, {Rix}, {Steidel}, {Adelberger},
  {Hunt} \& {Shapley}}]{pettini2002}
{Pettini} M, {Rix} SA, {Steidel} CC, {Adelberger} KL, {Hunt} MP, {Shapley} AE.
  2002.
\newblock \textit{\apj} 569:742--757

\bibitem[{{Pettini} et~al.(2001){Pettini}, {Shapley}, {Steidel}, {Cuby},
  {Dickinson} et~al.}]{pettini2001}
{Pettini} M, {Shapley} AE, {Steidel} CC, {Cuby} J, {Dickinson} M, et~al. 2001.
\newblock \textit{\apj} 554:981--1000

\bibitem[{{Pope} et~al.(2008){Pope}, {Chary}, {Alexander}, {Armus}, {Dickinson}
  et~al.}]{pope2008}
{Pope} A, {Chary} R, {Alexander} DM, {Armus} L, {Dickinson} M, et~al. 2008.
\newblock \textit{\apj} 675:1171--1193

\bibitem[{{Pozzetti} et~al.(2007){Pozzetti}, {Bolzonella}, {Lamareille},
  {Zamorani}, {Franzetti} et~al.}]{pozzetti2007}
{Pozzetti} L, {Bolzonella} M, {Lamareille} F, {Zamorani} G, {Franzetti} P,
  et~al. 2007.
\newblock \textit{\aap} 474:443--459

\bibitem[{{Prevot} et~al.(1984){Prevot}, {Lequeux}, {Prevot}, {Maurice} \&
  {Rocca-Volmerange}}]{prevot1984}
{Prevot} ML, {Lequeux} J, {Prevot} L, {Maurice} E, {Rocca-Volmerange} B. 1984.
\newblock \textit{\aap} 132:389--392

\bibitem[{{Quider} et~al.(2009){Quider}, {Pettini}, {Shapley} \&
  {Steidel}}]{quider2009}
{Quider} AM, {Pettini} M, {Shapley} AE, {Steidel} CC. 2009.
\newblock \textit{\mnras} 398:1263--1278

\bibitem[{{Quider} et~al.(2010){Quider}, {Shapley}, {Pettini}, {Steidel} \&
  {Stark}}]{quider2010}
{Quider} AM, {Shapley} AE, {Pettini} M, {Steidel} CC, {Stark} DP. 2010.
\newblock \textit{\mnras} 402:1467--1479

\bibitem[{{Ranalli}, {Comastri} \& {Setti}(2003)}]{ranalli2003}
{Ranalli} P, {Comastri} A, {Setti} G. 2003.
\newblock \textit{\aap} 399:39--50

\bibitem[{{Ravindranath} et~al.(2006){Ravindranath}, {Giavalisco}, {Ferguson},
  {Conselice}, {Katz} et~al.}]{ravindranath2006}
{Ravindranath} S, {Giavalisco} M, {Ferguson} HC, {Conselice} C, {Katz} N,
  et~al. 2006.
\newblock \textit{\apj} 652:963--980

\bibitem[{{Reddy} et~al.(2010){Reddy}, {Erb}, {Pettini}, {Steidel} \&
  {Shapley}}]{reddy2010}
{Reddy} NA, {Erb} DK, {Pettini} M, {Steidel} CC, {Shapley} AE. 2010.
\newblock \textit{\apj} 712:1070--1091

\bibitem[{{Reddy} et~al.(2005){Reddy}, {Erb}, {Steidel}, {Shapley},
  {Adelberger} \& {Pettini}}]{reddy2005}
{Reddy} NA, {Erb} DK, {Steidel} CC, {Shapley} AE, {Adelberger} KL, {Pettini} M.
  2005.
\newblock \textit{\apj} 633:748--767

\bibitem[{{Reddy} \& {Steidel}(2004)}]{reddy2004}
{Reddy} NA, {Steidel} CC. 2004.
\newblock \textit{\apjl} 603:L13--L16

\bibitem[{{Reddy} \& {Steidel}(2009)}]{reddysteidel2009}
{Reddy} NA, {Steidel} CC. 2009.
\newblock \textit{\apj} 692:778--803

\bibitem[{{Reddy} et~al.(2006){Reddy}, {Steidel}, {Fadda}, {Yan}, {Pettini}
  et~al.}]{reddy2006}
{Reddy} NA, {Steidel} CC, {Fadda} D, {Yan} L, {Pettini} M, et~al. 2006.
\newblock \textit{\apj} 644:792--812

\bibitem[{{Reddy} et~al.(2008){Reddy}, {Steidel}, {Pettini}, {Adelberger},
  {Shapley} et~al.}]{reddy2008}
{Reddy} NA, {Steidel} CC, {Pettini} M, {Adelberger} KL, {Shapley} AE, et~al.
  2008.
\newblock \textit{\apjs} 175:48--85

\bibitem[{{Rhoads} et~al.(2000){Rhoads}, {Malhotra}, {Dey}, {Stern}, {Spinrad}
  \& {Jannuzi}}]{rhoads2000}
{Rhoads} JE, {Malhotra} S, {Dey} A, {Stern} D, {Spinrad} H, {Jannuzi} BT. 2000.
\newblock \textit{\apjl} 545:L85--L88

\bibitem[{{Rieke} et~al.(2009){Rieke}, {Alonso-Herrero}, {Weiner},
  {P{\'e}rez-Gonz{\'a}lez}, {Blaylock} et~al.}]{rieke2009}
{Rieke} GH, {Alonso-Herrero} A, {Weiner} BJ, {P{\'e}rez-Gonz{\'a}lez} PG,
  {Blaylock} M, et~al. 2009.
\newblock \textit{\apj} 692:556--573

\bibitem[{{Rix} et~al.(2004){Rix}, {Pettini}, {Leitherer}, {Bresolin},
  {Kudritzki} \& {Steidel}}]{rix2004}
{Rix} SA, {Pettini} M, {Leitherer} C, {Bresolin} F, {Kudritzki} R, {Steidel}
  CC. 2004.
\newblock \textit{\apj} 615:98--117

\bibitem[{{Rodighiero} et~al.(2010){Rodighiero}, {Vaccari}, {Franceschini},
  {Tresse}, {Le Fevre} et~al.}]{rodighiero2010}
{Rodighiero} G, {Vaccari} M, {Franceschini} A, {Tresse} L, {Le Fevre} O, et~al.
  2010.
\newblock \textit{\aap} 515:A8+

\bibitem[{{Sajina} et~al.(2007){Sajina}, {Yan}, {Armus}, {Choi}, {Fadda}
  et~al.}]{sajina2007}
{Sajina} A, {Yan} L, {Armus} L, {Choi} P, {Fadda} D, et~al. 2007.
\newblock \textit{\apj} 664:713--737

\bibitem[{{Salpeter}(1955)}]{salpeter1955}
{Salpeter} EE. 1955.
\newblock \textit{\apj} 121:161--+

\bibitem[{{Sawicki}(2001)}]{sawicki2001}
{Sawicki} M. 2001.
\newblock \textit{\aj} 121:2405--2412

\bibitem[{{Sawicki} \& {Thompson}(2006)}]{sawicki2006a}
{Sawicki} M, {Thompson} D. 2006.
\newblock \textit{\apj} 642:653--672

\bibitem[{{Sawicki} \& {Yee}(1998)}]{sawicki1998}
{Sawicki} M, {Yee} HKC. 1998.
\newblock \textit{\aj} 115:1329--1339

\bibitem[{{Schechter}(1976)}]{schechter1976}
{Schechter} P. 1976.
\newblock \textit{\apj} 203:297--306

\bibitem[{{Shapiro} et~al.(2009){Shapiro}, {Genzel}, {Quataert}, {F{\"o}rster
  Schreiber}, {Davies} et~al.}]{shapiro2009}
{Shapiro} KL, {Genzel} R, {Quataert} E, {F{\"o}rster Schreiber} NM, {Davies} R,
  et~al. 2009.
\newblock \textit{\apj} 701:955--963

\bibitem[{{Shapley} et~al.(2004){Shapley}, {Erb}, {Pettini}, {Steidel} \&
  {Adelberger}}]{shapley2004}
{Shapley} AE, {Erb} DK, {Pettini} M, {Steidel} CC, {Adelberger} KL. 2004.
\newblock \textit{\apj} 612:108--121

\bibitem[{{Shapley} et~al.(2001){Shapley}, {Steidel}, {Adelberger},
  {Dickinson}, {Giavalisco} \& {Pettini}}]{shapley2001}
{Shapley} AE, {Steidel} CC, {Adelberger} KL, {Dickinson} M, {Giavalisco} M,
  {Pettini} M. 2001.
\newblock \textit{\apj} 562:95--123

\bibitem[{{Shapley} et~al.(2005){Shapley}, {Steidel}, {Erb}, {Reddy},
  {Adelberger} et~al.}]{shapley2005}
{Shapley} AE, {Steidel} CC, {Erb} DK, {Reddy} NA, {Adelberger} KL, et~al. 2005.
\newblock \textit{\apj} 626:698--722

\bibitem[{{Shapley} et~al.(2003){Shapley}, {Steidel}, {Pettini} \&
  {Adelberger}}]{shapley2003}
{Shapley} AE, {Steidel} CC, {Pettini} M, {Adelberger} KL. 2003.
\newblock \textit{\apj} 588:65--89

\bibitem[{{Shen} et~al.(2003){Shen}, {Mo}, {White}, {Blanton}, {Kauffmann}
  et~al.}]{shen2003}
{Shen} S, {Mo} HJ, {White} SDM, {Blanton} MR, {Kauffmann} G, et~al. 2003.
\newblock \textit{\mnras} 343:978--994

\bibitem[{{Siana} et~al.(2009){Siana}, {Smail}, {Swinbank}, {Richard},
  {Teplitz} et~al.}]{siana2009}
{Siana} B, {Smail} I, {Swinbank} AM, {Richard} J, {Teplitz} HI, et~al. 2009.
\newblock \textit{\apj} 698:1273--1281

\bibitem[{{Siana} et~al.(2008){Siana}, {Teplitz}, {Chary}, {Colbert} \&
  {Frayer}}]{siana2008}
{Siana} B, {Teplitz} HI, {Chary} R, {Colbert} J, {Frayer} DT. 2008.
\newblock \textit{\apj} 689:59--67

\bibitem[{{Smail}, {Ivison} \& {Blain}(1997)}]{smail1997}
{Smail} I, {Ivison} RJ, {Blain} AW. 1997.
\newblock \textit{\apjl} 490:L5+

\bibitem[{{Smail} et~al.(2007){Smail}, {Swinbank}, {Richard}, {Ebeling},
  {Kneib} et~al.}]{smail2007}
{Smail} I, {Swinbank} AM, {Richard} J, {Ebeling} H, {Kneib} J, et~al. 2007.
\newblock \textit{\apjl} 654:L33--L36

\bibitem[{{Somerville} et~al.(2008){Somerville}, {Hopkins}, {Cox}, {Robertson}
  \& {Hernquist}}]{somerville2008}
{Somerville} RS, {Hopkins} PF, {Cox} TJ, {Robertson} BE, {Hernquist} L. 2008.
\newblock \textit{\mnras} 391:481--506

\bibitem[{{Spergel} et~al.(2003){Spergel}, {Verde}, {Peiris}, {Komatsu},
  {Nolta} et~al.}]{spergel2003}
{Spergel} DN, {Verde} L, {Peiris} HV, {Komatsu} E, {Nolta} MR, et~al. 2003.
\newblock \textit{\apjs} 148:175--194

\bibitem[{{Springel} et~al.(2005){Springel}, {White}, {Jenkins}, {Frenk},
  {Yoshida} et~al.}]{springel2005}
{Springel} V, {White} SDM, {Jenkins} A, {Frenk} CS, {Yoshida} N, et~al. 2005.
\newblock \textit{\nat} 435:629--636

\bibitem[{{Steidel} et~al.(1998){Steidel}, {Adelberger}, {Dickinson},
  {Giavalisco}, {Pettini} \& {Kellogg}}]{steidel1998}
{Steidel} CC, {Adelberger} KL, {Dickinson} M, {Giavalisco} M, {Pettini} M,
  {Kellogg} M. 1998.
\newblock \textit{\apj} 492:428--+

\bibitem[{{Steidel} et~al.(1999){Steidel}, {Adelberger}, {Giavalisco},
  {Dickinson} \& {Pettini}}]{steidel1999}
{Steidel} CC, {Adelberger} KL, {Giavalisco} M, {Dickinson} M, {Pettini} M.
  1999.
\newblock \textit{\apj} 519:1--17

\bibitem[{{Steidel} et~al.(2005){Steidel}, {Adelberger}, {Shapley}, {Erb},
  {Reddy} \& {Pettini}}]{steidel2005}
{Steidel} CC, {Adelberger} KL, {Shapley} AE, {Erb} DK, {Reddy} NA, {Pettini} M.
  2005.
\newblock \textit{\apj} 626:44--50

\bibitem[{{Steidel} et~al.(2003){Steidel}, {Adelberger}, {Shapley}, {Pettini},
  {Dickinson} \& {Giavalisco}}]{steidel2003}
{Steidel} CC, {Adelberger} KL, {Shapley} AE, {Pettini} M, {Dickinson} M,
  {Giavalisco} M. 2003.
\newblock \textit{\apj} 592:728--754

\bibitem[{{Steidel} et~al.(2010){Steidel}, {Erb}, {Shapley}, {Pettini}, {Reddy}
  et~al.}]{steidel2010}
{Steidel} CC, {Erb} DK, {Shapley} AE, {Pettini} M, {Reddy} N, et~al. 2010.
\newblock \textit{\apj} 717:289--322

\bibitem[{{Steidel} et~al.(1996){Steidel}, {Giavalisco}, {Pettini}, {Dickinson}
  \& {Adelberger}}]{steidel1996b}
{Steidel} CC, {Giavalisco} M, {Pettini} M, {Dickinson} M, {Adelberger} KL.
  1996.
\newblock \textit{\apjl} 462:L17

\bibitem[{{Steidel} et~al.(2004){Steidel}, {Shapley}, {Pettini}, {Adelberger},
  {Erb} et~al.}]{steidel2004}
{Steidel} CC, {Shapley} AE, {Pettini} M, {Adelberger} KL, {Erb} DK, et~al.
  2004.
\newblock \textit{\apj} 604:534--550

\bibitem[{{Stolte} et~al.(2002){Stolte}, {Grebel}, {Brandner} \&
  {Figer}}]{stolte2002}
{Stolte} A, {Grebel} EK, {Brandner} W, {Figer} DF. 2002.
\newblock \textit{\aap} 394:459--478

\bibitem[{{Strateva} et~al.(2001){Strateva}, {Ivezi{\'c}}, {Knapp},
  {Narayanan}, {Strauss} et~al.}]{strateva2001}
{Strateva} I, {Ivezi{\'c}} {\v Z}, {Knapp} GR, {Narayanan} VK, {Strauss} MA,
  et~al. 2001.
\newblock \textit{\aj} 122:1861--1874

\bibitem[{{Swinbank} et~al.(2006){Swinbank}, {Chapman}, {Smail}, {Lindner},
  {Borys} et~al.}]{swinbank2006}
{Swinbank} AM, {Chapman} SC, {Smail} I, {Lindner} C, {Borys} C, et~al. 2006.
\newblock \textit{\mnras} 371:465--476

\bibitem[{{Swinbank} et~al.(2004){Swinbank}, {Smail}, {Chapman}, {Blain},
  {Ivison} \& {Keel}}]{swinbank2004}
{Swinbank} AM, {Smail} I, {Chapman} SC, {Blain} AW, {Ivison} RJ, {Keel} WC.
  2004.
\newblock \textit{\apj} 617:64--80

\bibitem[{{Swinbank} et~al.(2010){Swinbank}, {Smail}, {Chapman}, {Borys},
  {Alexander} et~al.}]{swinbank2010}
{Swinbank} AM, {Smail} I, {Chapman} SC, {Borys} C, {Alexander} DM, et~al. 2010.
\newblock \textit{\mnras} 405:234--244

\bibitem[{{Tacconi} et~al.(2010){Tacconi}, {Genzel}, {Neri}, {Cox}, {Cooper}
  et~al.}]{tacconi2010}
{Tacconi} LJ, {Genzel} R, {Neri} R, {Cox} P, {Cooper} MC, et~al. 2010.
\newblock \textit{\nat} 463:781--784

\bibitem[{{Tacconi} et~al.(2008){Tacconi}, {Genzel}, {Smail}, {Neri}, {Chapman}
  et~al.}]{tacconi2008}
{Tacconi} LJ, {Genzel} R, {Smail} I, {Neri} R, {Chapman} SC, et~al. 2008.
\newblock \textit{\apj} 680:246--262

\bibitem[{{Tacconi} et~al.(2006){Tacconi}, {Neri}, {Chapman}, {Genzel}, {Smail}
  et~al.}]{tacconi2006}
{Tacconi} LJ, {Neri} R, {Chapman} SC, {Genzel} R, {Smail} I, et~al. 2006.
\newblock \textit{\apj} 640:228--240

\bibitem[{{Taylor} et~al.(2010){Taylor}, {Franx}, {Glazebrook}, {Brinchmann},
  {van der Wel} \& {van Dokkum}}]{taylor2010}
{Taylor} EN, {Franx} M, {Glazebrook} K, {Brinchmann} J, {van der Wel} A, {van
  Dokkum} PG. 2010.
\newblock \textit{\apj} 720:723--741

\bibitem[{{Thomas} et~al.(2005){Thomas}, {Maraston}, {Bender} \& {Mendes de
  Oliveira}}]{thomas2005}
{Thomas} D, {Maraston} C, {Bender} R, {Mendes de Oliveira} C. 2005.
\newblock \textit{\apj} 621:673--694

\bibitem[{{Tinsley}(1968)}]{tinsley1968}
{Tinsley} BM. 1968.
\newblock \textit{\apj} 151:547--+

\bibitem[{{Toft} et~al.(2009){Toft}, {Franx}, {van Dokkum}, {F{\"o}rster
  Schreiber}, {Labbe} et~al.}]{toft2009}
{Toft} S, {Franx} M, {van Dokkum} P, {F{\"o}rster Schreiber} NM, {Labbe} I,
  et~al. 2009.
\newblock \textit{\apj} 705:255--260

\bibitem[{{Toft} et~al.(2007){Toft}, {van Dokkum}, {Franx}, {Labbe},
  {F{\"o}rster Schreiber} et~al.}]{toft2007}
{Toft} S, {van Dokkum} P, {Franx} M, {Labbe} I, {F{\"o}rster Schreiber} NM,
  et~al. 2007.
\newblock \textit{\apj} 671:285--302

\bibitem[{{Toft} et~al.(2005){Toft}, {van Dokkum}, {Franx}, {Thompson},
  {Illingworth} et~al.}]{toft2005}
{Toft} S, {van Dokkum} P, {Franx} M, {Thompson} RI, {Illingworth} GD, et~al.
  2005.
\newblock \textit{\apjl} 624:L9--L12

\bibitem[{{Tremonti} et~al.(2004){Tremonti}, {Heckman}, {Kauffmann},
  {Brinchmann}, {Charlot} et~al.}]{tremonti2004}
{Tremonti} CA, {Heckman} TM, {Kauffmann} G, {Brinchmann} J, {Charlot} S, et~al.
  2004.
\newblock \textit{\apj} 613:898--913

\bibitem[{{Treu} et~al.(2010){Treu}, {Auger}, {Koopmans}, {Gavazzi}, {Marshall}
  \& {Bolton}}]{treu2010}
{Treu} T, {Auger} MW, {Koopmans} LVE, {Gavazzi} R, {Marshall} PJ, {Bolton} AS.
  2010.
\newblock \textit{\apj} 709:1195--1202

\bibitem[{{Trujillo} et~al.(2007){Trujillo}, {Conselice}, {Bundy}, {Cooper},
  {Eisenhardt} \& {Ellis}}]{trujillo2007}
{Trujillo} I, {Conselice} CJ, {Bundy} K, {Cooper} MC, {Eisenhardt} P, {Ellis}
  RS. 2007.
\newblock \textit{\mnras} 382:109--120

\bibitem[{{van der Burg}, {Hildebrandt} \& {Erben}(2010)}]{vanderburg2010}
{van der Burg} RFJ, {Hildebrandt} H, {Erben} T. 2010.
\newblock \textit{\aap} 523:A74+

\bibitem[{{van der Wel} et~al.(2008){van der Wel}, {Holden}, {Zirm}, {Franx},
  {Rettura} et~al.}]{vanderwel2008}
{van der Wel} A, {Holden} BP, {Zirm} AW, {Franx} M, {Rettura} A, et~al. 2008.
\newblock \textit{\apj} 688:48--58

\bibitem[{{van Dokkum}(2008)}]{vandokkum2008a}
{van Dokkum} PG. 2008.
\newblock \textit{\apj} 674:29--50

\bibitem[{{van Dokkum} \& {Conroy}(2010)}]{vandokkum2010_nature}
{van Dokkum} PG, {Conroy} C. 2010.
\newblock \textit{\nat} 468:940--942

\bibitem[{{van Dokkum} et~al.(2003){van Dokkum}, {F{\" o}rster Schreiber},
  {Franx}, {Daddi}, {Illingworth} et~al.}]{vandokkum2003}
{van Dokkum} PG, {F{\" o}rster Schreiber} NM, {Franx} M, {Daddi} E,
  {Illingworth} GD, et~al. 2003.
\newblock \textit{\apjl} 587:L83--L87

\bibitem[{{van Dokkum} et~al.(2008){van Dokkum}, {Franx}, {Kriek}, {Holden},
  {Illingworth} et~al.}]{vandokkum2008b}
{van Dokkum} PG, {Franx} M, {Kriek} M, {Holden} B, {Illingworth} GD, et~al.
  2008.
\newblock \textit{\apjl} 677:L5--L8

\bibitem[{{van Dokkum}, {Kriek} \& {Franx}(2009)}]{vandokkum2009_nature}
{van Dokkum} PG, {Kriek} M, {Franx} M. 2009.
\newblock \textit{\nat} 460:717--719

\bibitem[{{van Dokkum} et~al.(2009){van Dokkum}, {Labb{\'e}}, {Marchesini},
  {Quadri}, {Brammer} et~al.}]{vandokkum2009_pasp}
{van Dokkum} PG, {Labb{\'e}} I, {Marchesini} D, {Quadri} R, {Brammer} G, et~al.
  2009.
\newblock \textit{\pasp} 121:2--8

\bibitem[{{van Dokkum} et~al.(2010){van Dokkum}, {Whitaker}, {Brammer},
  {Franx}, {Kriek} et~al.}]{vandokkum2010}
{van Dokkum} PG, {Whitaker} KE, {Brammer} G, {Franx} M, {Kriek} M, et~al. 2010.
\newblock \textit{\apj} 709:1018--1041

\bibitem[{{Venemans} et~al.(2007){Venemans}, {R{\"o}ttgering}, {Miley}, {van
  Breugel}, {de Breuck} et~al.}]{venemans2007}
{Venemans} BP, {R{\"o}ttgering} HJA, {Miley} GK, {van Breugel} WJM, {de Breuck}
  C, et~al. 2007.
\newblock \textit{\aap} 461:823--845

\bibitem[{{Wilkins}, {Trentham} \& {Hopkins}(2008)}]{wilkins2008}
{Wilkins} SM, {Trentham} N, {Hopkins} AM. 2008.
\newblock \textit{\mnras} 385:687--694

\bibitem[{{Williams} et~al.(2009){Williams}, {Quadri}, {Franx}, {van Dokkum} \&
  {Labb{\'e}}}]{williams2009}
{Williams} RJ, {Quadri} RF, {Franx} M, {van Dokkum} P, {Labb{\'e}} I. 2009.
\newblock \textit{\apj} 691:1879--1895

\bibitem[{{Williams} et~al.(2010){Williams}, {Quadri}, {Franx}, {van Dokkum},
  {Toft} et~al.}]{williams2010}
{Williams} RJ, {Quadri} RF, {Franx} M, {van Dokkum} P, {Toft} S, et~al. 2010.
\newblock \textit{\apj} 713:738--750

\bibitem[{{Wyder} et~al.(2005){Wyder}, {Treyer}, {Milliard}, {Schiminovich},
  {Arnouts} et~al.}]{wyder2005}
{Wyder} TK, {Treyer} MA, {Milliard} B, {Schiminovich} D, {Arnouts} S, et~al.
  2005.
\newblock \textit{\apjl} 619:L15--L18

\bibitem[{{Yan} et~al.(2007){Yan}, {Sajina}, {Fadda}, {Choi}, {Armus}
  et~al.}]{yan2007}
{Yan} L, {Sajina} A, {Fadda} D, {Choi} P, {Armus} L, et~al. 2007.
\newblock \textit{\apj} 658:778--793

\bibitem[{{Yan} et~al.(2010){Yan}, {Tacconi}, {Fiolet}, {Sajina}, {Omont}
  et~al.}]{yan2010}
{Yan} L, {Tacconi} LJ, {Fiolet} N, {Sajina} A, {Omont} A, et~al. 2010.
\newblock \textit{\apj} 714:100--114

\bibitem[{{Yoshikawa} et~al.(2010){Yoshikawa}, {Akiyama}, {Kajisawa},
  {Alexander}, {Ohta} et~al.}]{yoshikawa2010}
{Yoshikawa} T, {Akiyama} M, {Kajisawa} M, {Alexander} DM, {Ohta} K, et~al.
  2010.
\newblock \textit{\apj} 718:112--132

\bibitem[{{Young} \& {Scoville}(1991)}]{young1991}
{Young} JS, {Scoville} NZ. 1991.
\newblock \textit{\araa} 29:581--625

\bibitem[{{Yun}, {Reddy} \& {Condon}(2001)}]{yun2001}
{Yun} MS, {Reddy} NA, {Condon} JJ. 2001.
\newblock \textit{\apj} 554:803--822

\bibitem[{{Zirm} et~al.(2007){Zirm}, {van der Wel}, {Franx}, {Labb{\'e}},
  {Trujillo} et~al.}]{zirm2007}
{Zirm} AW, {van der Wel} A, {Franx} M, {Labb{\'e}} I, {Trujillo} I, et~al.
  2007.
\newblock \textit{\apj} 656:66--72

\end{thebibliography}

\clearpage
\section*{Acronyms/Definitions}

\begin{enumerate}
\item LBG: ``Lyman Break Galaxy." Star-forming $z\geq 3$ galaxy selected on the basis of
its rest-frame UV colors, which are indicative of a Lyman Break -- i.e. significant
absorption at wavelengths below 912~\AA.

\item UV selection: More general use of rest-frame UV photometry to select star-forming
galaxies not only at $z\geq 3$ using the Lyman Break, but also at $1.4 \leq z \leq 2.5$ using
different rest-frame UV color criteria.

\item DRG: ``Distant Red Galaxy." Galaxy selected on the basis of a red
observed-frame $J-K$ color, which corresponds to a Balmer or 4000~\AA\ break,
or significant dust extinction, for objects at $z\sim 2-4$. The majority
of DRGs are actively forming stars, but typically with
higher $M/L$  and stellar masses than UV-selected galaxies and sBzK galaxies.

\item sBzK: ``Star-forming BzK Galaxy." Galaxy selected on the basis of
its $B-z$ and $z-K$ colors to lie at $1.4\leq z \leq 2.5$ and be
actively star-forming. The sBzK criteria are tuned to find galaxies with Balmer
breaks, ongoing star formation, and a wide range of dust extinction properties.

\item pBzK: ``Passive BzK Galaxy." Galaxy selected on the basis of
its $B-z$ and $z-K$ colors to lie at $1.4\leq z \leq 2.5$ and be
devoid of star formation. The pBzK criteria are tuned to find galaxies with prominent
4000~\AA\ breaks and a lack of current star formation.

\item SMG: ``Submillimeter Galaxy." Galaxy selected on the basis of its powerful
luminosity in the submillimeter range of the spectrum. SCUBA 850~$\mu$m selection
has commonly been used to find such systems. The prolific IR luminosity is due to
reprocessed emission from dust (powered either by stars, an active nucleus, or both). 

\item LAE: ``Ly$\alpha$ Emitter." Galaxy selected on the basis of having
strong emission in the Ly$\alpha$ feature at rest-frame 1216~\AA. As a result
of the selection based on line strength, LAEs tend to be significantly fainter
in the rest-frame UV continuum than other star-forming sources at the same redshift.

\item LIRG: ``Luminous Infrared Galaxy." Galaxy characterized by 
$10^{11} L_{\odot} < L_{IR} \leq  10^{12} L_{\odot}$. Such systems,
along with ULIRGs, are much more common at $z\sim 2$ than in the local
Universe.

\item ULIRG: ``Ultra-luminous Infrared Galaxy." Galaxy characterized by 
$L_{IR} > 10^{12} L_{\odot}$. Such systems,
along with LIRGs, are much more common at $z\sim 2$ than in the local
Universe.

\end{enumerate}

\end{document}